\documentclass[longauth]{aa} 

\usepackage{natbib,twoopt}
\usepackage[breaklinks=true]{hyperref} 
\bibpunct{(}{)}{;}{a}{}{,}             
\makeatletter
   \newcommandtwoopt{\citeads}[3][][]{\href{http://adsabs.harvard.edu/abs/#3}%
    {\def\hyper@linkstart##1##2{}%
     \let\hyper@linkend\@empty\citealp[#1][#2]{#3}}}
   \newcommandtwoopt{\citepads}[3][][]{\href{http://adsabs.harvard.edu/abs/#3}%
    {\def\hyper@linkstart##1##2{}%
     \let\hyper@linkend\@empty\citep[#1][#2]{#3}}}
   \newcommandtwoopt{\citetads}[3][][]{\href{http://adsabs.harvard.edu/abs/#3}%
    {\def\hyper@linkstart##1##2{}%
     \let\hyper@linkend\@empty\citet[#1][#2]{#3}}}
   \newcommandtwoopt{\citeyearads}[3][][]%
    {\href{http://adsabs.harvard.edu/abs/#3}
    {\def\hyper@linkstart##1##2{}%
     \let\hyper@linkend\@empty\citeyear[#1][#2]{#3}}}
\makeatother
\usepackage{graphicx}
\usepackage{txfonts}
\usepackage{lscape}
\usepackage{xcolor}
\usepackage{subfig}
\usepackage{soul}
\usepackage{romannum}
\usepackage{placeins}
\AtBeginDocument{\pagenumbering{arabic}}

 \newcommand{\ms}{m\,s$^{-1}$}
 \newcommand{\teff}{$T_{\rm eff}$}
 \newcommand{\logg}{$\log g$}
 \newcommand{\vsini}{$v\sin i$}
 \newcommand{\feh}{[Fe/H]}
 \newcommand{\kms}{\hbox{km\,s$^{-1}$}}
 \newcommand{\prot}{$P_{\rm rot}=34.4\pm0.5\, \rm d\,$}
 \newcommand{\protv}{$34.4\pm0.5$\,}

\begin{document}

   \title{Optical and near-infrared stellar activity characterization of the early M dwarf Gl~205 with SOPHIE and SPIRou}


   \author{P. Cort\'es-Zuleta \inst{\ref{inst1},\ref{LSST},\thanks{pia.cortes@lam.fr}} 
          \and
          I. Boisse \inst{\ref{inst1}} 
          \and
          B. Klein \inst{\ref{inst2}} 
          \and
          E. Martioli \inst{\ref{inst3},\ref{inst11}} 
          \and
          P. I. Cristofari \inst{\ref{inst4}} 
          \and
          A. Antoniadis-Karnavas \inst{\ref{inst5},\ref{inst6}} 
          \and
          J-F. Donati \inst{\ref{inst4}} 
          \and
          X. Delfosse \inst{\ref{inst7}} 
          \and
          C. Cadieux \inst{\ref{inst8}} 
          \and
          N. Heidari \inst{\ref{inst14},\ref{inst15},\ref{inst1}} 
          \and
          \'E. Artigau \inst{\ref{inst8},\ref{inst12}} 
          \and
          S. Bellotti \inst{\ref{inst4},\ref{inst10}} 
          \and
          X. Bonfils \inst{\ref{inst7}} 
          \and
          A. Carmona \inst{\ref{inst7}} 
          \and 
          N. J. Cook \inst{\ref{inst8}} 
          \and
          R. F. D\'iaz \inst{\ref{inst9}} 
          \and
         R. Doyon \inst{\ref{inst8},\ref{inst12}} 
          \and
          P. Fouqu\'e \inst{\ref{inst4}} 
          \and
          C. Moutou \inst{\ref{inst4}} 
          \and
         P. Petit \inst{\ref{inst4}} 
          \and
          T. Vandal \inst{\ref{inst8}} 
          \and
          L. Acu\~na \inst{\ref{inst1}} 
          \and
          L. Arnold \inst{\ref{inst18}}
          \and
          N. Astudillo-Defru \inst{\ref{inst21}}
          \and
          V. Bourrier \inst{\ref{inst19}}
          \and
          F. Bouchy \inst{\ref{inst19}}  
          \and
         R. Cloutier \inst{\ref{inst16}} 
           \and
           S. Dalal \inst{\ref{inst22}}
           \and
           M. Deleuil \inst{\ref{inst1}}
           \and
           O. D. S. Demangeon \inst{\ref{inst5},\ref{inst6}}
           \and
           X. Dumusque \inst{\ref{inst19}}
           \and
         T. Forveille \inst{\ref{inst7}} 
          \and
         J. Gomes da Silva \inst{\ref{inst5}} 
          \and
            N. Hara \inst{\ref{inst19}}
            \and
         G. H\'ebrard \inst{\ref{inst11},\ref{inst13}} 
            \and
         S. Hoyer \inst{\ref{inst1}} 
          \and
          G. Hussain\inst{\ref{inst10}}
          \and
         F. Kiefer \inst{\ref{inst11},\ref{inst20}} 
         \and
         J. Morin \inst{\ref{inst17}}
         \and
       A. Santerne \inst{\ref{inst1}}
          \and
          N. C. Santos \inst{\ref{inst5},\ref{inst6}} 
          \and
          D. Segransan \inst{\ref{inst19}}
          \and
        M. Stalport \inst{\ref{inst19}}
          \and
          S. Udry \inst{\ref{inst19}}
        }

   \institute{\label{inst1} Aix Marseille Univ, CNRS, CNES, LAM, Marseille, France
                 \and
            \label{LSST} \textit{LSSTC DSFP} Fellow
         \and
          \label{inst2} Sub-department of Astrophysics, Department of Physics, University of Oxford, Oxford, OX1 3RH, UK
          \and
          \label{inst3} Laborat\'orio Nacional de Astrof\'isica, Rua Estados Unidos 154, 37504-364, Itajub\'a - MG, Brazil
          \and
          \label{inst4} Univ. de Toulouse, CNRS, IRAP, 14 av. Belin, 31400 Toulouse, France
          \and
          \label{inst5} Instituto de Astrof\'isica e Ciências do Espaço, Universidade do Porto, CAUP, Rua das Estrelas, 4150-762 Porto, Portugal
          \and
          \label{inst6} Departamento de F\'isica e Astronomia, Faculdade de Ciências, Universidade do Porto, Rua do Campo Alegre, 4169-007 Porto, Portugal
          \and
          \label{inst7} Univ. Grenoble Alpes, CNRS, IPAG, 38000 Grenoble, France
          \and
          \label{inst8} Universit\'e de Montr\'eal, D\'epartement de Physique, IREX, Montr\'eal, QC H3C 3J7, Canada
          \and
          \label{inst9} International Center for Advanced Studies (ICAS) and ICIFI (CON- ICET), ECyT-UNSAM, Campus Miguelete, 25 de Mayo y Francia, (1650), Buenos Aires, Argentina
          \and
          \label{inst10} Science Division, Directorate of Science, 
             European Space Research and Technology Centre (ESA/ESTEC),
             Keplerlaan 1, 2201 AZ, Noordwijk, The Netherlands
          \and
          \label{inst11} Institut d'Astrophysique de Paris, CNRS, UMR 7095, Sorbonne Universit\'{e}, 98 bis bd Arago, 75014 Paris, France
          \and
          \label{inst12} Observatoire du Mont-M\'egantic, Universit\'e de Montr\'eal, Montr\'eal, QC H3C 3J7, Canada
          \and
          \label{inst13} Observatoire de Haute-Provence, CNRS, Universit\'e d'Aix-Marseille, 04870 Saint-Michel-l'Observatoire, France
          \and
          \label{inst14} Department of Physics, Shahid Beheshti University, Tehran, Iran
          \and
          \label{inst15} Laboratoire J.-L. Lagrange, Observatoire de la C\^ote d’Azur (OCA), Universite de Nice-Sophia Antipolis (UNS), CNRS, Campus Valrose, 06108 Nice Cedex 2, France
          \and
         \label{inst16} Department of Physics \& Astronomy, McMaster University, 1280 Main Street West, Hamilton, ON, L8S 4K1, Canada
         \and
         \label{inst17}LUPM, Universit\'e de Montpellier, CNRS, Place Eugène Bataillon, F-34095 Montpellier, France
         \and
         \label{inst18} Canada-France-Hawaii Telescope, CNRS, Kamuela, HI 96743, USA
         \and
         \label{inst19} Observatoire Astronomique de l’Université de Genève, Chemin Pegasi 51b, CH-1290 Versoix, Switzerland
         \and
         \label{inst20} LESIA, Observatoire de Paris, Universit\'e PSL, CNRS, Sorbonne Universit\'e, Universit\'e Paris Cit\'e, 5 place Jules Janssen, 92195 Meudon, France
         \and
         \label{inst21} Departamento de Matemática y Física Aplicadas, Universidad Católica de la Santísima Concepción, Alonso de Rivera 2850, Concepción, Chile
         \and
         \label{inst22} Astrophysics Group, University of Exeter, Exeter EX4 2QL, UK
        }

   \date{Received October 18, 2022; accepted January 17, 2023}

 
  \abstract
 {The stellar activity of M dwarfs is the main limitation for discovering and characterizing exoplanets orbiting them, since it induces quasi-periodic radial velocity (RV) variations.}
   {We aim to characterize the magnetic field and stellar activity of the early, moderately active, M dwarf Gl~205 in the optical and near-infrared (nIR) domains.}
   {We obtained high-precision quasi-simultaneous spectra in the optical and nIR  with the SOPHIE spectrograph and SPIRou spectropolarimeter between 2019 and 2022. We computed the RVs from both instruments and the SPIRou’s Stokes V profiles. We used Zeeman-Doppler imaging (ZDI) to map the large-scale magnetic field over the time-span of the observations. We studied the temporal behavior of optical and nIR RVs and activity indicators with the Lomb-Scargle periodogram and a quasi-periodic Gaussian Process (GP) regression. In the nIR, we studied the equivalent width of Al \Romannum{1}, Ti \Romannum{1}, K \Romannum{1}, Fe \Romannum{1}, and He \Romannum{1}. We modeled the activity-induced RV jitter using a multi-dimensional GP regression with activity indicators as ancillary time series.}
 {The optical and nIR RVs have similar scatter but nIR shows a more complex temporal evolution. We observe an evolution of the magnetic field topology from a poloidal dipolar field in 2019 to a dominantly toroidal field in 2022. We measured a stellar rotation period of \prot in the longitudinal magnetic field. Using ZDI we measure the amount of latitudinal differential rotation (DR) shearing the stellar surface yielding rotation periods of $P_{\rm eq}=32.0\pm1.8$ d at the stellar equator and $P_{\rm pol}=45.5\pm0.3$ d at the poles. We observed inconsistencies in the activity indicators periodicities that could be explained by these DR values. The multi-dimensional GP modeling yields an RMS of the RV residuals down to the noise level of 3 \ms\, for both instruments, using as ancillary time series H$\alpha$ and the BIS in the optical, and the FWHM in the nIR.}
  {The RV variations observed in Gl~205 are due to stellar activity with a complex evolution and different expressions in the optical and nIR, revealed thanks to an extensive follow-up. Spectropolarimetry remains the best technique to constrain the stellar rotation period over standard activity indicators particularly for moderately active M dwarfs. }

    \titlerunning{Gl~205 by SOPHIE and SPIRou}
   \maketitle
%
\section{Introduction}
M dwarfs are the most abundant stars in the Milky Way \citep{Henry2006,Reyle2021} and they have become key-targets for exoplanetary surveys \citep[e.g.,][]{Morley2017}. Their low mass favors the detection of planets orbiting them as the Doppler variations are larger than for Solar-like stars for a given planetary mass and equilibrium temperature. Dedicated surveys to discover and characterize exoplanets around M dwarfs are currently being carried out using transit photometry (e.g., MEarth: \citealt{Nutzman2008}, TRAPPIST: \citealt{Gillon2011}, SPECULOOS: \citealt{Delrez2018}, Tierras: \citealt{Garcia2020}) and Doppler spectroscopy (e.g., HPF: \citealt{Mahadevan2012}, HARPS: \citealt{Bonfils2013}, HARPS-N: \citealt{Covino2013}, IRD: \citealt{Kotani2014}, CARMENES: \citealt{Quirrenbach2018}, MAROON-X: \citealt{Seifahrt2018}, SOPHIE: \citealt{Hobson2018}, SPIRou: \citealt{Donati2020}). 

Transit and radial velocity (RV) surveys have shown that M dwarfs are Earth-like planets hosts \citep{Bonfils2013,Kopparapu2013,Dressing2013,Dressing2015,Sabotta2021,Pinamonti2022}. However, despite the efforts to detect and characterize Earth-like planets around M dwarfs, the presence of stellar activity remains one of the main limitations in this endeavor. Cool stars such as M dwarfs are known to host magnetic fields causing the phenomena known as stellar activity \citep{Reiners2012}. 

Stellar magnetic activity produces quasi-periodic RV signals with amplitudes of a few meters per second that can easily lead to a false positive exoplanet detection \citep[e.g.,][]{Queloz2001,Desidera2004,Huelamo2008,Carolo2014,Bortle2021}. Features over the stellar surface, particularly dark spots and bright plages, can survive and evolve on time scales on a few stellar rotation cycles and thus, modulate the RVs at the period of the stellar rotation \citep{Boisse2011,Scandariato2017}.

Several techniques have been developed during the last decade to mitigate the effects of stellar activity in order to improve the detection of exoplanets around active stars. When the stellar rotation period is known, one can model the RV variations induced by spots by fitting sinusoidal signals and their harmonics \citep{Boisse2011} or by simulations of the active regions \citep{Dumusque2014}. Another approach is to use activity indicators built from spectral information. These can be proxies of the shape of the cross-correlation function (CCF), for example the full width at half maximum (FWHM) and the bisector inverse slope (BIS) \citep{Queloz2001,Boisse2011}, or of the chromospheric emission in spectral lines sensitive to activity, such as the Ca II H\&K and $\rm H\alpha$ \citep{Boisse2009}. A third approach, among others, is to model the stellar surface structures generating RV variations \citep[e.g.,][]{Hebrard2016,Klein2021,Klein2022}.

The activity-induced behavior of the spectral lines in M dwarf spectra has been studied only for a handful of lines, such as $\rm H\alpha$, Ca II H\&K and IRT, Na \Romannum{1} D, and He \Romannum{1}, with the $\rm H\alpha$ and CaII H\&K  lines the most extensively explored. \citep[e.g.,][]{Gomes2011,Newton2017,Fuhrmeister2019,Lafarga2021}. The majority of the well-known activity tracers are located at optical wavelengths. Recently, some near-infrared lines have been studied, for example the He \Romannum{1} triplet \citep{Schofer2019,Fuhrmeister2019,Fuhrmeister2020} and the K \Romannum{1} line \citep{Fuhrmeister2022,Terrien2022}. The He \Romannum{1} triplet in absorption is not detected over the range of the whole M dwarf spectral type and disappears for stars later than M5. When in emission, it could be related to flaring events \citep{Fuhrmeister2019}. Moreover, \cite{Fuhrmeister2020} found that the variability in the He \Romannum{1} triplet may be correlated with the $\rm H\alpha$ variability only for active M dwarfs. In the particular case of the active M dwarf AU Mic, the He \Romannum{1} flux correlated well with the RVs \citep{Klein2021}. Regarding the K \Romannum{1} line emission, it has been found that it is rarely correlated or anti-correlated with $\rm H\alpha$ \citep{Fuhrmeister2022}. Moreover, \cite{Terrien2022} found clear signals of Zeeman broadening in this line modulated by the stellar rotation in Gl~699 and Teegarden's star data.

More recently the use of data-driven techniques namely Gaussian Processes regression (GPR) \citep[e.g.,][]{Haywood2014,Rajpaul2015,Jones2017,Gilbertson2020,Klein2021,Barragan2022,Delisle2022} or Principal Component Analysis (PCA) \citep[e.g.,][]{Davis2017,Cretignier2022} have shown good results on modelling the contribution of stellar activity in RVs. Nowadays the use of GPs has become the standard procedure for modeling and filtering stellar activity in RVs curves and transits exoplanet searches, since the GPR can easily fit the quasi-periodic activity signal. This technique is particularly successful when the information from activity indicators is used when fitting the GPs on the RV time series along with the Keplerian signal \citep{Rajpaul2015,Barragan2019,SuarezMascareno2020,Faria2022,Zicher2022}.

Spectropolarimetry data has been widely used to measure and constrain the properties of the large-scale magnetic field of M dwarfs \citep{Donati2008,Morin2008,Morin2010}. In particular, it has been shown that the longitudinal magnetic field $B_{\ell}$ is a reliable magnetic activity tracer for M dwarfs and therefore can be used to determine the stellar rotation period \citep[e.g.,][]{Morin2008,Morin2010,Folsom2016,Hebrard2016,Martioli2022}.

Another way to mitigate stellar activity is to observe in near-infrared wavelengths as for M dwarfs, the spot-induced RV jitter decreases as a function of wavelength \citep{Martin2006,Desort2007,Reiners2010,Mahmud2011}. The flux contrast between the dark spots and the stellar surface is smaller at near-infrared wavelengths than in the optical, generating less activity RV jitter. However, this effect strongly depends on the spectral type and the spot configuration on the stellar surface \citep{Reiners2010,Andersen2015}. On the contrary, the RV jitter due to the Zeeman effect increases at longer wavelengths as the induced variation is proportional to the wavelength and the magnetic field strength \citep{Hebrard2014}. In cases of low-temperature contrast, the expected gain of the near-infrared is hampered by the increasing Zeeman effect \citep{Reiners2013,Klein2020}. Simultaneous observations at the optical and near-infrared can help to disentangle the activity contribution at both wavelength domains \citep{Reiners2013,Robertson2020}. Moreover, multiwavelength observations have been crucial to reject the planetary nature of RV signals (e.g., TW Hydrae: \citealt{Huelamo2008}; AD Leo: \citealt{Carleo2020}, \citealt{Carmona2022})

Gl~205 is an early (M1.5), nearby (5.7 pc), slow-rotating M1.5 dwarf with moderate levels of activity. It has a mass of $0.55\pm0.03\,M_{\odot}$ and a radius of $0.56\pm0.03\,R_{\odot}$ \citep{Schweitzer2019}. More stellar parameters are listed in Table~\ref{tab:stellarparams}. Previously, Gl~205 has been monitored with HARPS \citep{Bonfils2013}, and with HARPS-Pol and NARVAL \citep{Hebrard2016}. Using the $ \rm H\alpha $ and Ca II H\&K indices, \citet{Bonfils2013} found the stellar rotational period to be close to 33 d, and using spectropolarimetric data, \citealt{Hebrard2016} determined a rotation period of $P_{\rm rot} = 33.63\pm0.37$ d. Photometry data in the $V$ band showed a periodic signal of 33 d and a possible long-trend magnetic cycle of $\sim1500$ d \citep{Kiraga2007}. The discrepancies between the periodicities of several activity indicators could be explained by differential rotation shearing the stellar surface and may induce a difference of $\sim10$ d in the rotation period between the equator and the pole \citep{Hebrard2016}.

In this work, we analyze the magnetic field and stellar activity of the early-M dwarf Gl~205, intensively monitored over a 2-yr period with the SOPHIE optical spectrograph and with the SPIRou near-infrared spectropolarimeter. This paper is structured as follows. In Section ~\ref{sec:datareduction} we describe the observations and the reduction of the SOPHIE and SPIRou data. Section~\ref{sec:stellarcaract} is dedicated to the stellar characterization of Gl~205 and Section~\ref{sec:phometry} to the analysis of TESS photometry. We describe the magnetic field properties using the SPIRou spectropolarimetric data in Section~\ref{sec:magfield}. In Section~\ref{sec:radialvelocities} we compare the optical and near-infrared RVs and in Section~\ref{section:activity} we describe and analyse the activity indicators. We filter the RV variations due to activity using a multi-dimensional GP framework in Section~\ref{GPfiltering}. In Section~\ref{sec:planet} we discuss our RV limit detection of Keplerian signals. In Section~\ref{sec:discussion} we discuss our results and present the principal conclusions of this work.

\begin{table}
\centering
\footnotesize
 \caption[]{\label{tab:stellarparams} Stellar parameters of Gl~205. }
 
 
\begin{tabular}{l l | c c}
 \hline \hline
  Parameter &
  Units&
  Value &
  References \\
 \hline
 RA  & h m s &  05 31 27.395785 & 1\\
 dec & d m s &  -03 40 38.024004 & 1 \\
 Parallax & mas &  $175.31 \pm 0.02$ & 1\\ 
Distance &pc  & $5.7040 \pm 0.0006$ & 1\\
Spectral type& &  M1.5V & 2 \\
$V$ & mag & 7.968 & 3\\
$B$ & mag & 9.443 & 3\\
$J$ & mag & 4.83 & 4 \\
$K$ & mag & 3.90 & 4 \\
Mass &\(M_\odot\) & $0.549 \pm 0.029$ & 5\\
Radius &\(R_\odot\) & $0.556 \pm 0.033 $ & 5\\
$ \rm L_{*}$& \(L_\odot\) & $0.061 \pm 0.006$ & 6\\

Inclination& $^{\circ}$ & $60\pm10$ & 7 \\
\vsini &$\rm km~s^{-1}$ & $0.7 \pm0.1$ & this work\\
$P_{\rm rot}$& d & \protv & this work\\
\logg &dex & $4.70 \pm 0.1$ & this work \tablefootmark{a}\\
$ \rm T_{eff,~SOPHIE}$& K & $3878 \pm 81$& this work\tablefootmark{b} \\
$ \rm T_{eff,~SPIRou}$& K & $3770 \pm 30$ & this work\tablefootmark{a} \\
$\rm [Fe/H]$ & dex  & $0.21 \pm 0.06 $ & this work\tablefootmark{b}\\
$\rm [M/H]$ & dex & $0.43\pm0.1$ & this work\tablefootmark{a} \\
$\rm [\alpha/Fe]$ & dex & $-0.08 \pm 0.04$ & this work\tablefootmark{a} \\
Age & Gyr & $3.6^{5.0}_{1.7}$ & this work\tablefootmark{a} \\

\hline

\end{tabular}
\tablebib{(1)~\citealt{GaiaDR3}; (2)~\citealt{Kiraga2007}; (3)~\citealt{Koen2010}; 
(4)~\citealt{Ducati2002};
(5)~\citealt{Schweitzer2019}; (6)~\citealt{Baraffe2015}; (7)~\citealt{Hebrard2016}}
\tablefoot{
\tablefoottext{a}{Derived from SPIRou spectra.}
\tablefoottext{b}{Derived from SOPHIE spectra.}
}
\end{table}

\section{Observations and data reduction}\label{sec:datareduction}

\begin{figure*}
\centering
    \includegraphics[width=0.99\textwidth]{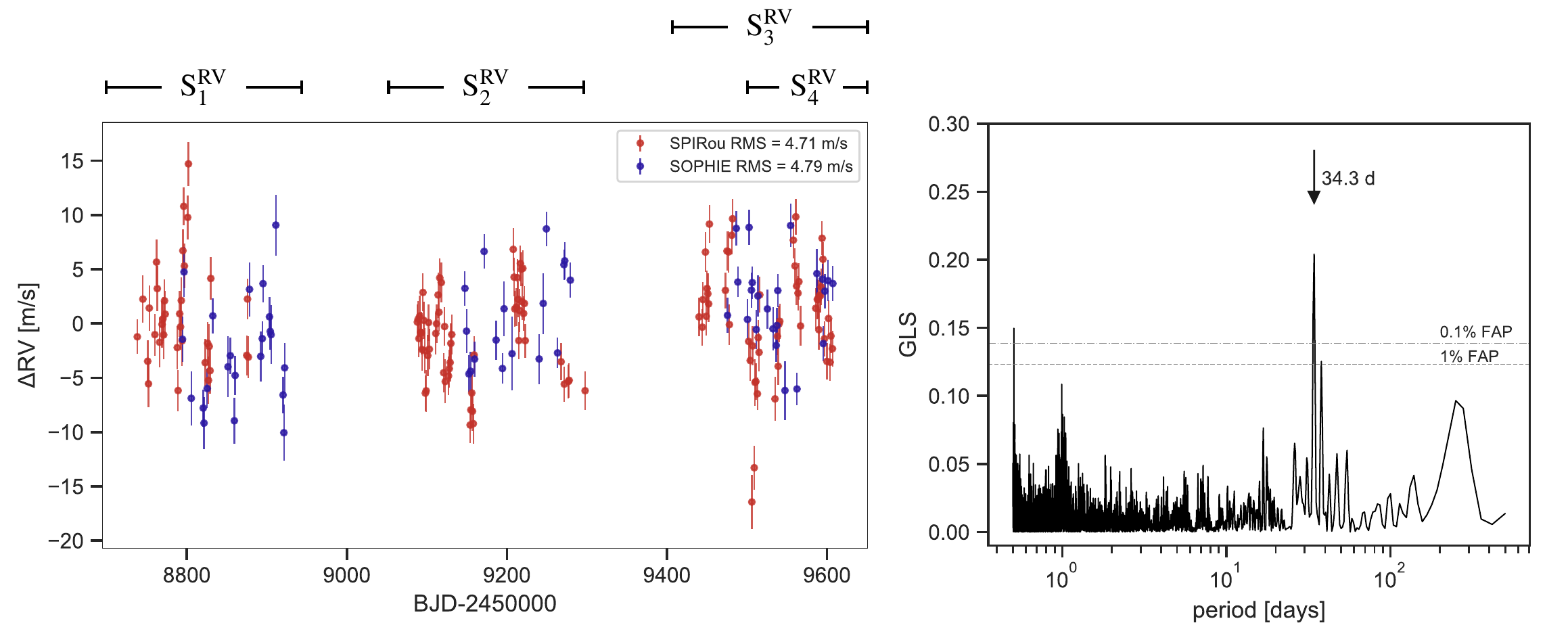}
    \caption{\emph{Left:} Radial velocity time series of Gl~205 from SOPHIE in blue and SPIRou in red with an offset equal to the mean RV of each instrument, $\rm RV_{\rm SOPHIE} = 8727.8$\,\ms\, and $\rm RV_{\rm SPIRou} = 9345.5$\,\ms. The scatter in the time series is 4.7 \ms\, in the SPIRou data set and 4.8 \ms\, in SOPHIE. The labels above the time series depict the four sub sets of observations described in Section~\ref{sec:radialvelocities}. \emph{Right:} GLS periodogram of the full RVs time series, combining SOPHIE and SPIRou observations. The highest peak is marked with an arrow at 34.3 d.}
\label{fig:radialvelocities}
\end{figure*}

\subsection{SOPHIE}
SOPHIE is a high-resolution fiber-fed, cross-dispersed échelle spectrograph mounted on the 1.93m telescope at the Observatoire de Haute-Provence (OHP) \citep{Perruchot2008,Bouchy2013}. It covers a wavelength domain from 3870 to 6940~\AA ~across 39 spectral orders.

Observations of Gl~205 were carried out between November 2019 and January 2022 as a part of the Sub-program 3 (SP3) of the SOPHIE exoplanet consortium, which is dedicated to hunt for exoplanets around M dwarfs. So far, the main results of this program include the detection of the exoplanets Gl~96b \citep{Hobson2018}, Gl~378b \citep{Hobson2019}, and Gl~411b \citep{Diaz2019}, and the independent confirmation of Gl~617Ab \citep{Hobson2018}

The observations were gathered using the high-resolution (HR) mode, on which the spectrograph reaches a resolving power of $ \rm \lambda / \Delta \lambda \sim 75000 $. In order to measure the instrumental drift, simultaneous calibrations with a Fabry-Pérot (FP) étalon were performed. In total, we gathered 74 spectra of the star, with a median exposure time of 930 s and a median signal-to-noise ratio (SNR) per pixel at 550 nm of 95. After removing the observations with airmass>1.6, SNR<80, or affected by moonlight pollution, a total of 62 observations remained. 


We used the SOPHIE Data Reduction Software \citep[DRS,][]{Bouchy2009a} to reduce and extract the spectra. The SOPHIE observations are corrected for the charge transfer inefficiency (CTI) effect of the CCD following \citet{Bouchy2009b} and \citet{Hobson2018}. The DRS automatically computes the radial velocities using the CCF technique, which is obtained by cross-correlation of the spectra with an empirical M2V mask. The DRS also uses the CCF to deliver stellar activity indicators, such as the depth of the CCF defined as CCF contrast, CCF FWHM, and BIS.

However, in the case of M dwarfs spectra the CCF method is not an optimal approach due to the large number of absorption lines. To use most of the Doppler information available, we used a template-matching algorithm to obtain high-precision radial velocities \citep[NAIRA,][]{Astudillo2015,Astudillo2017b}. First, all the available spectra are normalized by the blaze functions and scaled to the unity by their median. Second, the spectra are shifted using the RVs computed by the DRS. Third, these spectra are co-added to build a high SNR stellar template. Finally, the maximum-likelihood RV is the minimum of the Chi-square profile obtained by shifting the stellar template over an array of RVs. 

A list of standard stars was used to correct the long-term variations of the RV zero-point in SOPHIE, an effect described by \citet{Courcol2015}. To do so, we systematically monitored each possible night a group of G-type stars: HD185144, HD9407, HD89269A, and a group of M dwarfs: Gl~514, Gl~15A, and Gl~686 \citep{Hobson2018}.

The final radial velocities from NAIRA used in this work with their error bars are listed in the Table~\ref{table:SOPHIEdata}, along with the activity indicators: CCF FWHM, CCF contrast, BIS, the S index, and $\rm H\alpha$\, line (see Section~\ref{section:activity}). Figure~\ref{fig:radialvelocities} shows the SOPHIE RV time series whose average error bars are 1.9 \ms\, and the scatter is 4.8 \ms.

\subsection{SPIRou}

The Spectro-Polarimetre InfraRouge \citep[SPIRou,][]{Donati2020} is a high-resolution near-infrared spectropolarimeter and velocimeter mounted at the Canada-France-Hawaii Telescope (CFHT) in Hawaii. With a nominal spectral range from 9800 to 23500~\AA, it covers the \textit{Y}, \textit{J}, \textit{H}, and \textit{K} bands of the infrared spectrum at a spectral resolving power of $ \rm \lambda / \Delta \lambda \sim 70\,000 $. The observations of Gl~205 are part of the Planet Search program (WP1) of the SPIRou Legacy Survey \citep[SLS,][]{Donati2020}, whose main goal is to perform a systematic RV monitoring of nearby M dwarfs.

SPIRou operates as a spectropolarimeter as well as spectrograph. A spectropolarimetric sequence consists of four sub-exposures, each one with a different rotation angle of the half-wave Fresnel rhombs in the polarimeter. We used the spectropolarimetric mode in order to obtain a set of Stokes I (unpolarized) and Stokes V (circularly polarized) profiles spectra per spectropolarimetric sequence of 4 sub-exposures (see Section~\ref{sec:magfield}).

The star was observed from September 2019 to January 2022, collecting a total of 156 sequences of 4 sub-exposures. The median exposure time of the observations is 61 s per rhomb position (244 s for a complete sequence) and the median SNR per pixel at 1670 nm is 290. We removed from the analysis 11 sequences with SNR<150 or airmass>1.7. 

The data were reduced using the SPIRou data reduction software APERO\footnote{\url{https://github.com/njcuk9999/apero-drs}} v0.7.194 \citep{Cook2022}. APERO performs an automatic reduction of the 4096$\times$4096 pixels raw images, including correction for detector effects, and identification and removal of bad pixels and cosmic rays. The data are calibrated by performing flat and blaze corrections, then are optimally extracted \citep{Horne1986} from both science channels (fibers A and B that carry orthogonal polarimetric states of the incoming light) and from the calibration channel (fiber C). 

The pixel-to-wavelength calibration is done using an UNe hollow cathode lamp and a Fabry-Pérot etalon, following \citet{Hobson2021}, in order to obtain the wavelength at the observatory rest-frame. Then APERO utilizes the \texttt{barycorrpy}\footnote{\url{https://github.com/shbhuk/barycorrpy}}\citep{Kanodia2018} Python code to compute the Barycentric Earth Radial Velocity (BERV) and Barycentric Julian Date (BJD) of each exposure. APERO performs a telluric and night emission correction in two steps. The first step is to obtain an atmospheric absorption model built with the TAPAS \citep[Transmissions of the AtmosPhere for AStronomical data,][]{Bertaux2014} which is applied to pre-clean the science frames. This model only leaves percent-level residuals in deep ($>50$\%) lines of H$_2$O and dry absorption molecules (e.g., CH$_4$, O$_2$, CO$_2$, N$_2$O, and O$_3$). This procedure of building a TAPAS model is also done for a set of rapid rotating hot stars observed at different atmospheric conditions, in order to built a library of telluric residual models. This grid of telluric residual models have 3 degrees of freedom (optical depths of H$_2$O and dry components, and a constant). The second step is to subtract this telluric residual model to the pre-cleaned data to obtain a telluric corrected spectra.

In a standard procedure, APERO computes automatically the radial velocities by cross-correlation of the telluric-corrected spectra with a given binary mask of stellar absorption lines. However, in this work, we made used of the RVs computed by the line-by-line (LBL) method based on the \citet{Bouchy2001} framework and optimized for SPIRou data. The LBL method is fully described in \citet{Artigau2022}. This algorithm exploits the radial velocity content per line on the spectra to obtain one single RV measurement. Usually, an M dwarf observed by SPIRou will have $\sim$16\,000 individual spectral lines. A finite-mixture model approach deals with the high-sigma outliers in the RVs of the lines that come from cosmic rays or errors in the correction for tellurics. After outlier removal, the final RV is the mean of a Gaussian distribution containing the individual RVs of all the lines, and its uncertainty is derived following \citet{Bouchy2001}. This method is also applied to the simultaneous calibration fiber to correct for the instrumental drift. The LBL RVs are corrected for the instrumental drift and for the long-term zero point using a Gaussian Process regression with data of the most observed stars in the SPIRou Legacy Survey. The details of this procedure will be described in Vandal et al. (in prep). The log of the observations and the RV measurements from the LBL method are listed in the Table~\ref{table:SPIRoudata}. Figure\ref{fig:radialvelocities} shows the SPIRou RV time series whose average error bar is 1.9 \ms\, and the scatter is 4.4 \ms.

\subsection{TESS}
Currently, the Transiting Exoplanet Survey Satellite (TESS; \citealt{Ricker2014}) is in its extended mission, after successfully completing an all-sky survey of bright stars during its primary mission. TESS observed Gl~205 with 2-minute cadence during Sector 6, between December 12, 2018 and January 6, 2019, and again in Sector 32, from November 19 to December 16, 2020. We obtained the Presearch Data Conditiong (PDC) flux time series, processed by the TESS Science Processing Operations Center (SPOC), from the Mikulski Archive for Space Telescopes (MAST)\footnote{\url{mast.stsci.edu}}. The light curves are shown in Figure~\ref{fig:TESS}. We used the quality flags given by the pipeline to remove bad regions of the light curves thus, we only kept data points with quality flag equals zero.

\section{Stellar characterization}\label{sec:stellarcaract}
We use the high-resolution spectra from SOPHIE and SPIRou, independently, to derive the atmospheric stellar parameters of Gl~205. For the SOPHIE optical part we made use of the \texttt{ODUSSEAS} \citep{Antoniadis2020} code to compute the effective temperature, \teff, and the metallicity, [Fe/H]. For the SPIRou near-infrared spectra we follow \citet{Cristofari2022} to derive \teff, surface gravity (\logg), overall metallicity ([M/H]), and alpha-enhancement ([$\alpha$/H]). The \teff\, measured from the optical and near-infrared spectra are in good agreement. The values of stellar parameters derived for Gl~205 are listed in Table~\ref{tab:stellarparams}. In this section we describe in detail both techniques.

\subsection{SOPHIE spectra}\label{SOPHIE_spectra}

A detailed description of the machine learning tool \texttt{ODUSSEAS} can be found in \citet{Antoniadis2020}. The method is based on measuring the pseudo equivalent widths (pEWs) of absorption lines and blended lines in the range between 5300 Å and 6900 Å. The line list consists of 4104 absorption features, the same as used by \citet{Neves2014}.

\texttt{ODUSSEAS} receives 1D spectra and their resolutions as input. The tool contains a supervised machine learning algorithm based on the \texttt{scikit-learn} Python package \citep{scikit-learn}, in order to determine the \teff\, and [Fe/H] of M dwarf stars. 

Applied to new spectra, \texttt{ODUSSEAS} measures the pEWs of their lines and compares them to the models generated from the reference HARPS spectra sample, convolved to the respective resolution of the new spectra to be analyzed. The reference data set is built with spectra taken from the HARPS M dwarf sample. In the case of the SOPHIE spectrum of star Gl~205, the HARPS reference spectra are convolved from their original resolution of 115000 to the SOPHIE resolution of 75000. 

For the analysis of Gl~205, the new reference data set of the upgraded \texttt{ODUSSEAS} version has been applied and includes spectra from 47 M dwarfs. The references for training and testing the models are the pEWs of the 47 HARPS spectra, used together with interferometry-based \teff\, \citep{Rabus2019,Khata2021} and [Fe/H] derived by applying the method by \citet{Neves2012} using updated values of parallaxes from Gaia DR3.

The resulting stellar parameters of the star are calculated from the mean values of 100 determinations obtained by randomly shuffling and splitting each time the training (80\% of the reference sample, i.e. 37 stars) and testing groups (remaining 20\%, i.e. 10 stars). We report parameter uncertainties derived by quadratically adding the dispersion of the resulting stellar parameters and the mean absolute errors of the machine learning models at this resolution. We obtained an $T_{\rm eff} = 3878\pm81\,$K and $\rm [Fe/H] = 0.21 \pm 0.06\,$dex.

\subsection{SPIRou spectra}\label{SPIRou_spectra}

We estimate the  \teff, \logg, [M/H], and [$\alpha$/H] from a high resolution template spectrum built from the over 500 spectra from sub-exposures acquired with SPIRou. The process relies on the direct comparison of the template spectrum to a grid of synthetic spectra computed from MARCS model atmospheres \citep{Cristofari2022,Cristofari2022b}. The comparison is performed on carefully selected spectral windows containing about 20 atomic lines and 40 molecular lines. 

Prior to the comparison, synthetic spectra are broadened to account for instrumental effects, and the local continuum of the models is adjusted on windows built around the selected lines. Comparing the template spectrum to a grid of synthetic spectra with various \teff, \logg, [M/H] and [$\alpha$/H] results in the computation of 3 dimensional $\chi^2$ grid on which we fit a 3D 2nd degree polynomial to retrieve a minimum.

With this model, we derive $T_{\rm eff}=3770 \pm 30$~K, $\log{g}=4.70 \pm 0.1$~dex, $\rm{[M/H]} = 0.43 \pm 0.1$~dex, and $\rm [\alpha/Fe] = -0.08 \pm 0.04$. 

\subsection{Age and galactic population}

We computed the age of Gl~205 with the \texttt{stardate}\footnote{\url{https://github.com/RuthAngus/stardate}}\citep{Angus2019} Python package which combines isochrones fitting with gyrochronology. The inputs of this code are the stellar parameters derived in Sections~\ref{SPIRou_spectra} and \ref{SOPHIE_spectra}: \teff, \logg, \feh, and the parallax from Table~\ref{tab:stellarparams}. Since we obtained two estimations of \teff, we tested both values to see if we obtain results in agreement. Using the \teff\, from SOPHIE spectra we measure a stellar age of $3.7^{+5.0}_{-1.6}$ Gyr. In the other hand, using the SPIRou \teff\, we estimate the age at $3.6^{+5.0}_{-1.7}$ Gyr. Both estimations are in complete agreement. 

The rotation period of Gl~205 is higher than that of stars of similar mass in the 4 Gyr-old open cluster M67 sequence \citep{Dungee2022}. Calibrating the gyrochronology relationship with this sequence results in an age determination of $5.2\pm0.7$ \citep{Fouque2023}, within the error bars of our independent estimate. 

In other to know to which galactic population (thin disk, thick disk, or halo) Gl~205 belongs, we followed \citet{Reddy2006} to obtain the probabilities of belonging to the three populations, based on the galactic velocities of the star. For Gl~205 the galactic velocities are (U, V, W) = (49, 13, 31) \kms, computed using the position, proper motion, and parallax from GAIA DR3 \citep{GaiaDR3}. We found a $99\%$ probability that Gl~205 belongs to the thin disk of the Milky Way.

The ages of the stars in the thin disk have a wide range. However, it has been stated that most of them have ages less than 5 Gyr, reaching up to 14 Gyr \citep{Reddy2006,Haywood2008,Holmberg2009}. Our estimation of the age of Gl~205 is in agreement with these results. Regarding the metallicity, \citet{AllendePrieto2004} found that the mean metallicity of the stars in the thin disk is <[Fe/H]>=$-0.09\pm0.19$. We estimated a high-metallicity for Gl~205 of $0.21\pm0.06$ which is located within 2$\sigma$ of the mean value.

\subsection{Comparison with the literature}

We compared our results with previous studies of Gl~205. \citet{Schweitzer2019} derived the photospheric parameters \teff, \logg, and [Fe/H] from CARMENES VIS spectra and the stellar radius using the Stefan-Boltzmann's law. Using the \logg\, and the stellar radius, the authors derived the stellar mass. Their results of stellar mass and radius are listed in Table~\ref{tab:stellarparams}. They found  $T_{\rm eff}=3891 \pm 51$~K, $\log{g}=4.64 \pm 0.07$~dex, and $\rm [Fe/H] = 0.23 \pm 0.16\,$dex. Our results of [Fe/H] and \teff\, computed from SOPHIE spectra and the \logg\, from SPIRou spectra, are in particular good agreement within $1\sigma$ with the values of CARMENES. The \teff\, derived using SPIRou spectra is within $3\sigma$ to the CARMENES result.

\citet{Neves2014} obtained [Fe/H] and \teff\, of Gl~205 from HARPS spectra with values of $\rm [Fe/H] = 0.19\pm0.09$ and $T_{\rm eff}=3670\pm110$~K. Our estimation of the metallicity from SOPHIE spectra is in good agreement with the one from HARPS and our \teff\, value lies within $3\sigma$. However, it is in better agreement with our SPIRou \teff.

Using spectra of Gl~205 from the APOGEE survey, \citet{Souto2022} determined $T_{\rm eff}=3820\pm110$~K, $\log{g}=4.67 \pm 0.2$~dex, and $\rm [Fe/H] = 0.29 \pm 0.10\,$dex. Their result of \teff\, is in particular good agreement with our value derived from SPIRou spectra. \citet{Maldonado2015} used HARPS and HARPS-N spectra to derive stellar parameters of early-M dwarfs, finding a metallicity of Gl~205 of $\rm [Fe/H] = -0.03 \pm 0.19\,$dex. Our estimation of \feh\, is within 2$\sigma$ to their value.

\section{Light curve analysis}\label{sec:phometry}
In this section we describe the analysis of the TESS light curves of Gl~205 to identify stellar flares and search for exoplanet transits. Since each TESS sector has a duration of $\sim27$ d, the rotation period of Gl~205 (see Table~\ref{tab:stellarparams}) is not covered in one single sector and thus, the determination of the rotation period from the TESS data is not possible. Moreover, there is a time gap of 684 d between the sector 6 and sector 32.

\subsection{Flares identification}

For the identification of stellar flares we used the Python package \texttt{stella}\footnote{\url{https://github.com/afeinstein20/stella}} \citep{Feinstein2020}. This open-source code uses convolutional neural networks (CNN) to identify flares events in the 2-minute cadence TESS light curves and delivers the probability of such event for a given time. The reported flare probability is the average prediction of the 10 models available in \texttt{stella}. After identification, \texttt{stella} uses an empirical flare model \citep{Walkowicz2011,Davenport2014} to obtain the best-fit parameters through a $\chi^2$-fit. The flare model includes a sharp Gaussian rise and an exponential decay \citep{Feinstein2020b}. Candidates events with probability higher than 50\% of being a flare were considered for modeling. In the modeling process, the part of the light curves that includes the flare is detrended to account for stellar variability.

\begin{figure}
    \centering
    \includegraphics[width=\linewidth]{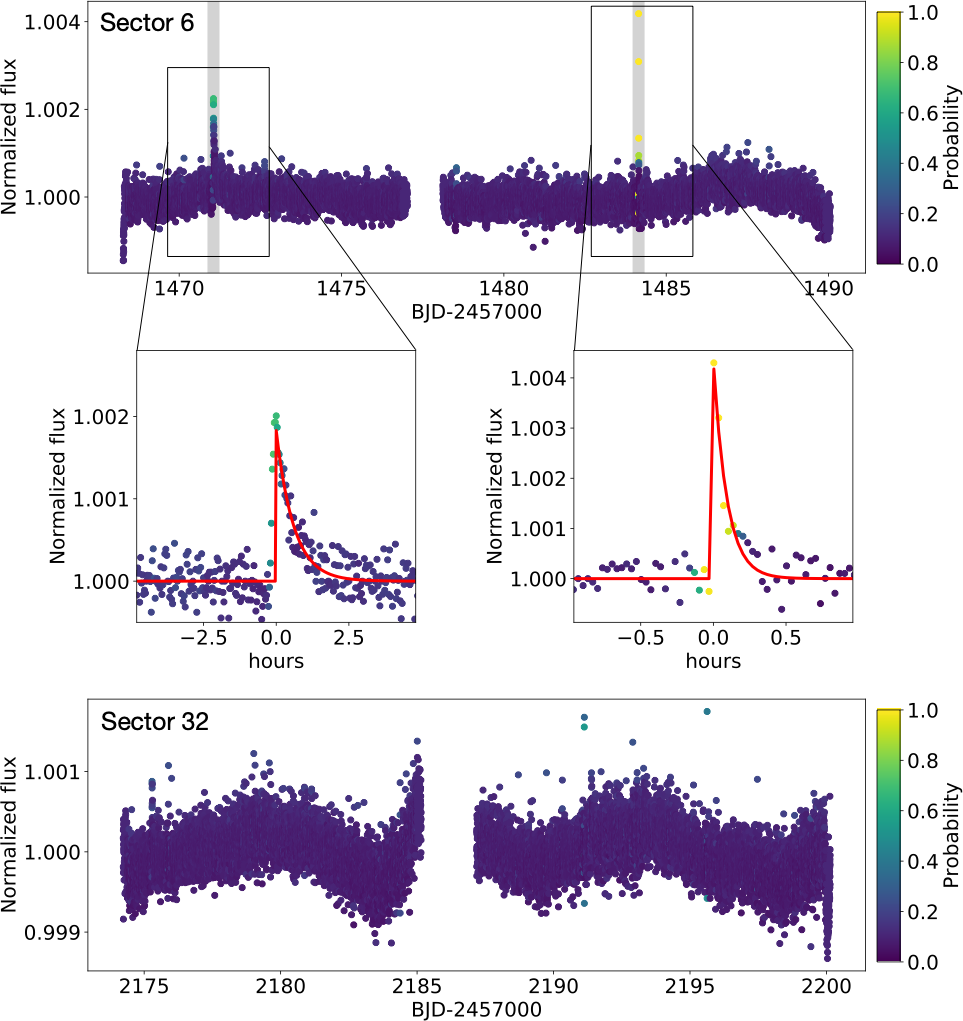} \
    \caption{TESS light curves of Gl~205 during Sector 6 (top panel) and Sector 32 (bottom panel). The middle panels show a zoom in around the region of the two flares identified in Sector 6 using the \texttt{stella} package. The red line depicts the best-fit flare model.}
    \label{fig:TESS}
\end{figure}

We applied the algorithm in the available TESS sectors of Gl~205 using the available trained CNNs from \citet{Feinstein2020b}. We identified two flares events during sector 6 and none during sector 32 (see Figure~\ref{fig:TESS}).  We report in Table~\ref{tab:flares} the time of the flare's peak, its amplitude, the equivalent duration (ED) which is measured as the area of the flare event, the rise and fall parameters from the flares' model, and the probability.

The second flare identified seems more prominent than the
first, reaching higher flux amplitude. However, both events have the same equivalent duration of about 10 hours meaning that the energy of these events are similar but with different time-scales. While the first flare lasted ~2.5 hours, the second flare lasted ~0.5 hours. 

\citet{Gunther2020} performed a study of stellar flares in the first data release of TESS. They found that mid to late M dwarfs show the highest fraction of flaring stars. However, the authors warned that this may be influenced by the TESS target selection. Only 10$\%$ of the early M dwarfs in their sample are flaring stars. Moreover, fast rotators (P$<$5 d) may have higher flares rates than slow rotators. Our findings in Gl~205 agree with the results of \citet{Gunther2020} since we observed a low rate of flares for this star which is expected for early and slow-rotators M dwarfs.


\begin{table}
\footnotesize
    \caption{Parameters of the flares events identified in the TESS light curves of Gl~205.}
    \label{tab:flares}
    \centering
    \begin{tabular}{cccccc}
    \hline
    \hline
    $\rm t_{peak}$ & Amplitude & ED & Rise & Fall  &  Probability  \\
    BJD & [rel. flux] & [hour] & -- & -- & [\%]  \\  \hline
    2458471.056 & 0.0018 & 9.97 & 0.0002 & 0.025 & 67 \\
    2458484.149 & 0.0041 & 10.01 & 0.0002 & 0.004 & 99 \\

    \hline
    \end{tabular}
    \tablefoot{Time of the flare's peak $\rm t_{peak}$, equivalent duration of the flare event ED, Gaussian rise of the flare and its exponential decay.}
\end{table}

\subsection{Planet transit search}
The TESS data validation report of Gl~205 identifies a planet candidate with an orbital period of 22.2 d. However, the transit event candidate proposed by the automatic pipeline is clearly affected by the flare of 2458471.056~BJD. We conduct our own analysis in order to identify possible transits in the light curves.

\begin{figure}[ht]
    \centering
    \includegraphics[width=1.0\hsize]{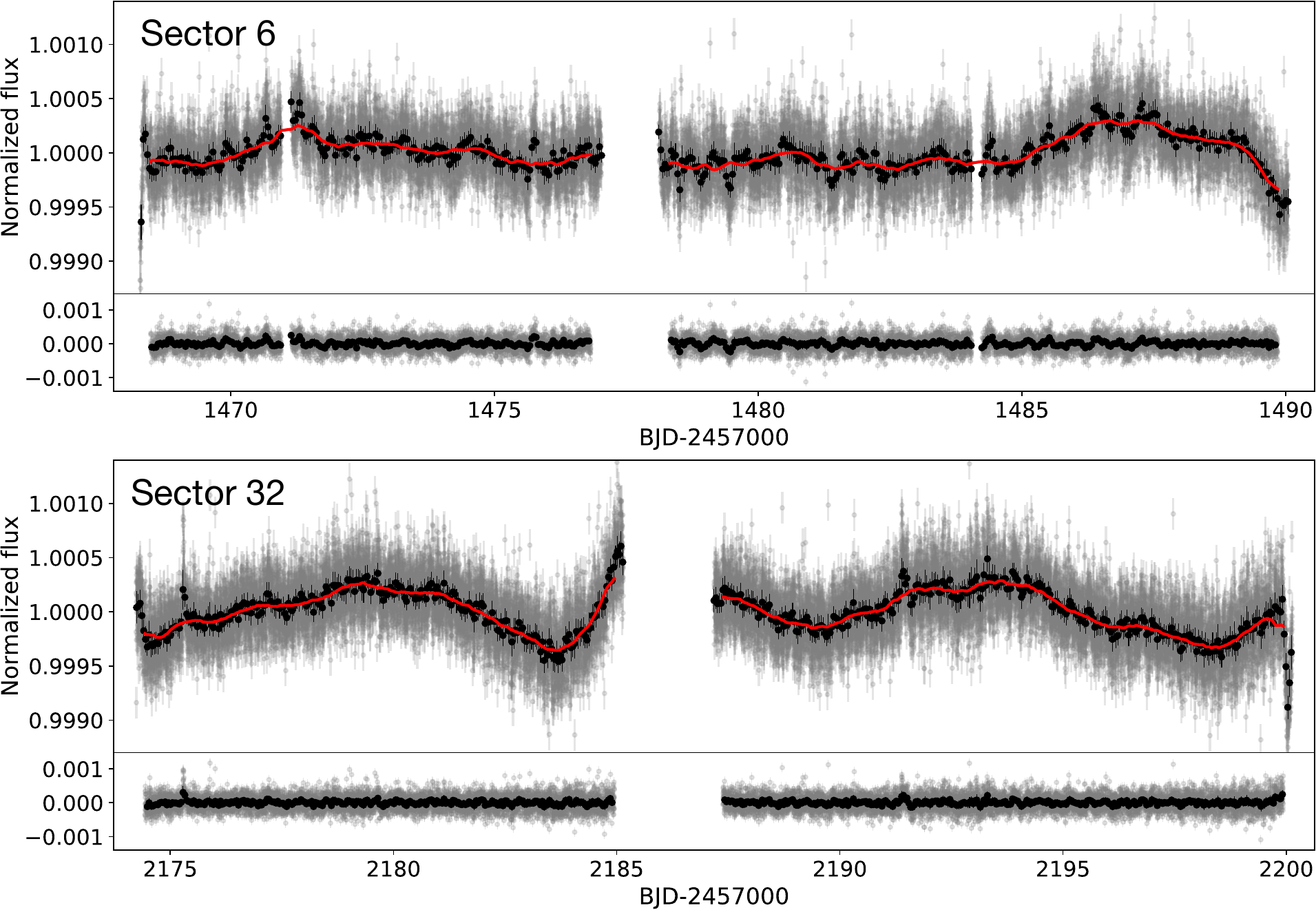}
    \caption{TESS light curves of Gl~205 during sector 6 (top) and sector 32 (bottom). The red line depicts the applied detrending model to remove the stellar variability using the \texttt{$\rm w\bar{o}tan$} package. Below each panel are the residuals of the model with an standard deviation of 210 ppm. No planetary transits were found in this data set.}

    \label{fig:TESS_detrending}
\end{figure}

First, we removed outliers using a $3\sigma$-clipping procedure and excluded the data affected by the flares. Since the light curves are affected by stellar variability we used the Python package \texttt{ w$\bar{o}$tan}\footnote{\url{https://github.com/hippke/wotan}}\citep{Hippke2019b} to remove the trends. This code includes several methods to perform light curve detrending. We applied a method based on a time-windowed sliding filter with an iterative robust location estimator following the results of \citet{Hippke2019b}. For the detrending model, we excluded the edges of the light curves since these zones are usually affected by strong systematics. Figure~\ref{fig:TESS_detrending} shows the detrending model of the light curves and the residuals which have a standard deviation of 210 ppm. 

After detrending the light curves, we used the \texttt{transit least squares}\footnote{\url{https://github.com/hippke/tls}} algorithm \citep{Hippke2019} to search for periodic transit events. The algorithm searches for transit features in an automatically built grid of orbital periods and transit durations. The grid of periods depends on the time-span of the data and the transit durations come from an empirical relation detailed in \citet{Hippke2019}.

We first applied the algorithm in the whole data set, including sector 6 and sector 32. The grid of periods was automatically set between 0.6 to 365~d. The maximum signal detection efficiency (SDE) is found at an orbital period of 11.86 d but it has low significance (SDE = 8) and the periodogram is highly degenerate. The SNR of this stacked transit signal is 2.5 with only three events in the data set, one in sector 6 and two in sector 32. To confirm or rule-out this period we applied the algorithm in sector 32, independently, since two transits were identified in this sector. No transits were found in this sector thus, we discarded transit events in the two available TESS sectors of Gl~205.

\section{Magnetic field analysis}\label{sec:magfield}

The SPIRou spectropolarimetric products, in particular the Stokes $I$ and $V$ profiles were obtained using the \texttt{Libre-ESpRIT} pipeline described in \citet{Donati1997}.
We applied least-squares deconvolution \citep[LSD;][]{Donati1997} to compute the average Stokes $I$ (unpolarized) and Stokes $V$ (circularly-polarized) line profiles for all our SPIRou observations. We used a mask of atomic lines, spanning SPIRou spectral domain, computed from a \texttt{ATLAS9} local thermodynamical equilibrium model of atmosphere \citep{kurucz1993}, assuming an effective temperature of 3750\,K and a surface gravity of $\log$\,g\,=\,5.0. Note that we only selected lines with a relative absorption larger than 3\% (from the unpolarized continuum) to avoid an over-representation of weak lines in our final line list. Lines affected by strong tellurics, i.e. of relative absorption deeper than 20\% within $\pm$30\,\kms\ from the line center, are masked out in the LSD process. The extracted Stokes $I$ and $V$ profiles feature a mean central wavelength of 1700\,nm, an effective Land\'e factor of 1.25 and a relative depth of 12\% with respect to the continuum.

\subsection{Stellar rotation period}\label{sec:prot}

The disk-integrated longitudinal magnetic field $B_{\ell}$ was computed using the Stokes V and Stokes I profiles, following the method of \citet{Donati1997}:
\begin{equation}
    B_{\ell}~[G] = \frac{-2.14\cdot10^{11}}{\lambda_{0}\cdot g_{\rm eff}\cdot c}\frac{\int u V ( u)d u}{\int(I_{c}-I)d u}
\end{equation}
where $I$ and $V$ are the unpolarized and circularly-polarized LSD profiles, $I_c$ is the continuum level, $ u$ is the velocity in $\rm km~s^{-1}$, $\lambda_{0}$ is the mean wavelength, $g_{\rm eff}$ is the effective Land\'e factor and $c$ the speed of light in $\rm km s^{-1}$.

The $B_{\ell}$ time series of Gl~205 is listed in Table~\ref{table:SPIRoudata} of the Appendix~\ref{App:RVs}. The mean values of the $B_{\ell}$ time series is 1.6 G, the standard deviation is 2.9 G and the mean of the error bars is 1.0 G. It is expected that the longitudinal magnetic field is modulated by the rotational period of the star, since the large-scale magnetic topology at the surface of the star is expected to evolve on a longer time scale. To measure the stellar rotation period, we first computed the generalized Lomb-Scargle (GLS) periodogram \citep{Lomb1976,Scargle1982,Zechmeister2009} implemented in the \texttt{astropy} \citep{astropy:2013,astropy:2018} Python package. The GLS periodogram of the $B_{\ell}$ time series shows a strong peak of periodicity at 32.7 d with false alarm probability (FAP) below 1\% (see Figure~\ref{SPIROU_activity}). 

\begin{figure*}
    \centering
    \includegraphics[width=1.0\textwidth]{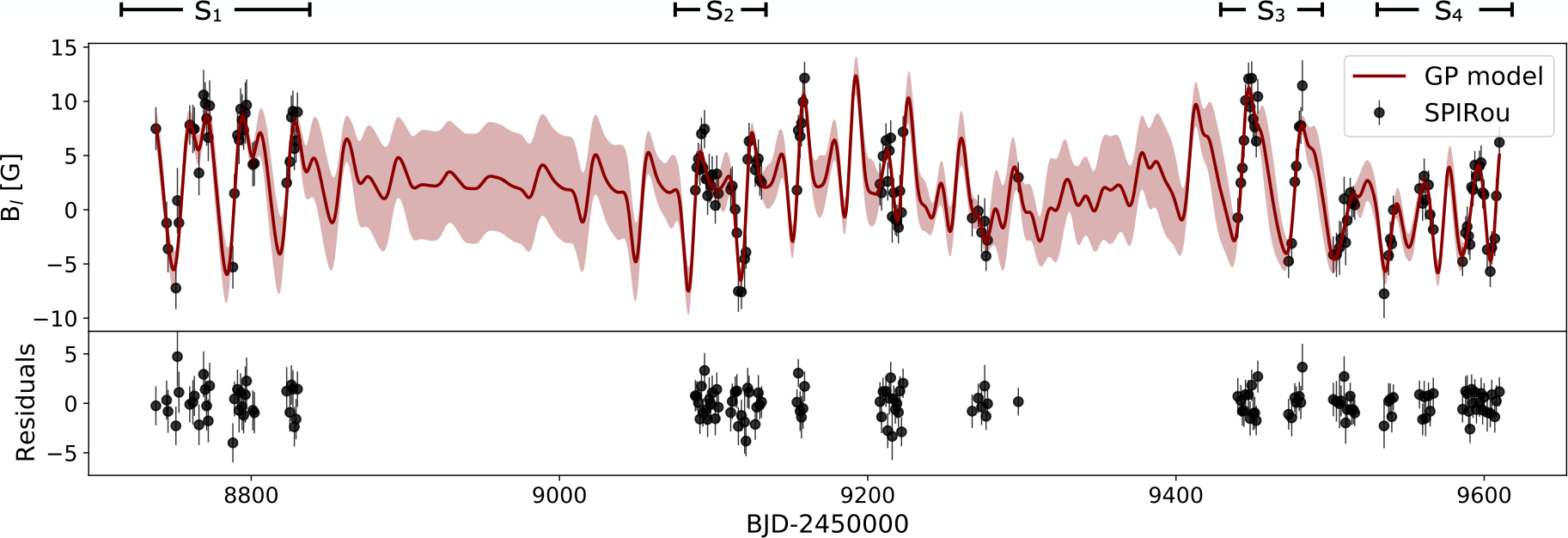}
    \caption{Time series of the SPIRou longitudinal magnetic field $B_{\ell}$. In red color is the GP model using a quasi-periodic kernel. The residuals of this model have an scatter of 0.97 G and they are shown in the bottom panel. At the top of the figure are marked the four subset of data used in Section~\ref{sec:ZDI}.}
    \label{Blong_timeseries}
\end{figure*}

However, it has been shown that stellar activity follows a quasi-periodic behavior \citep[e.g.,][]{Haywood2014,Angus2018} rather than a single sinusoidal signal, as is the case of the periodicity searched in a GLS periodogram. Moreover, the magnetic field of Gl~205 is known to evolve on a timescale of a few rotation cycles \citep{Hebrard2016}. Thus, we take advantage of the flexibility of the GPs to constrain the rotational period of the star using the $B_{\ell}$ time series. For this purpose, it is important to highlight that the three well defined observational seasons of the $B_{\ell}$ (see Figure \ref{Blong_timeseries}) are long enough to cover a few stellar rotation cycles.

The kernel of the GP regression is set to be the quasi-periodic as defined in \citet{Roberts2013}:

\begin{equation}\label{eq:kernel2}
    k_{i,j} = A^{2} \text{exp}\bigg[-\frac{(x_{i}-x_{j})^2}{2l^2}- \frac{1}{\beta^2}\text{sin}^2\bigg(\frac{\pi(x_{i}-x_{j})}{P_{\rm rot}}\bigg)\bigg] + \sigma^2\delta_{i,j}
\end{equation}

\noindent where $x_i$ and $x_j$ are two observation dates, $A$ is the amplitude of the covariance, $l$ is the decay time, $\beta$ is the smoothing factor, $P_{\rm rot}$ is the stellar rotation period, and $\sigma$ is the uncorrelated white noise also known as the jitter term. 

\begin{figure}
    \centering
    \includegraphics[width=0.99\hsize]{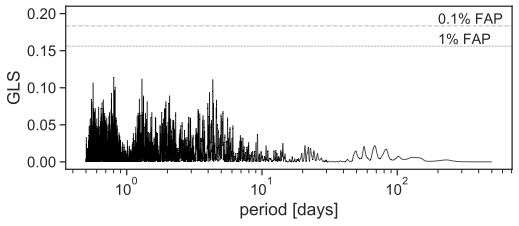}
    \caption{GLS periodogram of the residuals of the GP regression applied in the longitudinal magnetic field $B_{\ell}$. }
    \label{Blong_GP_residuals}
\end{figure}

The posterior distributions of the hyper-parameters were sampled from a Markov-chain Monte Carlo (MCMC) routine using the package \texttt{emcee} \citep{emcee}. We set up 50 walkers and 5000 steps after a burn-in phase of 500 steps. The priors used and the final posterior distributions are listed in Table~\ref{table:GP} and in Figure~\ref{Blong_GP_corner} is displayed the corner plot of the posterior distributions. The best-fitting model of the GP regression is illustrated in Figure~\ref{Blong_timeseries} which has a reduced $\chi^2$ of 0.92. As a sanity check, we plot the GLS periodogram of the GP model residuals in Figure~\ref{Blong_GP_residuals} where we see that there is no periodic signal left.

We measured a stellar rotation period of \prot and a decay time of $l=62^{+15}_{-12}$\,d, which implies that the active features could evolve relatively fast, on a time scale of about two rotation cycles, as the GP decay time has been proved as a good indicator of the average time scale evolution of the active features \citep{Nicholson2022}. With this new estimation of the rotation period, we can derive the rotational velocity \vsini. Assuming a stellar radius of $0.556\pm0.033M_{\odot}$ and inclination of $60^{\circ}\pm10^{\circ}$ (see Table~\ref{tab:stellarparams}), we obtained a \vsini\, of $0.7\pm0.1$ \kms. \citep{Hebrard2016} obtained a \vsini\, of $1.0\pm0.5$ \kms from their ZDI analysis, which is in agreement with our results. 

\begin{table}[ht]
\footnotesize
\caption{Priors and best fit hyper-parameters of the GP model using a quasi-periodic kernel in the SPIRou longitudinal magnetic field data $B_{\ell}$.}         
\label{table:GP}      
\centering

\begin{tabular}{c c| c c}     
\hline\hline       
Parameter & Units & Prior & Posterior  \\ 
\hline                    
   mean value $\mu $ & G        & $ \rm \mathcal{U}(-\infty,+\infty) $ & $1.3\pm0.9$\\  
   jitter $\sigma $ & G    & $ \rm \mathcal{U}(0,+\infty) $ & $0.5\pm0.2$  \\ 
   amplitude $\alpha $ & G       & $ \rm \mathcal{U}(0,+\infty) $ & $3.1^{+0.6}_{-0.4}$    \\
   decay time $l$ & d          & $ \rm \mathcal{U}(30,300) $ & $62^{+15}_{-12}$  \\
   smoothing factor $\beta$ & & $ \rm \mathcal{U}(0.25,1.25) $ & $0.6\pm0.1$ \\
   rotation period $P_{\rm rot}$& d   & $ \rm \mathcal{U}(1,100) $       & $34.4\pm0.5$  \\
   RMS of residuals & G & & 0.97 \\
   $\chi^2$ & & & 0.92 \\
\hline                  
\end{tabular}
\tablefoot{The symbol $\mathcal{U}(a,b)$ defines an uniform prior with $a$ and $b$ the minimum and maximum limits, respectively.}
\end{table}

\subsection{Zeeman-Doppler Imaging}\label{sec:ZDI}

We use Zeeman-Doppler imaging \citep{Semel1989,Brown1991,Donati1997B} to invert the observed Stokes $V$ LSD profiles into distributions of the large-scale magnetic field at the surface of Gl~205. As described in \citet{Donati2006}, ZDI decomposes the large-scale field vector into its poloidal and toroidal components, both expressed as weighted sums of spherical harmonics. For a given field distribution, local Stokes $V$ profiles are computed for each resolution element of the visible hemisphere of the star using analytical expressions from the Unno-Rachkovski’s solution to the radiative transfer equation in a plane-parallel Milne-Eddington atmosphere \citep{Unno1956}. These profiles are then (i)~shifted to the local projected rotational velocity, (ii)~weighted according to stellar inclination and limb-darkening law \citep[assumed linear with a coefficient of 0.3][]{claret2011}, and (iii)~combined into global profiles at the times of the observations. ZDI uses a conjugate gradient algorithm to iteratively compare the synthetic profiles to the observed Stokes $V$ LSD profiles down to a given reduced $\chi^{2}$. The degeneracy between multiple magnetic maps is lifted by imposing a maximum-entropy regularization condition to the fit, assuming that the map with the minimum amount of information is the most reliable \citep{Skilling1984}

The intrinsic evolution of the magnetic field is not yet included in our ZDI code, despite notable progress over the last few years \citep[][]{Yu2019,Finociety2022}. To prevent the code from focusing on the intrinsic evolution of the field rather than on its rotational modulation, we divide our observations into four subsets of data, namely S$_{1}$, S$_{2}$, S$_{3}$, S$_{4}$, containing respectively 36, 31, 19 and 31 observations. The time windows spanned by these four data sets are indicated in Table~\ref{tab:results_ZDI} and depicted in Figure~\ref{Blong_timeseries}.  We then apply ZDI independently to each of these data sets. Note that these four seasons were defined so that the longitudinal field remains roughly periodic on time scales larger than a rotation period, ensuring that the magnetic topology does not dramatically evolve during each season and that the variation of the Stokes V profiles reflects primarily the stellar modulation rather than the intrinsic evolution of the field.


Our ZDI reconstruction includes the modelling of latitudinal differential rotation (DR) shearing the large-scale field at the surface of Gl~205 \citep{Donati2000,Petit2002}. The stellar rotation rate $\Omega$ is assumed to vary as a function of the colatitude $\theta$ such that

\begin{equation}
    \Omega (\theta) = \Omega_{\rm{eq}} - \left(  \cos \theta \right)^{2} \rm{d}\Omega \text{,}
    \label{eq:DR}
\end{equation}

\noindent
where $\Omega_{\rm{eq}}$ and $\rm{d}\Omega$ stand respectively for the rotation rate at the stellar equator and the difference in rotation rate between the equator and the pole. The best DR parameters explaining the observations are estimated using the method described in \citet{Donati2000}. For a wide range of DR parameters, we apply ZDI to a fixed level of entropy. The best parameters with error bars are estimated by fitting a 2D paraboloid to the resulting $\chi^{2}$ distribution in the $\left(\Omega_{\rm{eq}}, \rm{d}\Omega \right)$ space.

\begin{figure*}
    \centering
    \includegraphics[width=\linewidth]{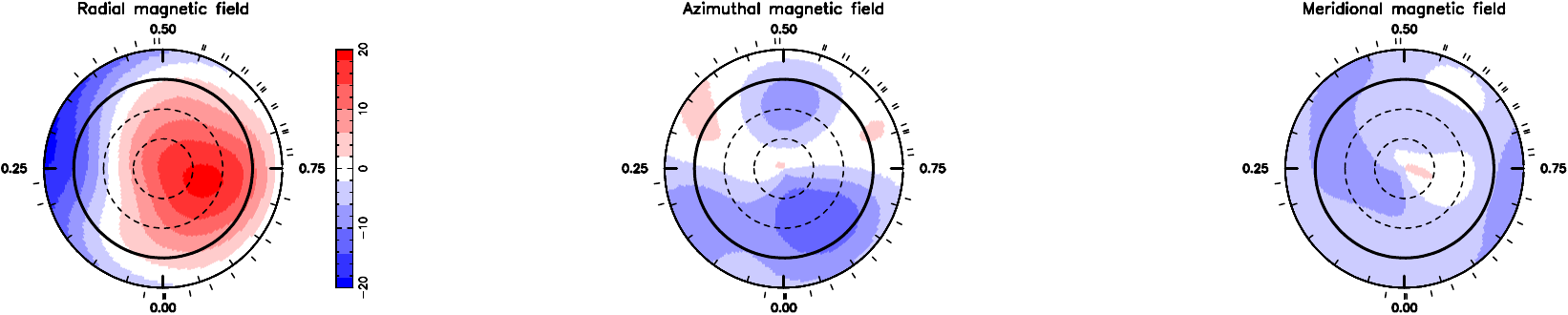} \
    \includegraphics[width=\linewidth]{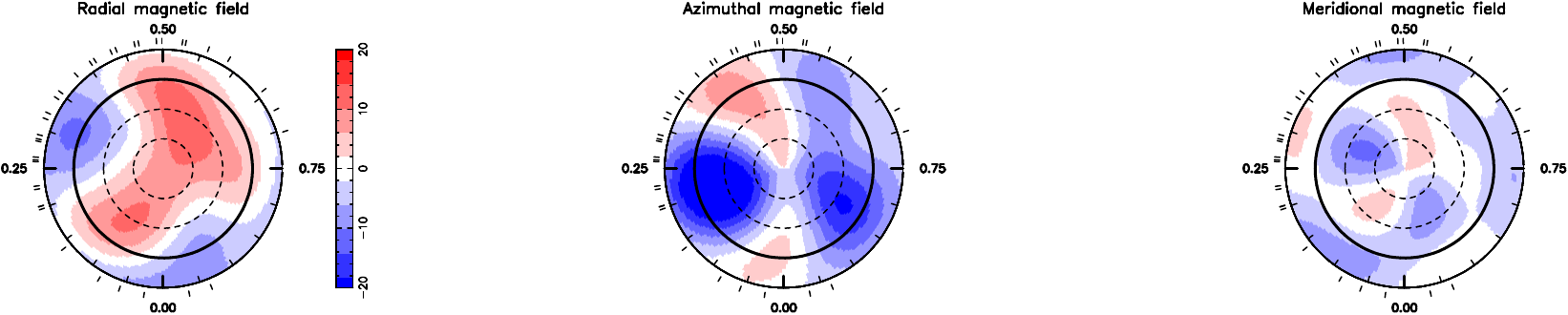} \
    \includegraphics[width=\linewidth]{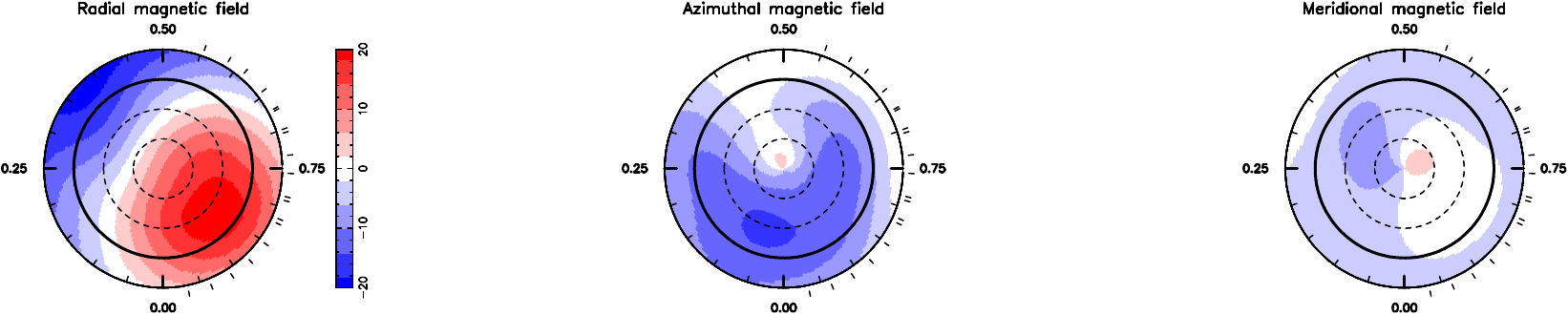}\
    \includegraphics[width=\linewidth]{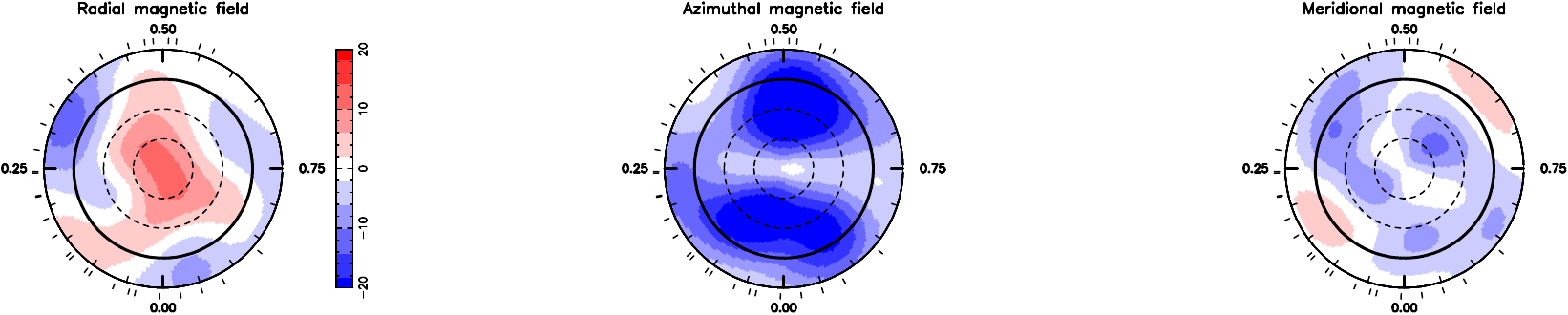}
    \caption{Radial, azimuthal and meridional components of the large-scale magnetic field at the surface of Gl~205 for the four subsets of data defined in Section~\ref{sec:ZDI}. From top to bottom, the rows show the results for Seasons S$_{1}$ (September - December 2019), S$_{2}$ (August - October 2020), S$_{3}$ (August - September 2021), and S$_{4}$ (November 2021 - January 2022). Each panel is a flattened polar view of the stellar surface. The circles indicate the equator (black solid line) and the -30, 30 and 60$^{\circ}$ parallels (dashed lines) and the ticks around the star mark the spectropolarimetric observations in unit of stellar rotation phase. The color bar depicts the magnetic flux in Gauss.}
    \label{fig:mag_maps}
\end{figure*}

\begin{table*}
    \caption{Main results of the spectropolarimetric analysis of each of subset of data of the Gl~205 SPIRou observations with ZDI. }
    \label{tab:results_ZDI}
    \centering
    \begin{tabular}{ccccccccccc}
    \hline
    \hline
     Name & Start date & End date & $\Delta T$ & $<B>$ & $f_{\rm{pol}}$  &  $f_{\rm{axi}}$ & $\Omega_{\rm{eq}}$ & d$\Omega$ & $P_{\rm{eq}}$ & $P_{\rm{pol}}$ \\ 
     -- & -- & -- & [d] & [G] & [\%] & [\%] & [rad\,d$^{-1}$] & [rad\,d$^{-1}$] & [d] & [d] \\
    \hline
   S$_{1}$ & 11-09-19 & 12-12-19 & 92 & 13.2 & 90 & 67 & 0.205\,$\pm$\,0.010 & 0.046\,$\pm$\,0.021 & 30.6\,$\pm$\,1.4 & 39.5\,$\pm$\,5.8 \\
   S$_{2}$ &  26-08-20 & 08-10-20 & 43 & 8.9 & 72 & 45 & 0.19\,$\pm$\,0.01 & 0.050\,$\pm$\,0.046 & 33.1\,$\pm$\,1.8 & 44.8\,$\pm$\,15.2 \\
   S$_{3}$ &   13-08-21 & 24-09-21 & 42 & 14.8 & 77 & 46 & 0.207\,$\pm$\,0.008 & 0.077\,$\pm$\,0.020 & 30.3\,$\pm$\,1.2 & 48.4\,$\pm$\,8.1 \\
   S$_{4}$ & 16-11-21 & 30-01-22 & 75 & 12.3 & 35 & 77 & 0.190\,$\pm$\,0.005 & 0.061\,$\pm$\,0.011 & 33.0\,$\pm$\,0.8 & 48.5\,$\pm$\,4.6 \\
    \hline
    \end{tabular}
    \tablefoot{Total time span of each subset of data $\Delta T$, average large-scale magnetic strength $<B>$, fraction of poloidal field $f_{\rm{pol}}$, fraction of axisymmetric field $f_{\rm{axi}}$, best fitting DR parameters $\Omega_{\rm{eq}}$ and d$\Omega$, equatorial and polar rotation periods, $P_{\rm{eq}}$ and $P_{\rm{pol}}$.}
\end{table*}

The best-fitting maps of the large-scale field vector for the four subsets of data are shown in Figure~\ref{fig:mag_maps}, and the associated magnetic properties are listed in Table~\ref{tab:results_ZDI}. Note that the best-fits to the observed Stokes $V$ LSD profiles are shown in Appendix~\ref{App:StokesV_fits} for the four seasons. Data sets S$_{1}$ to S$_{4}$ have been fitted to reduced $\chi^{2}$ of 1.0, 1.08, 1.1 and 1.05, starting from reduced $\chi^{2}$ of 2.4, 1.5, 4.1 and 1.4. As expected from the time series of longitudinal magnetic field shown in Figure~\ref{Blong_timeseries}, we observe a significant fluctuation in the field topology throughout our observations. In Seasons S$_{1}$ and S$_{3}$, the field is found to be a dipole of 13-15\,G \citep[consistent with the topology found in][]{Hebrard2016}, respectively tilted at 36$^{\circ}$ and 56$^{\circ}$ to the rotation axis towards phases 0.78 and 0.84. A weaker and more complex field is found in Season S$_{2}$. Though dominantly poloidal, the magnetic energy budget is much more spread into the dipolar (45\%), quadripolar (20\%) and octupolar (31\%) than in Seasons S$_{1}$ and S$_{3}$ (both dipolar at more than 90\%). Between Seasons S$_{3}$ and S$_{4}$, the field quickly evolves, on a 2-month time scale, from a poloidal dipole to a dominantly toroidal topology\footnote{Note that removing a small fraction of randomly-selected Stokes $V$ profiles in each season leads to fully-consistent magnetic properties, which confirms that field topology remains globally consistent throughout each season. In contrast, seasons S$_{2}$, S$_{3}$, S$_{4}$ cannot be extended without dramatically affecting the resulting magnetic topologies, which confirms that the season lengths are relatively well-tuned for the ZDI analysis.}.

\begin{figure}
    \centering 
    \includegraphics[width=0.8\linewidth]{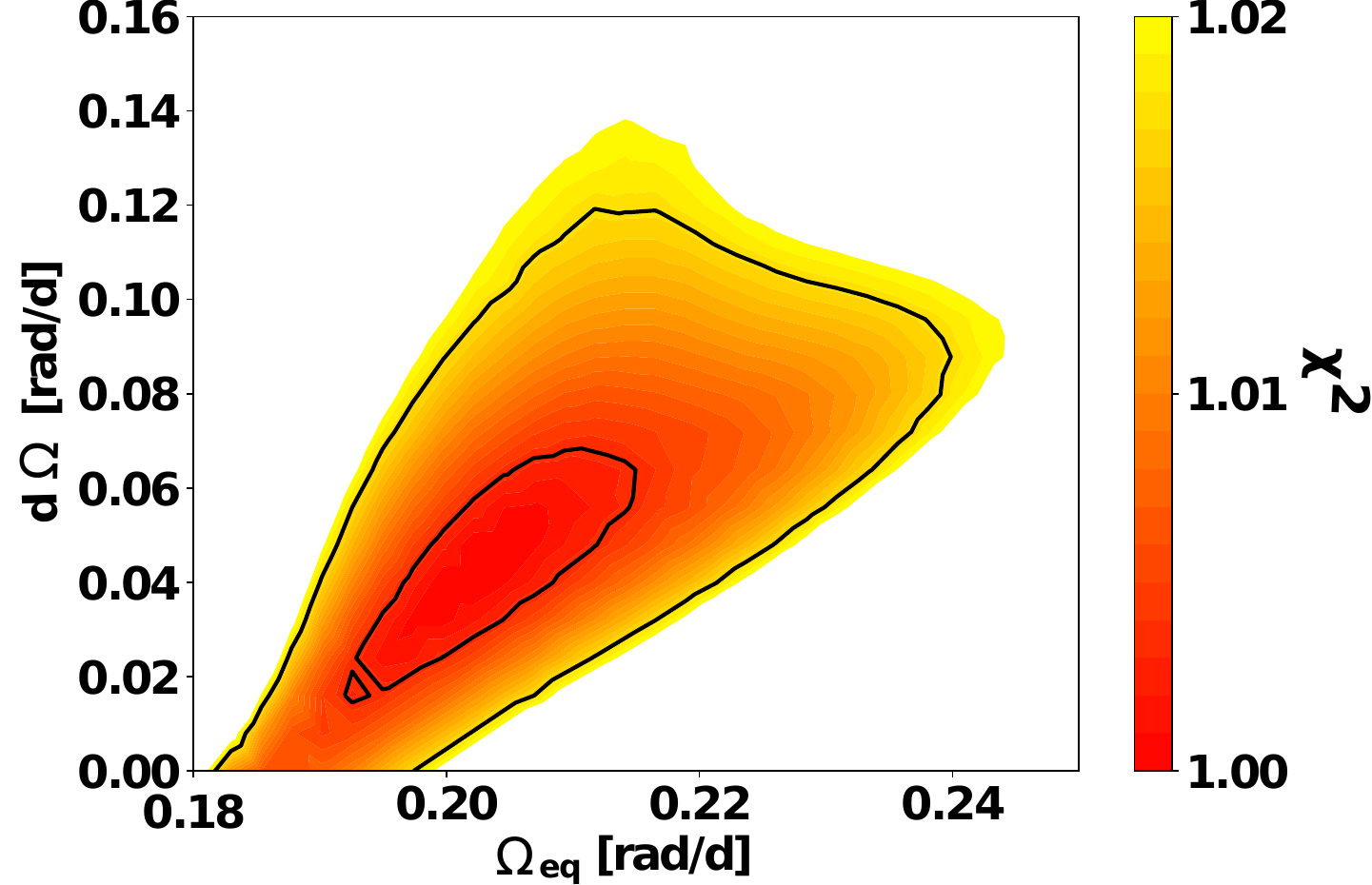} \\ 
    \includegraphics[width=0.8\linewidth]{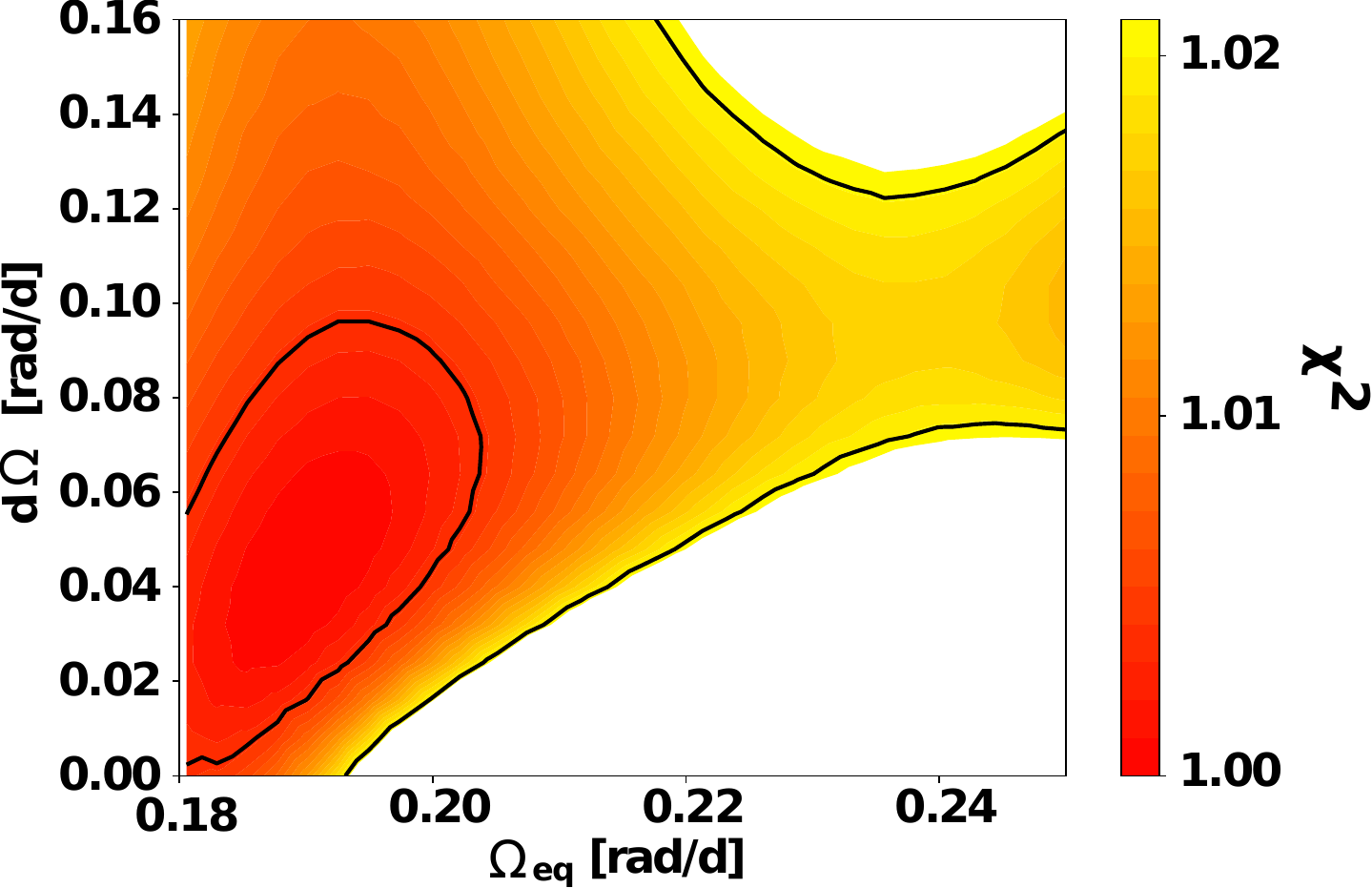} \\
    \includegraphics[width=0.8\linewidth]{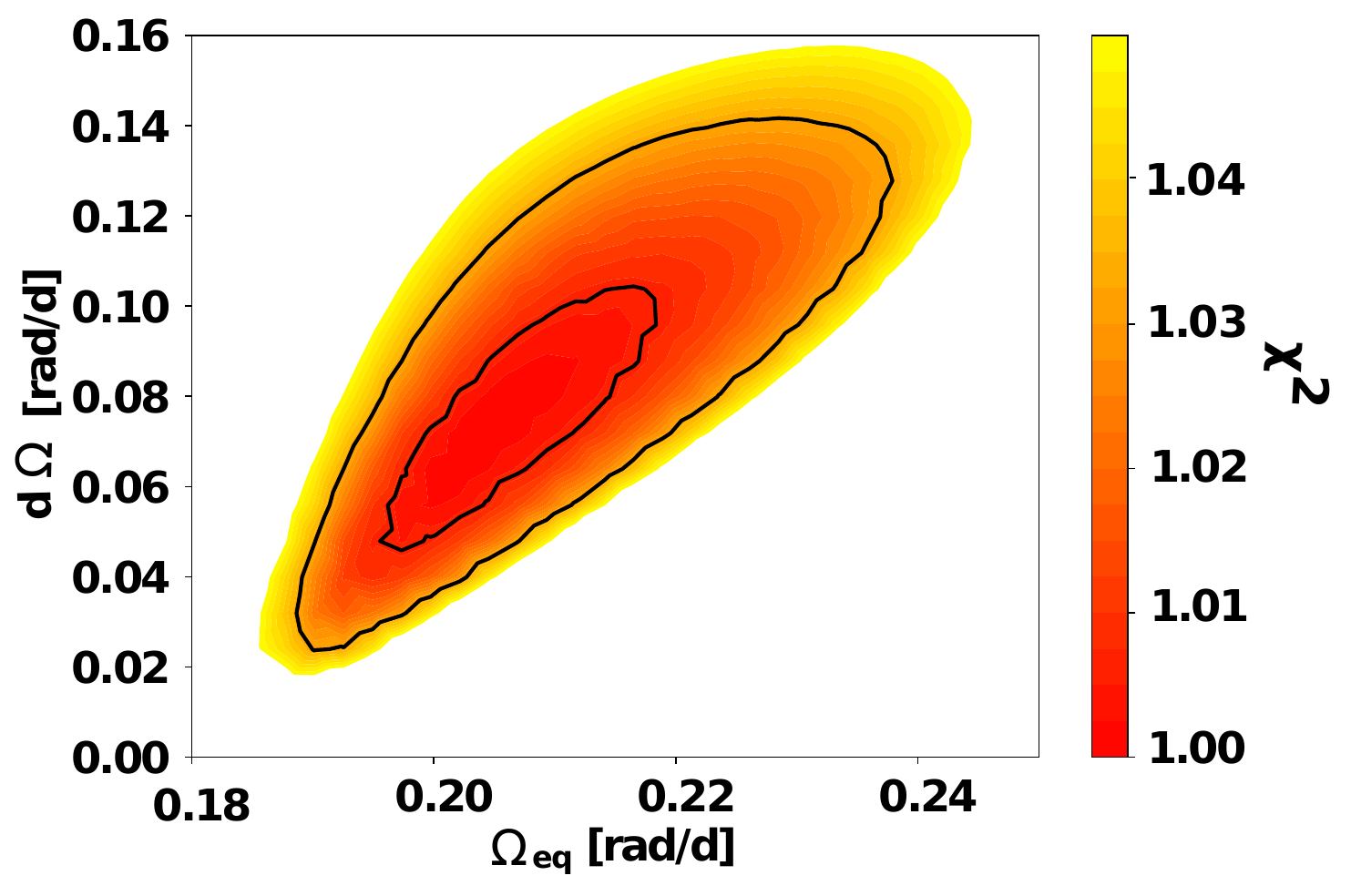} \\
    \includegraphics[width=0.8\linewidth]{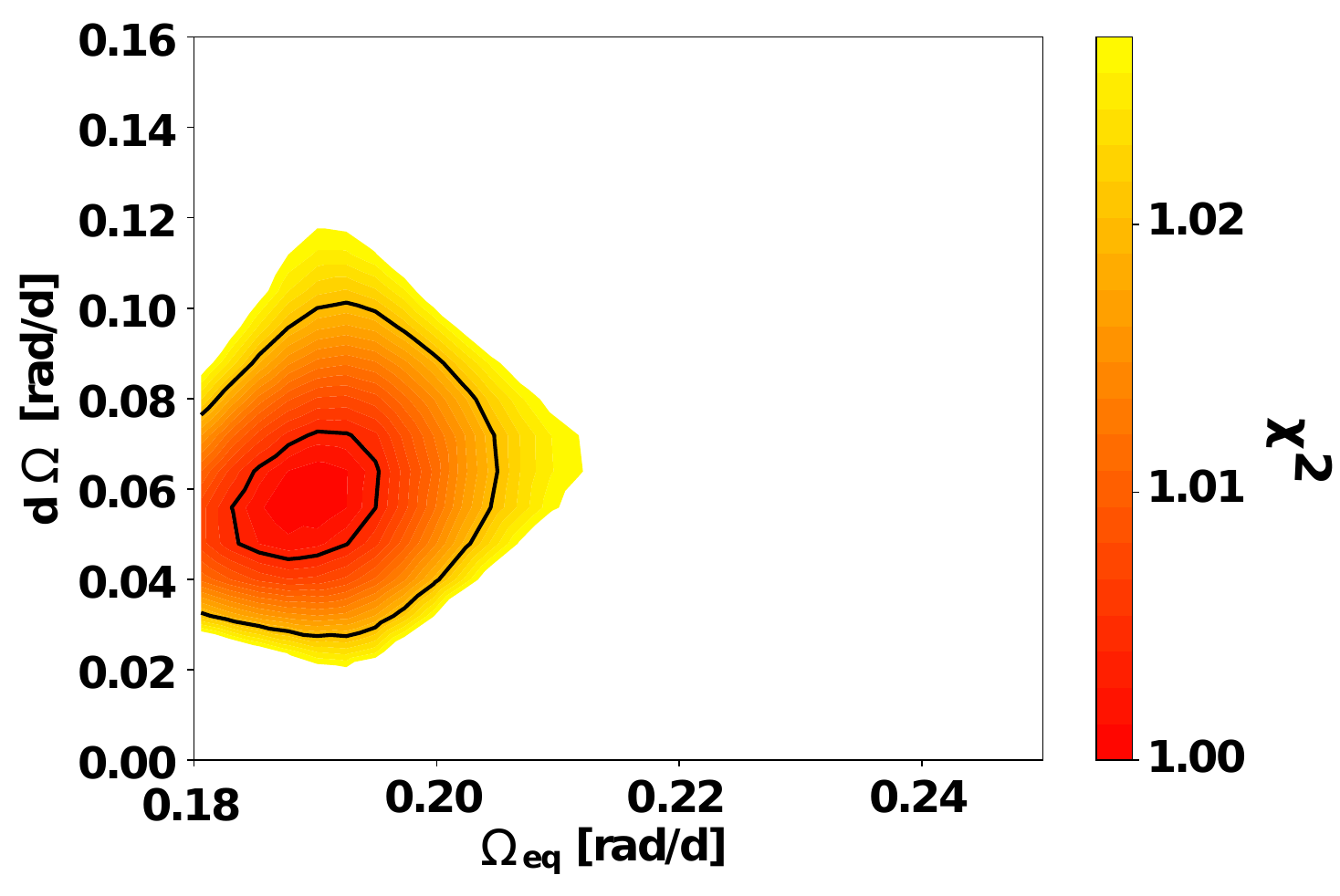}
    \caption{Distribution of reduced $\chi^{2}$ as a function of $\Omega_{\rm{eq}}$ and d$\Omega$ extracted from the Stokes $V$ LSD profiles in Season S$_{1}$ to S$_{4}$, defined in Table~\ref{tab:results_ZDI}. In each map the 1 and 3 $\sigma$ contours are indicated by the black solid lines.}
    \label{fig:chi2_map}
\end{figure}

The best-fitting DR parameters and associated equatorial and polar stellar rotation periods are listed in Table~\ref{tab:results_ZDI}, whereas the 2D $\chi^{2}$ maps associated to the fit in the ($\Omega_{\rm{eq}}$, d$\Omega$) space are shown in Figure~\ref{fig:chi2_map}. Consistent latitudinal DRs of 0.05-0.08\,rad\,d$^{-1}$ are detected in Seasons S$_{1}$, S$_{3}$, and S$_{4}$, whereas solid-body rotation cannot be firmly excluded in Season S$_{2}$. These values of latitudinal DRs are the same order of magnitude as for other early-M dwarfs with slightly faster rotation \citep{Donati2008}. 

We note that more complex magnetic topologies, such as seen in Seasons S$_{2}$ and S$_{4}$, are associated with larger equatorial rotation periods than simpler dipolar fields. This could reflect the fact that the field is dominated by smaller-scale structures mostly located at mid-to-high latitudes and, thus, only partially covering the stellar equator. However, the accuracy of our DR parameters is strongly limited by the relatively low latitudinal precision of our reconstruction, due to the very low \vsini\, of the star. As a consequence, no conclusion can be drawn on a potential evolution of the DR throughout the observations. We report an inverse-variance weighted average of the DR values from Season S$_{1}$ to S$_{4}$ of $P_{\rm eq} = 32.0\pm1.2$\,d for the rotation period at the stellar equator, and $P_{\rm pol} = 45.5\pm0.3$\,d for the rotation period at the poles.


\section{Optical and near-infrared radial velocities}\label{sec:radialvelocities}

\subsection{Periodogram analysis}
\begin{figure*}
    \centering
    \includegraphics[width=1.0\textwidth]{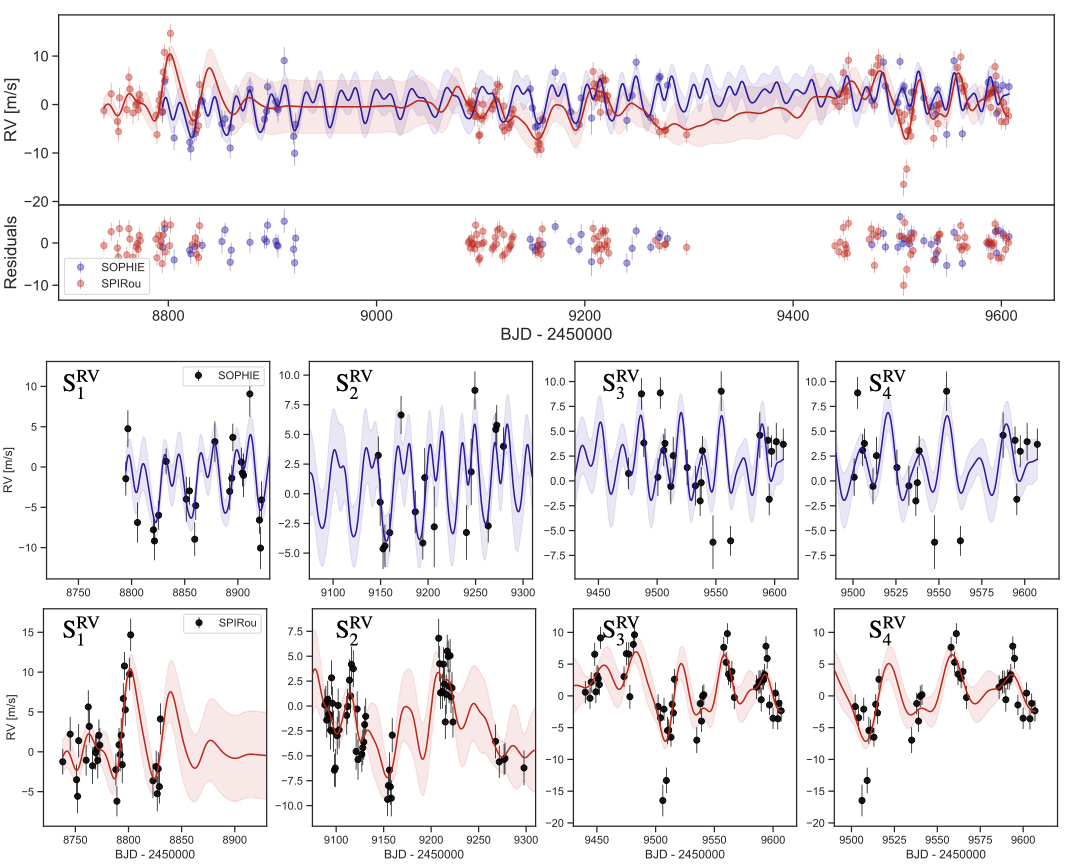}
    \caption{Time series of the SOPHIE and SPIRou radial velocities. In blue and red colors are the best fitting GP models using a quasi-periodic kernel in the SOPHIE and SPIRou data, respectively. The light-blue and light-red colors depict the 3$\sigma$ level of the uncertainties. The residuals of this model have an scatter of 2.6 \ms\, for SOPHIE, and 2.4 \ms\, for SPIRou and they are shown below the time series (\emph{top}). Zoom in of the SOPHIE and SPIRou radial velocities divided in the seasons defined in Table \ref{table:season_GP_RV} with their best fitting GP models (\emph{middle and bottom}).}
    \label{GP_RVs}
\end{figure*}

The almost simultaneous acquisition of SOPHIE and SPIRou RVs allows us to quantify the stellar activity jitter in the optical and near-infrared domains. We computed the GLS periodogram to look for periodicities in the RVs. The whole RVs time series are shown in Figure~\ref{fig:radialvelocities} along with its GLS periodogram. We identified a highly significant (FAP $<<0.1$\%) period at 34.3\,d and a second peak with also low FAP at 37.9\,d. There is also a third peak with low FAP at $\sim$ 0.5 d, which is an alias of the 1-d periodicity due to the daily sampling.

The SOPHIE and SPIRou data sets do not show the same periodicities. In the SOPHIE RVs, there is no significant periodicity found by the GLS periodogram, however, the strongest peak occurs close to the $P_{\rm rot}/2$ harmonic at $\sim17$\,d (see the top-right panel of Figure~\ref{SOPHIE_activity}). The SPIRou RVs on the contrary, show a clear peak of periodicity at 34.4 d (see the top-left panel of Figure~\ref{SPIROU_activity}) which is the expected stellar rotation period. In terms of the scatter of the RVs, the data sets have a root-mean-square (RMS) of 4.7\,\ms\, for SPIRou and 4.8\,\ms\, for SOPHIE.

\subsection{Global modeling}

In order to obtain a first approximation of the RV jitter magnitude, we applied a GP with a quasi-periodic kernel in the SOPHIE and SPIRou RV time series, following Section \ref{sec:prot}. The MCMC procedure to obtain the posterior distribution is set up with 50 walkers and 5000 steps after a burn-in phase of 500 steps. The priors applied in the hyper-parameters are listed in Table \ref{table:GP_RVs}. Since we measured the stellar rotation period from the spectropolarimetric data, we applied a more constraint prior to the GP period of the radial velocities. 

The best-fit model of the GP regressions are shown in Figure \ref{GP_RVs} and the final posterior distribution of the GP hyper-parameters are listed in Table \ref{table:GP_RVs}. The period found in the SOPHIE and SPIRou RVs of $34.2^{+2.8}_{-0.9}$ d and $39.2\pm2.4$ d, respectively, are consistent at a 3$\sigma$ level with the period of the longitudinal magnetic field of \protv d (see Section \ref{sec:prot}. The decay time found for the SOPHIE RVs is larger than for SPIRou and consistent with more than two stellar rotations ($l$ = $96^{+41}_{-47}$ d). In the case of SPIRou, however, the decay time of $38^{+4}_{-2}$ d is comparable with the rotation period of $39.2\pm2.4$ d. This effect is problematic for the quasi-periodicity of the signal. When there is a short evolution time scale, $l \leqslant P_{\rm rot}$, the periodicity of the signal is meaningless \citep[see e.g.,][]{Rajpaul2015,Barragan2022}. 

We tested applying a different prior in the decay time for the SPIRou RVs, keeping it uniform but with limits between 70 to 150 d, to ensure at least two rotation cycles. As results, we obtained consistent posterior distributions, within error bars, with the values from Table \ref{table:GP_RVs}, except for the decay time. With a different prior, we obtained a decay time of $73^{+5}_{-2}$ d. However, the log-likelihood of the first model (log-$L$ = -374) is slightly higher than the model with a new prior in the decay time (log-$L$ = -384), meaning that the first model fits better the data.

\begin{table}[ht]
\footnotesize
\caption{Priors and best fit hyper-parameters of the GP model using a quasi-periodic kernel in the SOPHIE and SPIRou radial velocities.}         
\label{table:GP_RVs}      
\centering

\begin{tabular}{c c| c | c c}     
\hline\hline       
          &        &     &  SOPHIE & SPIRou \\
Parameter & Units & Prior & Posterior & Posterior \\ 
\hline                    
   mean value $\mu $ & m/s        & $ \rm \mathcal{U}(-\infty,+\infty) $ & $0.6\pm1.4$ & $0.5\pm1.7$\\  
   RV jitter $\sigma $ & m/s    & $ \rm \mathcal{U}(0,+\infty) $ & $2.5\pm0.8$ & $2.0\pm0.3$ \\ 
   amplitude $\alpha $ & m/s       & $ \rm \mathcal{U}(0,+\infty) $ & $4.1^\pm 1.1$ & $5.4\pm1.2$   \\
   decay time $l$ & d          & $ \rm \mathcal{U}(30,300) $ & $96^{+41}_{-47}$ & $38^{+4}_{-2}$ \\
   smoothing factor $\beta$ & & $ \rm \mathcal{U}(0.25,1.25) $ & $0.5\pm0.2$ & $0.9\pm0.3$\\
   rotation period $P_{\rm rot}$& d   & $ \rm \mathcal{U}(30,50) $       & $34.2^{+2.8}_{-0.9}$ & $39.2\pm2.4$ \\
   RMS of residuals & m/s & & 2.6 & 2.4 \\
   $\chi^2$ & & & 2.1 & 1.8 \\
\hline                  
\end{tabular}
\tablefoot{The symbol $\mathcal{U}(a,b)$ defines an uniform prior with $a$ and $b$ the minimum and maximum limits, respectively.}
\end{table}

\subsection{Seasonal analysis}

The RVs observation time-span is clearly divided in three seasons (see Figure~\ref{fig:radialvelocities}): the first epoch from 58738 BJD to 58922 BJD (September 2019 - March 2020), the second epoch from 59088 BJD to 59297 BJD (August 2020 - March 2021), and the third epoch from 59440 BJD to 59607 BJD (August 2021 - January 2022). These subsets are longer than the subsets defined in Section~\ref{sec:ZDI} since in order to look for periodic variability it is required a longer time-span than for the ZDI analysis. The details of the RV subsets such as the start and ending dates as well as the number of data points included are listed in Table~\ref{table:season_GP_RV}. Note that these RV subset are defined as S$^{RV}_{1}$, S$^{RV}_{2}$, and S$^{RV}_{3}$ in order to differentiate them from the seasons defined in Section~\ref{sec:ZDI}. These RV subsets are depicted in Figure~\ref{fig:radialvelocities}.

Even though the seasons S$^{RV}_{1}$, S$^{RV}_{2}$ overlaps at some extent with S$_{1}$ and S$_{2}$, we can not directly compare S$_{3}$ with the RV data since only three SOPHIE observations are within this season. Therefore we defined a fourth season of RVs observations S$^{RV}_{4}$, from 59500 BJD to 59607 BJD (October 2021 - January 2022) that overlaps with S$_{4}$ and it will allow us to measure the differences between seasons S$_{3}$ and S$_{4}$ of the $B_\ell$ in the RVs data sets.

To have some insight about the variability of the RV signal, we define the amplitude $A$ as the peak-to-valley difference and we listed it alongside the RMS of the model residuals and the reduced $\chi^2$ in Table~\ref{table:season_GP_RV}, for all the data set and each RV season.

It is clearly seen in Figure \ref{GP_RVs} that the best-fitting GP model for the SOPHIE RVs is more stable in time than the one for the SPIRou RVs, and it is similar to a double sinusoidal model at $P_{\rm rot}$ and its first harmonic. As seen in the middle and bottom panels of Figure \ref{GP_RVs}, the signal in the SOPHIE RVs remains more or less consistent between S$^{RV}_{1}$ and S$^{RV}_{2}$, with similar shapes and amplitudes of $10.8\pm2.9$ and $9.8\pm3.0$ \ms, respectively. The goodness of the model is similar for the season S$^{RV}_{1}$ through S$^{RV}_{4}$, with RMS between 2.3 and 2.8 \ms. However, during S$^{RV}_{3}$ and S$^{RV}_{4}$ the reduced $\chi^2$ gets slightly worse than the other seasons. This may be due to the presence of outliers in the sub-data set.

On the contrary, the SPIRou RV signal is not consistent between one season to the other (see bottom panels of Figure \ref{GP_RVs}), exhibiting high variability. The amplitude of the RV signal goes slightly up between S$^{RV}_{1}$ and S$^{RV}_{4}$, from $13.7\pm2.3$ to $15.2\pm2.2$ \ms, but it not significant and the difference is within the error bars. As in the SOPHIE RVs, the there no important differences in the RMS or reduced $\chi^2$ of each season, however, the goodness of the model gets worse for S$^{RV}_{3}$. As for the SOPHIE RVs, this may be due to the presence of outliers. If these outliers are removed, the RMS improves to 2.4 \ms.

At this point, we can not tell if there are significant differences between S$^{RV}_{3}$ and S$^{RV}_{4}$ of the SOPHIE and SPIRou RVs, in order to measure a possible impact in the RVs due to the changes in the magnetic field topology (see Section \ref{sec:ZDI}). As listed in Table \ref{table:GP_RVs}, there is little discrepancy between these two seasons in the amplitudes, RMS, and reduced $\chi^2$. Despite the fact that the $B_{\ell}$ shows high variability at this epochs, the radial velocities do not increase their scatter. This may mean that the variability in the longitudinal magnetic field is not directly expressed in high RV dispersion.

\begin{table}[]
\centering
\footnotesize
\setlength{\tabcolsep}{1pt}
\caption{Results of the RV analysis in the observational seasons of the SOPHIE and SPIRou.}
\renewcommand{\arraystretch}{1.5}
\begin{tabular}{lccc|ccc|ccc}
\hline \hline
    Name & Start & End & N & \multicolumn{3}{c}{SOPHIE}     & \multicolumn{3}{|c}{SPIRou}    \\
    & &  &  & $A$ & RMS  & $\chi^2$ & $A$ & RMS & $\chi^2$ \\
      & & & & [m/s] & [m/s] & - & [m/s] & [m/s] &  - \\
        \hline

\hline
All & 09-09-19 & 27-01-22 & 201 & $13.6\pm2.5$ & 2.6 & 2.1  & $21.3\pm4.5$ & 2.4  & 1.8   \\ 
S$^{\rm RV}_{1}$ & 11-09-19 & 13-03-20 & 53 & $10.8\pm2.9$ & 2.6  & 1.9  & $13.7\pm2.3$  & 2.4 & 1.8   \\
S$^{\rm RV}_{2}$  & 26-08-20 & 04-03-21 & 72 & $9.8\pm3.0$  & 2.3  & 2.1   & $12.0\pm3.0$  & 2.0  & 1.5 \\
S$^{\rm RV}_{3}$  & 13-08-21 & 27-01-22 & 76 & $8.6\pm3.3$  & 2.7  & 3.4  & $15.2\pm2.2$   & 3.0  & 2.9   \\
S$^{\rm RV}_{4}$ & 12-10-21 & 27-01-22 & 58 & $8.6\pm2.8$ & 2.8 & 4.0 & $15.2\pm2.2$ &  2.9 & 2.8 \\
\hline
\label{table:season_GP_RV}
\end{tabular}
\tablefoot{Number of data points $N$, the amplitude of the RV variation is defined as the peak-to-valley difference of the GP model. The goodness of the fit is described with the RMS of the residuals alongside the reduced $\chi^2$.}
\end{table}

\section{Activity indicators}\label{section:activity}

\subsection{Optical activity indicators} 

Using the SOPHIE data we computed standard activity indicators that quantify distortions in the CCF profile, such as the BIS, FWHM, and contrast. The bisector inverse slope (BIS) \citep{Queloz2001} is defined as the velocity span between the top (55\% < CCF depth < 80\%) and the bottom (20\% < CCF depth < 40\%) of the CCF. The FWHM and the contrast are direct measurements of the width and the depth of the CCF, respectively.

\begin{figure*}[h!]
\centering
\includegraphics[width=\hsize]{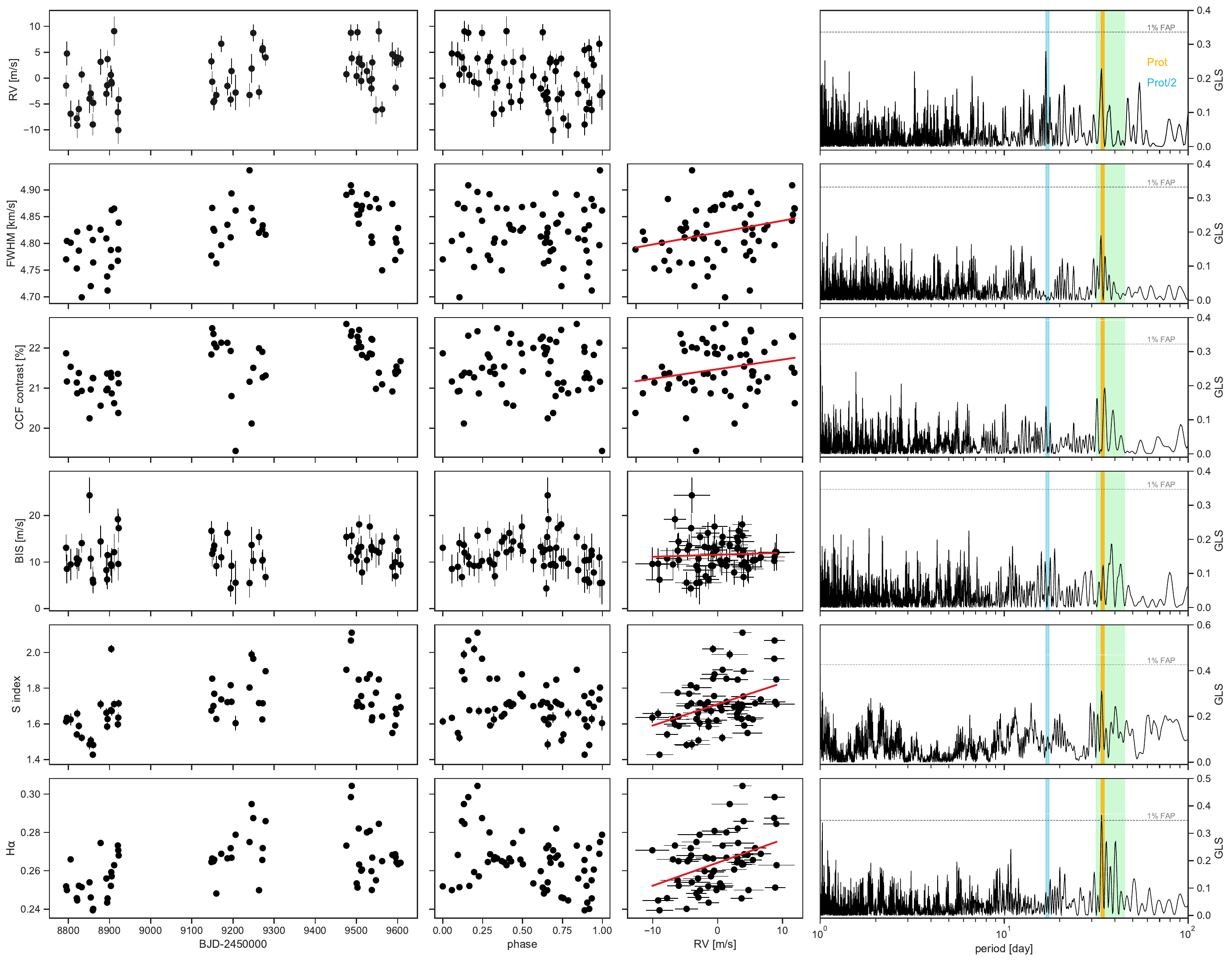}
\caption{SOPHIE RVs and optical activity indicators time series (\emph{left}), data folded at the $P_{\rm rot}$ of 34.4\,d (\emph{middle left}), correlation with the RVs where the red line depicts the linear correlation between the both quantities (\emph{middle right}), and GLS periodogram (\emph{right}) where the horizontal dashed black line indicates the FAP level at 1\% and the vertical color lines mark the $P_{\rm rot}$ at 34.4\,d in yellow, the first harmonic at $P_{\rm rot}/2$ in light blue and the time range of differential rotation in light green. }
\label{SOPHIE_activity}
\end{figure*}

The time series of the BIS, FWHM, and contrast of the CCF from the SOPHIE data are shown in Figure~\ref{SOPHIE_activity}, with their corresponding GLS periodograms. No significant periodicity close to the expected stellar rotation period of \prot is found in their periodograms.  

Within the SOPHIE spectral domain is located the well-known $\rm H\alpha$ line \citep{Kurster2003,Bonfils2007,Boisse2009}, and CaII H\&K lines that are measured to compute the Mt. Wilson S index  \citep{Wilson1968,Baliunas1995}. 
The so-called S index is a measure of the emission at the core of the CaII H\&K lines, located at 3922\r{A} and 3968\r{A} and for the $\rm H\alpha$ is the flux at 6562\r{A}. To compute these indices we follow \cite{Boisse2009}, where it is defined as the ratio between the flux in a defined range of continuum and the core of the absorption lines.

In Figure~\ref{SOPHIE_activity} are shown the time series of the $\rm H\alpha$ and S indices of our SOPHIE data along with their GLS periodograms. The time series of $\rm H\alpha$ shows a peak of periodicity at 33.7 d and in the case of the S index, long-term trends dominate the periodogram. 

Usually the correlation or anti-correlation between the RV and the activity indicators is a hint of the stellar origin of the signal \citep[e.g.,][]{Queloz2001,Boisse2011}. However, the lack of correlation can be due to phase shifts between the RVs and the activity indicators \citep[e.g.,][]{Bonfils2007,Dumusque2014a}. For example, \citet{CollierCameron2019} measured a temporal lag of 1 and 3 days between the maxima in the RVs and the maxima of the FWHM and BIS, respectively. In the present work we do not explore the possibility of the phase lags in our data set but we rather warn the reader about it. 

The SOPHIE BIS, FWHM, and contrast are not correlated or anti-correlated with the RVS, and the S index and $\rm H\alpha$ are correlated with the RVs with Pearson's coefficients of 0.4. We computed the p-value associated to this correlation in order to prove the significance. The p-value is defined as the probability that the observed correlation is a false positive, under a true null-hypothesis. In this case, the null-hypothesis is that there is no correlation between the variables. The correlation between the RVs and the $\rm H\alpha$ and the S index has a p-value of $8\cdot10^{-4}$ and $9\cdot10^{-4}$, respectively. This means that the found correlations are statistically significant.


We applied a GP regression following the same procedure as in Section \ref{sec:prot} to study the periodicity of the signal, since we have seen how the longitudinal magnetic field and the radial velocities follow a quasi-periodic trend. This behavior could have an effect in the classical GLS periodogram and hide the real period of the signal. Since the SOPHIE DRS do not have implemented yet the error bars derivation for the FWHM and the contrast, we assumed equal uncertainties for all the data points of 0.1 $\%$CCF for the contrast, and 0.01 \kms for the FWHM.

We used the same priors as in Section \ref{sec:radialvelocities} which are listed in Table \ref{table:GP_RVs}. To obtain the posterior distribution from the MCMC we 50 walkers and 5000 steps after a burn-in phase of 500 steps. The best-fitted GP model for the SOPHIE activity indicators are shown in Figure \ref{fig:GP_SOPHIE_activity}. The mean of the posterior distributions of the GP hyper-parameters are listed in Table \ref{table:GP_activity} and the corner plot of the posterior distributions are shown in Figure \ref{fig:GP_SOPHIE_activity}.

In the case of the SOPHIE activity indicators, the quasi-periodic GP fits a model with a periodicity in agreement with the expected rotation period of \prot. This is in particular interesting for the BIS, FWHM, and CCF contrast since their GLS periodogram do not exhibit significant peaks of periodicity at the rotation period. The decay time of the GP model of the FWHM, BIS, and H$\alpha$ are consistent with an average time scale decay of at least two rotation cycle, as also seen in the longitudinal magnetic field B$_\ell$. 



\subsection{Near-infrared activity indicators}

Within the SPIRou's LBL framework, we can obtain the differential line width \citep[dLW,][]{Zechmeister2018} and the chromatic velocity slope. The differential line width corresponds to the second derivative of the spectral profile and can be expressed in units of FWHM. We follow the definition by \citet{Artigau2022} using this quantity as $\rm FWHM_{\rm LBL}$ . The chromatic velocity slope is defined as the RV gradient as a function of wavelength \citep{Zechmeister2018}. 

The time series and the GLS periodograms of $\rm FWHM_{\rm LBL}$ and the chromatic velocity slope are shown in Figure~\ref{SPIROU_activity}, along with their phase-folded and correlation with RVs plots. When phase-folding $\rm FWHM_{\rm LBL}$ by the stellar rotation period of 34.4\,d, it seems to be modulated in time. Moreover, $\rm FWHM_{\rm LBL}$ is slightly correlated with the RVs with a Pearson's coefficient of 0.3 and a p-value of $4\cdot10^{-5}$, which means that the found correlation is significant. The highest peak of the $\rm FWHM_{\rm LBL}$ GLS periodogram is at 34.4\,d but with FAP slight above 1\%, which corresponds to the stellar rotation period. The chromatic velocity slope indicator is not correlated or anti-correlated with the RVs, however the most significant peak of periodicity in its periodogram is located at 31.3\,d, close to the rotation period.



As for the SOPHIE activity indicators, we also computed a GP model in the SPIRou activity indicators including the FWHM, chromatic velocity slope, and pEW of the near-infrared spectral lines. The best-fit GP model for the FWHM and the chromatic velocity slope are shown in Figure \ref{fig:GP_SPIRou_activity} and the mean of the posterior distribution of the hyper-parameters are listed in Table \ref{table:GP_activity}. The corner plots of the posterior distributions are shown in Figures \ref{fig:GP_SPIRou_activity_corner1} and \ref{fig:GP_SPIRou_activity_corner2}. The period measured in the FWHM and the chromatic velocity slope are in agreement, within error bars, with the expected rotation period of \protv. Similar to the activity indicators in the optical, the measured decay time of the signal are consistent with two stellar rotation cycles. The results of the near-infrared spectral lines are discussed in the next subsection.

\subsubsection{Diagnostics on near-infrared spectral lines}

Observations in the infrared domain introduce two challenges. On the one hand, we lack well-identified and characterized spectral lines in this domain that can be used as stellar activity tracers. On the other hand, M dwarf spectra do not always show a clear flux continuum within the whole near-infrared wavelength range, since their spectra have a large number of absorption lines and hamper the proper measurement of equivalent widths.

We selected near-infrared absorption lines of chemical species that have been proved as activity indicators in optical or near-infrared wavelengths for Sun-like stars or M dwarfs, to test their potential for near-infrared M dwarfs data. These absorption lines are the titanium (Ti \Romannum{1}) line at 10499\AA~\citep{Spina2020}, the helium triplet (He \Romannum{1}) at 10833\AA~\citep{Schofer2019,Fuhrmeister2019,Fuhrmeister2020}, the iron (Fe \Romannum{1}) line at 11692\AA~ \citep{Yana2019,Cretignier2020}, the potassium line (K \Romannum{1}) at 12435\AA~\citep{Barrado2001,Robertson2016,Terrien2022}, and the aluminum (Al \Romannum{1}) line at 13154\AA~\citep{Spina2020}. The specific central wavelength of these lines is listed in Table~\ref{table:nearIR_lines}.

The equivalent width (EW) of spectral lines can be used as a proxy for the strength of the chromospheric activity and it is defined as following \citet{Gray2008_book}:

\begin{equation}\label{ew_eq}
    EW = \int\bigg( 1-\frac{F_\lambda}{F_0}\bigg) ~d\lambda
\end{equation}

\noindent where $F_{\lambda}$ is the flux of the spectral line and $F_0$ is the continuum flux surrounding the spectral line.

As the continuum in the M dwarfs spectra is unknown, we used a pseudo-continuum as in \cite{Schofer2019}, and therefore we measured the pseudo-equivalent width (pEW). To this, we first stacked the four consecutive exposures of each SPIRou observation to obtain one single high signal-to-noise spectra per epoch. Then, we used a Python re-implementation of the continuum fitting routine from IRAF\footnote{Image Reduction and Analysis Facility, \url{https://iraf-community.github.io}} to obtain the pseudo-continuum. This pseudo-continuum is defined as the local continuum of a spectral zone of 2 to 5 \AA~ where the spectral line is at the center. The details of the wavelength range that we used to compute the pseudo-continuum are listed in Table~\ref{table:nearIR_lines}. 

To each spectrum per epoch, we fitted a Voigt profile in the absorption lines within an MCMC framework using the \texttt{emcee} \citep{emcee} Python package. The parameters of the Voigt profile are the Gaussian width, the Cauchy-Lorentz scale, the amplitude, and the center of the line. The fitted model of the Voigt profile corresponds to $F_\lambda$ in Equation~\ref{ew_eq}. We used the posterior distributions of the fitted parameters to compute the uncertainties of the pEW using the bootstrap method. To look for periodicities in the time series of the tested spectral lines, we computed the GLS periodogram. We then considered as significant a peak with its FAP below 1\%. In Table~\ref{table:nearIR_lines} are shown the highest peaks of periodicities in the GLS periodograms for each line within a range of 1 d to 100 d, to exclude periodicities coming from the 1-d alias and long-term variability.

\begin{figure*}[ht]
\centering
\includegraphics[width=0.96\hsize]{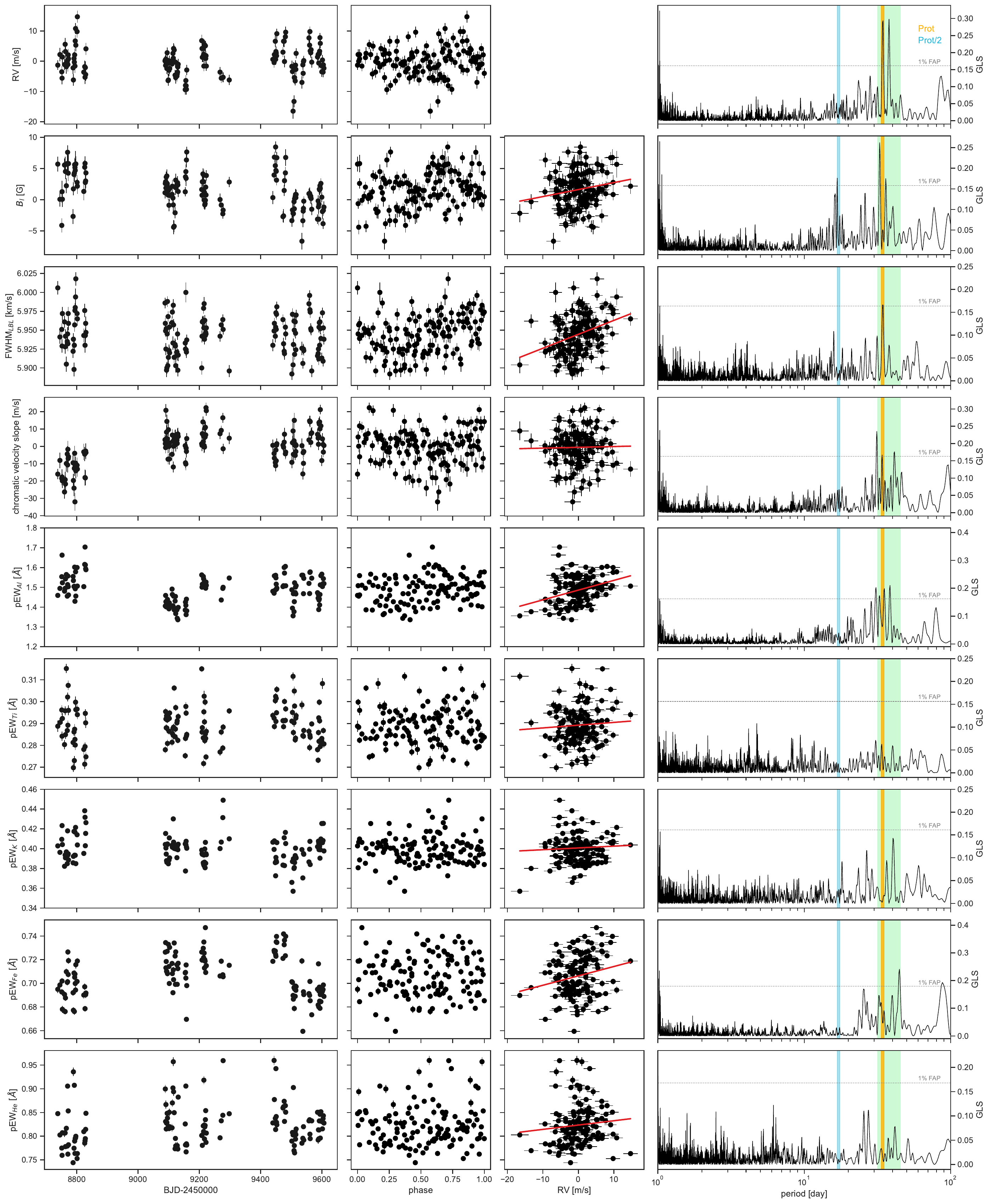}
\caption{Same as Figure~\ref{SOPHIE_activity} for SPIRou data.}
\label{SPIROU_activity}
\end{figure*}

In Figure~\ref{SPIROU_activity} are displayed the pEW time series of the Al \Romannum{1}, Ti \Romannum{1}, K \Romannum{1}, Fe \Romannum{1}, and He \Romannum{1} lines, along with their GLS periodograms, phase-folded plot, and correlation plot with RVs. The highest peak of periodicity of each line is listed in Table~\ref{table:nearIR_lines}, for all the data and for each of the RV seasons. Only Al \Romannum{1} exhibits peaks of periodicities with a FAP below 1\% at 35 d\,, close to the expected stellar rotation of 34.4 d. Moreover, Al \Romannum{1} shows several peaks between 30 to 40 d that may be related to the differential rotation. The lines best correlated with the RVs are Al \Romannum{1} and Fe \Romannum{1} with a Pearson's coefficient of 0.3 and 0.2, respectively. However, the correlation between the RVs and Al \Romannum{1} is more significant with a p-value of $4\cdot10_{-4}$ versus the p-value of 0.01 in the correlation with the Fe \Romannum{1}. 

Excluding Al \Romannum{1}, the rest of the spectral lines tested Fe \Romannum{1}, Ti \Romannum{1}, K \Romannum{1}, and He \Romannum{1} do not seem to be modulated by the stellar rotation and show small or none correlation with the RVs, at least from the analysis of their periodograms. The Fe \Romannum{1} line in particular, shows a peak of periodicity at 44.7 d\, which is similar to the expected stellar rotation at the poles ($P_{\rm pol} = 45.5$\,d). This line could be tracing active surface features at latitudes close to the stellar pole.



The best-fitting GP models of the tested spectral lines are shown in Figure \ref{fig:GP_SPIRou_activity} and the final posterior distribution of the hyper-parameters are listed in Table \ref{table:GP_activity}. Most of the spectral lines have periodicities longer than the measured rotation period of \protv, except for the He \Romannum{1}. However, as we have seen in previous results within this work, it seems that the He \Romannum{1} line is not sensitive to stellar activity. Moreover, it is clear that the GP do not fully modeled the variability in the test spectral lines, and we observed high values of the reduced $\chi^2$. The underestimation of the uncertainties of the pEW computation could play a role in this, nevertheless, we will investigate it further in future works.

Thanks to the analysis of the GLS periodograms and the GP regression, we can see that the periodicity of the Al \Romannum{1} and Fe \Romannum{1} lines are consistent between these two techniques, however, the decay time is shorter than expected and do not covers two complete cycles. 

\begin{table*}
\caption{Properties and periodicities of the near-infrared spectral lines within the SPIRou domain.}         
\label{table:nearIR_lines}      
\centering
\footnotesize

\begin{tabular}{c| c c | c c c c}     
\hline\hline       
& \multicolumn{2}{c}{Line information} & \multicolumn{4}{c}{Periodicities} \\ 
Line & Central wavelength  & Continuum range  & All data & S$^{RV}_{1}$ & S$^{RV}_{2}$ & S$^{RV}_{3}$ \\ 
     & [\AA] & [\AA] & [d] & [d] & [d] & [d]\\
\hline                    
    Ti \Romannum{1} & 10498.99 &  10490 - 10510 & 4.7 & 39.6 & 8.2 & 9.4 \\
    He \Romannum{1} & 10833.25 &  10820 - 10845 & 6.2 & 1.6 & 27.2 & 25.3  \\
    Fe \Romannum{1} & 11693.18 & 11684 - 11704 & \textbf{44.7} & 30.8 & 39.5 & 33.4 \\
    K \Romannum{1}  & 12435.58 & 12410 - 12460 & 40.5 & 12.3 & 39.5 & 39.5\\
    Al \Romannum{1} & 13154.32 & 13140 - 13170 & \textbf{35.0} & 2.0 & 35.5 & \textbf{41.4}\\
\hline                  
\end{tabular}
\tablefoot{Wavelength values are in vacuum. In bold are the peaks of the GLS periodograms with FAP<1\%. We searched for periodicities related to the stellar rotation period in a range between 1 d and 100 d}
\end{table*}

\subsection{Seasonal analysis}

Since stellar activity has a quasi-periodic behavior, the activity signals should be consistent in short-time scales and therefore, their periodicities could be measured with the periodograms in a seasonal analysis. We observed that the strongest peaks of the GLS periodograms of the optical activity indicators are not consistent through the three seasons of data previously defined as S$^{RV}_{1}$, S$^{RV}_{2}$, and S$^{RV}_{3}$. The GLS periodograms for each one of the three subsets are displayed in Figure~\ref{SOPHIE_seasons}. For the CCF contrast and the BIS, the periodicity close to the rotation period gains significance towards the second subset but remains with a high FAP. On the other hand, the peak at the stellar rotation period decreases in significance towards the 2021/2022 observations. During S$^{RV}_{1}$, the H$\alpha$ time series shows an important power excess at a period slightly longer than the expected rotation period. During S$^{RV}_{2}$ and S$^{RV}_{3}$, H$\alpha$ keeps some power excess around the rotation period but is not significant.

The SPIRou activity indicators, the FWHM$_{\rm LBL}$, and the chromatic velocity slope, do not show the same season behavior. As shown in Figure~\ref{SPIRou_seasons}, the FWHM$_{\rm LBL}$ have a peak of periodicity not related to the stellar rotation in S$^{RV}_{1}$, during S$^{RV}_{2}$ and S$^{RV}_{3}$ we can see some power excess close to the rotation period but with FAP above 10\%. On the other hand, the chromatic velocity slope has a high-significant peak of periodicity slightly shorter than the rotation period in S$^{RV}_{3}$, and nothing during S$^{RV}_{1}$ or S$^{RV}_{3}$. The periodicity seen in S$^{RV}_{3}$ may be affected by the window function of that particular season (see Figure~\ref{SPIRou_seasons}).

The tested near-infrared spectral lines also do not keep consistent periodicities across the three subsets of RVs observations S$^{RV}_{1}$, S$^{RV}_{2}$, and S$^{RV}_{3}$. The strongest peaks of the GLS periodograms of Al \Romannum{1}, Ti \Romannum{1}, K \Romannum{1}, and Fe \Romannum{1} change from season to season and they are close to the rotation period at least during one season (see Table~\ref{table:nearIR_lines}. However, this periodicity observed during at least one season is $\sim39$\,d instead of the expected $P_{\rm rot}$. This may be related to active features in the stellar surface at mid-latitudes where the rotation should the larger than at the equator. The Fe \Romannum{1} line in particular, even though when considering the whole data set its periodicity is larger than $P_{\rm rot}$, the periodicities of each RV season is between 30 to 40 d\, which could be coming from the stellar rotation modulation. This variability in the periodicities observed for the spectral lines may be a hint of the high temporal variability in the stellar activity manifest as highly variable active features.

Although we can see some hints of seasonal behaviors in the optical and near-infrared activity indicators in this work, most of their periodicities are not significant and therefore, we can not describe trends of the stellar activity during each season. For example, H$\alpha$ is clearly modulated by the stellar rotation during S$^{RV}_{1}$ which could be due to high chromospheric activity levels. However, we do not see this behavior in other activity indicators.

\section{Stellar activity modeling}\label{GPfiltering}

\begin{figure*}[h]
    \centering
    \includegraphics[width=0.8\linewidth]{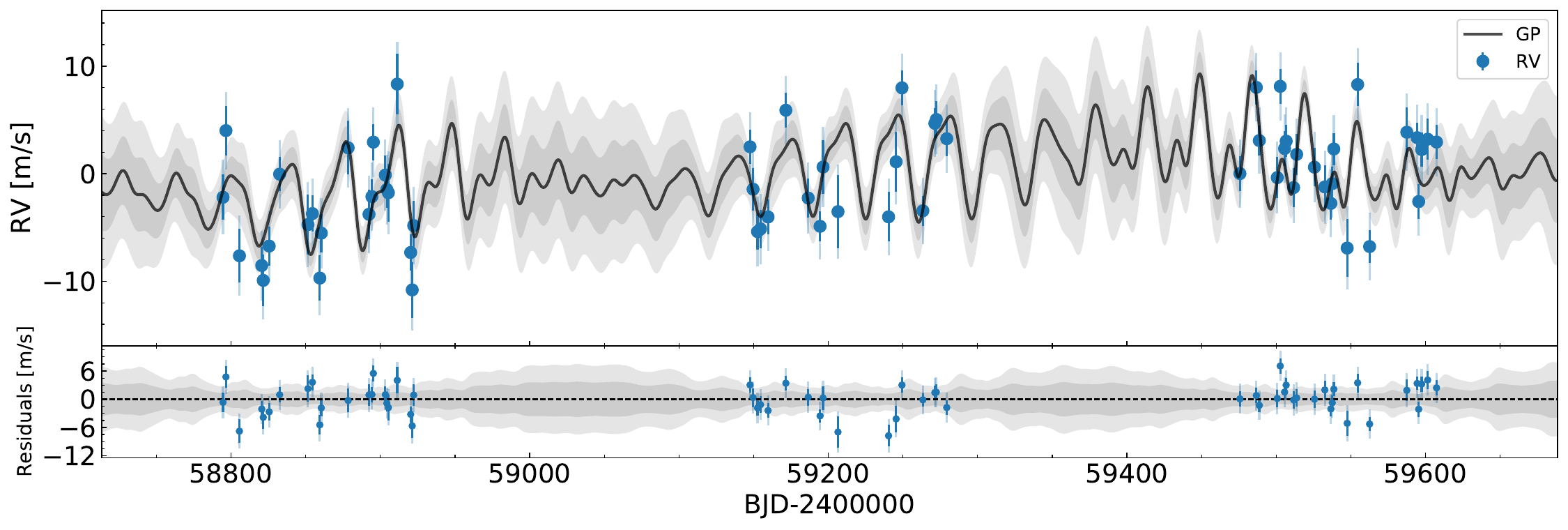} \\
    \includegraphics[width=0.8\linewidth]{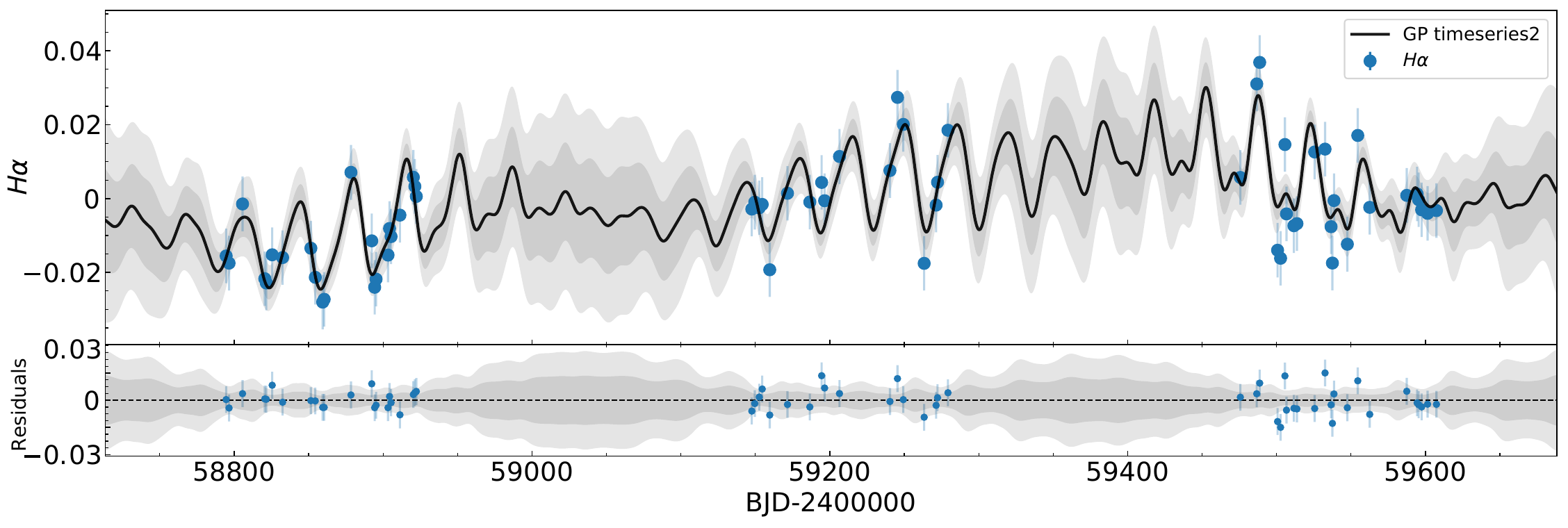} \\
    \includegraphics[width=0.8\linewidth]{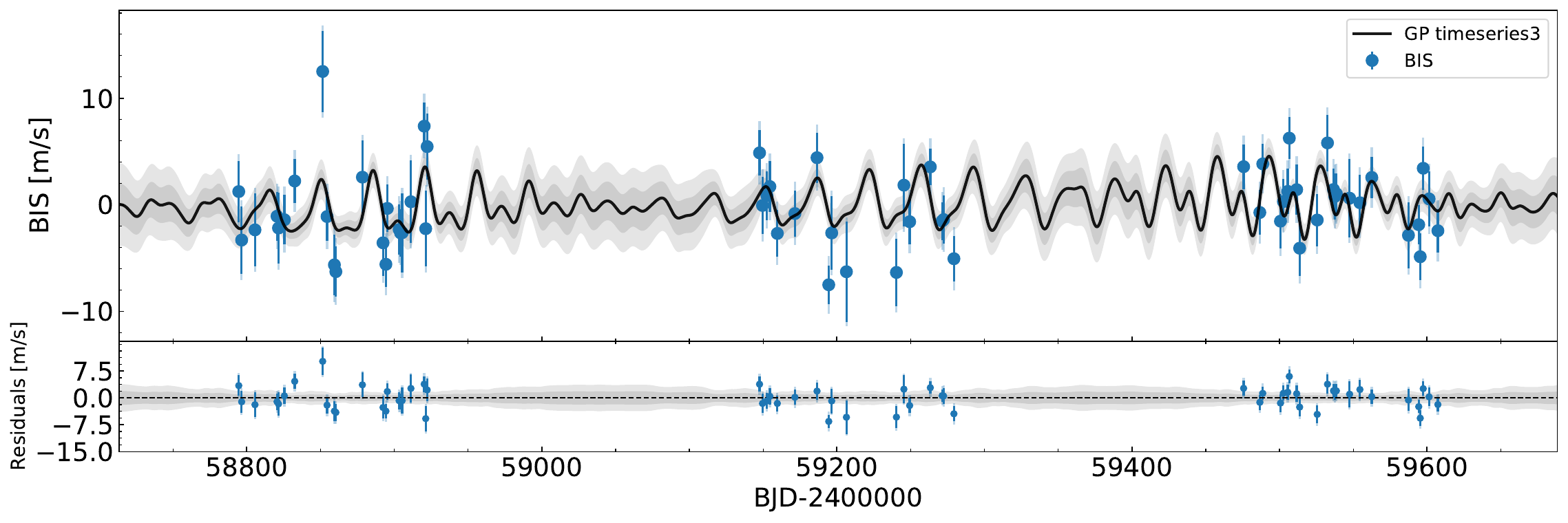} \\
    \caption{Multi-dimensional GP regression of the SOPHIE RVs using as ancillary time series the $\rm H\alpha$ index and the BIS. The black line indicates the GP model and the grey areas are the $1\sigma$ and $2\sigma$ levels. Below each time series are the residuals of the GP model. For all the time series the mean value has been removed.}
    \label{fig:SOPHIE_multiGP}
\end{figure*}

\begin{figure*}[h]
    \centering
    \includegraphics[width=0.8\linewidth]{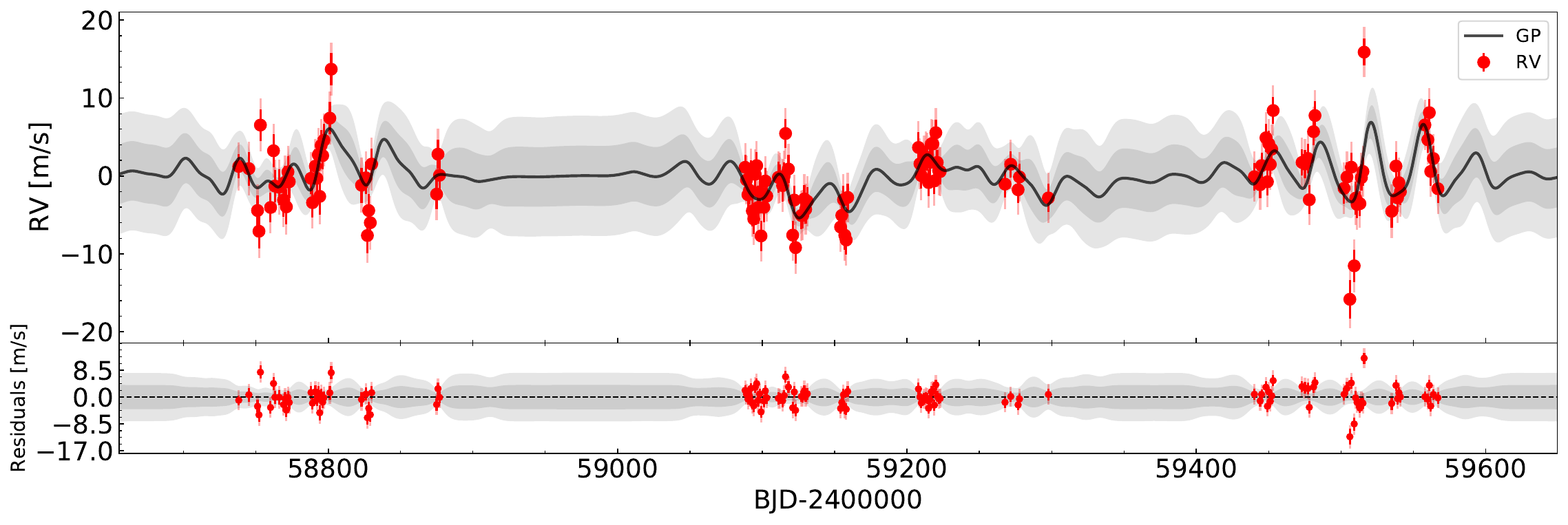} \\
    \includegraphics[width=0.8\linewidth]{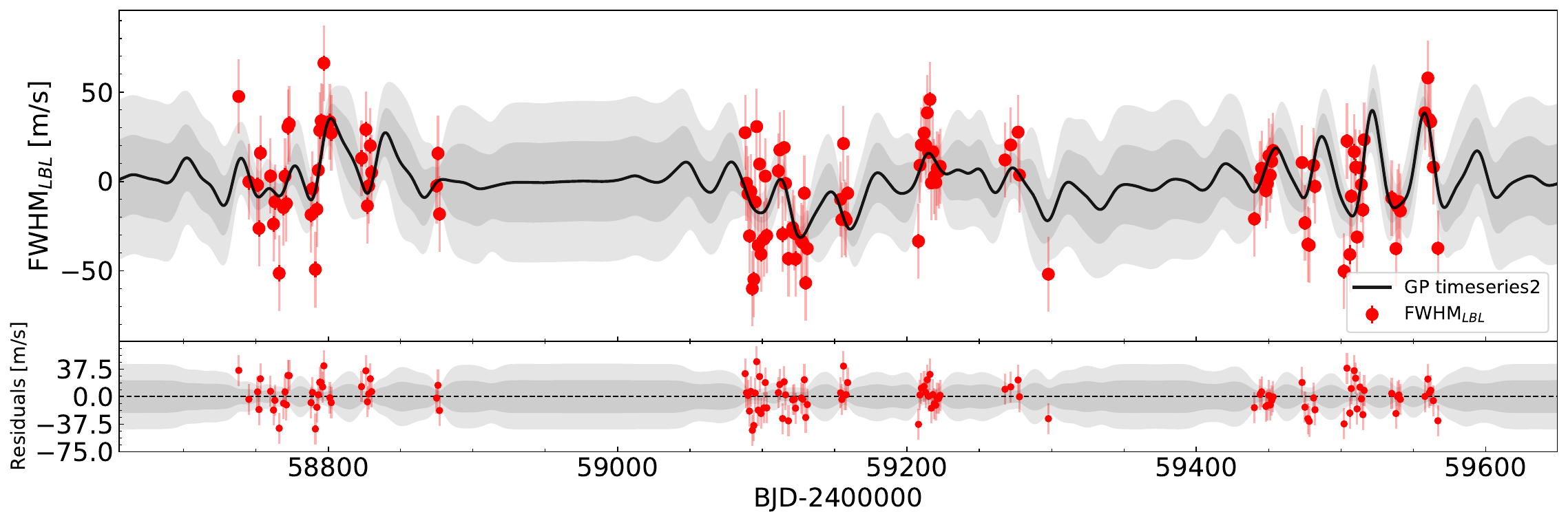} \\
    \caption{Multi-dimensional GP regression of the SPIRou RVs using as ancillary time series the $\rm FWHM_{LBL}$. The black line indicates the GP model and the grey areas are the $1\sigma$ and $2\sigma$ levels. Below each time series are the residuals of the GP model. For all the time series the mean value has been removed.}
    \label{fig:SPIRou_multiGP}
\end{figure*}

\citet{Rajpaul2015} describes a multi-dimensional GP framework on which the stellar activity in the RVs time series can be modeled simultaneously with the Keplerian signal using the information from activity indicators. In summary, this framework assumes that the stellar activity signals and the RVs can be modeled by the same latent function, namely $G(t)$, and its time derivative, $\dot{G}(t)$. $G(t)$ is related to the area of the visible stellar disc covered by active regions and $\dot{G}(t)$ describes the evolution in time of these active regions. Therefore, the RV variations induced by activity can be expressed as a linear combination of $G(t)$ and $\dot{G}(t)$. Physically, this linear combination will account for the convective blueshift suppression and the evolution of spots on the stellar surface.  

In principle, the ancillary time series, such as the activity indicators, should be also defined as linear combinations of $G(t)$ and $\dot{G}(t)$ if they are affected by both processes: the covered area by active regions and its time evolution in the stellar surface (e.g., BIS). However, some activity indicators, such as log$R'_{HK}$ and $S_{HK}$,  can be described only by $G(t)$ as they do not account for the time evolution of spots (see \citealt{Rajpaul2015} and \citealt{Barragan2022} for more details).

We used the Python code \texttt{pyaneti}\footnote{\url{https://github.com/oscaribv/pyaneti}}\citep{pyaneti,Barragan2022} to apply the multi-dimensional GP framework in our SOPHIE and SPIRou data. As the code was built for planet RV and transit modeling and there is no planet signal detected in this system, we set up a planet signal with an amplitude of $K=0$ m/s then only the stellar activity signal is considered in the modeling. The kernel of the GP regression is set to be the same as in Section~\ref{sec:magfield}, Equation~\ref{eq:kernel2}. However, the amplitude $A$ in Equation~\ref{eq:kernel2} in the multi-dimensional framework is replaced by the amplitudes $C_{i}$ as the following:

\begin{align*}
    \Delta RV &= C_0 G(t) + C_1 \dot{G(t)} \\
    \Theta_1 &= C_2 G(t) + C_3 \dot{G(t)} \\
    \Theta_2 &= C_4 G(t) + C_5 \dot{G(t)} \\
    \;\; \vdots  \notag\\
    \Theta_N &= C_i G(t) + C_{i+1} \dot{G(t)} \\
\end{align*}
 \noindent
where $\Theta_{1,\hdots,N}$ are the activity indicators. In this way it is guaranteed that the hyper-parameters of the kernel will be shared except for the amplitudes $C_i$. We follow the procedure as in Section~\ref{sec:magfield} to obtain the posterior distributions.

We modelled the stellar activity of the SOPHIE RVs using as ancillary time series the $\rm H\alpha$ index and the BIS as the following:
\begin{align*}
    \Delta RV_{\rm SOPHIE} &= C_0 G(t) + C_1 \dot{G(t)} \\
    H\alpha &= C_2 G(t)  \\
    {\rm BIS} &= C_4 G(t) + C_5 \dot{G(t)} \\
\end{align*}

We chose the $\rm H\alpha$ instead of the S index because the former shows a peak of periodicity at the stellar rotation period and is clearly correlated with the RVs (see Figure~\ref{SOPHIE_activity}). The details about the computation of the activity indicators are in Section~\ref{section:activity}.

The model obtained from the multi-dimensional GP regression is shown in Figure~\ref{fig:SOPHIE_multiGP} for the RVs, the $\rm H\alpha$, and the BIS time series. The priors and posteriors are listed in Table~\ref{table:multiGP_SOPHIE_RV} and in Figure \ref{fig:multiGP_SOPHIE_corner} is the corner plot of the posterior distributions. The RV jitter is $2.7\pm0.6$ m/s and the rotation period is determined at $35.4\pm1.0$\, d. The residuals of the RVs have a scatter of 2.9 m/s and do not show any significant periodicity in the GLS periodogram.

In the case of the SPIRou data set, we used as ancillary time series the $\rm FWHM_{LBL}$ as it is the activity indicator that is best correlated to the RVs and shows a significant peak at the stellar rotation period (see Figure~\ref{SPIROU_activity}). The model is the following:

\begin{align*}
    \Delta RV_{\rm SPIRou} &= C_0 G(t) + C_1 \dot{G(t)} \\
    {\rm FWHM}_{\rm LBL} &= C_2 G(t)  \\
\end{align*}

We only used $\rm FWHM_{LBL}$ because there is no analogous to the BIS or any other activity indicator that could trace the temporal evolution of the active regions in the SPIRou data set. The model obtained for the RVs and the $\rm FWHM_{LBL}$ is shown in Figure~\ref{fig:SPIRou_multiGP} and the priors and posteriors are listed in Table~\ref{table:multiGP_SOPHIE_RV}. In Figure \ref{fig:multiGP_SPIRou_corner} is the corner plot of the posterior distributions. The rotation period was found at $37.7_{-2.8}^{+3.0}$ d, the RV jitter is $2.7\pm0.3$ \ms and the RMS of the RV residuals is 3.1 m/s, and  The GLS periodogram of the RV residuals does not show any periodicity. 


We tested the $B_\ell$ as ancillary time series together with the $\rm FWHM_{LBL}$, since it has been shown how it traces the magnetic activity, we could filter more activity signal in the RVs. The stellar rotation period obtained from this test is in better agreement with the value obtained in Section~\ref{sec:magfield} than the one from only using $\rm FWHM_{LBL}$ as ancillary time series, with a value of $P_{\rm rot} = 34.6\pm0.8$\, d. However, the scatter of the residuals is higher with a value of 3.7\,\ms. 

In Figure~\ref{fig:SOPHIE_multiGP} we see the complexity in the structures describing the SOPHIE RVs which is due to the high harmonic complexity of the signal. We can probe this with the smoothing factor $\beta$ value from the GP model. The smoothing factor $\beta$ is higher for the SOPHIE RVs ($\beta=0.81^{+0.29}_{-0.20}$) than for SPIRou ($\beta=0.62^{+0.31}_{-0.18}$). 

The high harmonic complexity seen in the optical can be empirically proven with the amplitudes of $G(t)$ and $\dot{G}(t)$ \citep{Barragan2022b}. The function $\dot{G}(t)$ describes the area covered by active regions on the stellar surface as a function of time \citep{Aigrain2012,Rajpaul2015}. We see that the amplitude of $G(t)$, $C_{0}=3.6^{+1.6}_{-1.0}$ is smaller than the amplitude of $\dot{G}(t)$, $C_{1}=24.2^{+12.3}_{-8.9}$, which explains the observed complexity.

However, we see a different result in the near-infrared. The amplitudes of $G(t)$ and $\dot{G}(t)$ are similar with values of $C_{0}=3.8^{+1.5}_{-0.8}$ and $C_{1}=3.2^{+4.7}_{-2.3}$. This could be an indication that the function $\dot{G}(t)$ is not fully describing the origin of the RV variations and might be evidence of a different process dominating the stellar activity in the near-infrared, such as the Zeeman effect. 

\begin{table}[ht]
\caption{Priors and best-fit hyper-parameters of the multi-dimensional GP model using a quasi-periodic kernel in the SOPHIE and SPIRou data.}         
\label{table:multiGP_SOPHIE_RV}      
\centering
\footnotesize
\setlength{\tabcolsep}{1.5pt}
\renewcommand{\arraystretch}{1.4}
\begin{tabular}{c|c c c c}     
\hline\hline
& & SOPHIE & SPIRou \\
Parameter & Prior & Posterior  & Posterior \\
\hline
Mean RV [m/s]  & $\rm \mathcal{U}(-\infty,+\infty)$ & $8729.6\pm1.5$  & $9345.9\pm1.0$\\
RV jitter [m/s] & $\rm \mathcal{U}(0,100)$& $2.7 \pm0.6$  & $2.7 \pm 0.3$\\
Mean H$\alpha$ & $\rm \mathcal{U}(-\infty,+\infty)$ & $0.267\pm0.009$ &  \\
H$\alpha$ jitter & $\rm \mathcal{U}(0,100)$ &$0.007\pm0.001$ & \\
Mean BIS [m/s] & $\rm \mathcal{U}(-\infty,+\infty)$ & $1.8\pm1.0$ & \\
BIS jitter [m/s] & $\rm \mathcal{U}(0,100)$& $2.0\pm0.5$ & \\
Mean FWHM$_{\rm LBL}$ [m/s] & $\rm \mathcal{U}(-\infty,+\infty)$&  & $6070\pm7$ \\
FWHM$_{\rm LBL}$ jitter  [m/s] & $\rm \mathcal{U}(0,100)$& & $20.7\pm1.7$ \\
$C_0$ & $\rm \mathcal{U}(0,100)$ & $3.6 _{ - 1.0 } ^ { + 1.6 }$ & $3.8 _{ - 0.8 } ^ { + 1.5 }$\\
$C_1$ & $\rm \mathcal{U}(0,100)$ & $24.2 _{ - 8.9 } ^ { + 12.3 }$  & $3.2 _{ - 2.3 } ^ { + 4.7 }$\\ 
$C_2$ & $ \rm \mathcal{U}(0,100)$ & $17.1 _{ - 4.0 } ^ { + 6.5 }$  & $22.4_{ - 5.3 } ^ { + 7.8 }$\\
$C_4$ &$ \rm \mathcal{U}(0,100)$ & $9.5 _{ - 0.7 } ^ { + 0.8}$\\
$C_5$ & $ \rm \mathcal{U}(0,100)$& $-17.5 _{ - 12.3 } ^ { + 7.9 }$\\
decay time $l$ [d] & $ \rm \mathcal{U}(30,150)$ & $86 _{ - 15 } ^ { + 10 }$  & $37 _{ - 5 } ^ { + 11 }$\\
smoothing factor $\beta$ & $ \rm \mathcal{U}(0.1,2.0)$ & $0.81 _{ - 0.20 } ^ { + 0.29 }$ & $0.62 _{ - 0.18 } ^ { + 0.31 }$\\
rotation period $P_{\rm rot}$ [d] & $ \rm \mathcal{U}(20,50)$ & $35.4 \pm1.0$ & $37.7 _{ - 2.8 } ^ { + 3.0 }$\\
RMS of RV residuals [m/s]  & & 2.9 & 3.1 \\
$\chi^2$  &  &  1.02 & 1.01 \\
\hline
\end{tabular}
\tablefoot{The symbol $ \rm \mathcal{U}(a,b)$ defines an uniform prior with $a$ and $b$ the minimum and maximum limits, respectively.}
\end{table}

\section{Detection of RV planet's signal}\label{sec:planet}

Up to date, no planet has been confirmed orbiting Gl~205. From our RV analysis of Section~\ref{sec:radialvelocities}, we do not observe periodicities that could be attributable to Keplerian signals. Considering the total time span of our SOPHIE and SPIRou observations of almost 900 days, we can discard long-period planets with orbital periods longer than 450 days. In particular, the detection of long-period planets would be highly affected due to the two gaps of observations of $\sim$100 days each (see Figure~\ref{fig:radialvelocities}). Further observation campaigns of Gl~205 may help on the discovery of long-period planets around this star.

Each well-defined season of RVs observations in Figure~\ref{fig:radialvelocities} last 184, 209, and 167 days, therefore our highest detection sensitivity is given for planets with periods of less than 100 days and with amplitudes greater than 4.8 \ms, which corresponds to the scatter in the RVs. Moreover, we do not observe periodicity peaks in the GLS periodograms of the SOPHIE and SPIRou RVs that are not related to the stellar rotation period.

The observed peaks of periodicity in the RVs GLS periodogram are not consistent during the three seasons of observations and they are always correlated with one or more activity indicators. Furthermore, the procedure to filter the stellar activity described in Section~\ref{GPfiltering} allows us to discard remnant Keplerian periodic signals on the RVs. In both cases, for the SOPHIE and SPIRou data sets, the multi-dimensional GP absorbs most of the activity-induced RV signal leaving an RMS of the residuals of $\sim$3 \ms\,. The GLS periodogram of these RVs residuals does not exhibit any periodicity left with low FAP that could be attributed to a Keplerian signal.

Furthermore, we tested the coherence and consistency of the periodicities using the Stacked Bayesian generalized Lomb-Scargle (Stacked-BGLS) periodogram from \citet{Mortier2017}. The main idea behind this algorithm to disentangle signals is that stellar activity will produce short-lived incoherent signals while a Keplerian is a long-lived consistent signal.

\begin{figure*}[h]
    \centering
    \includegraphics[width=1.0\hsize]{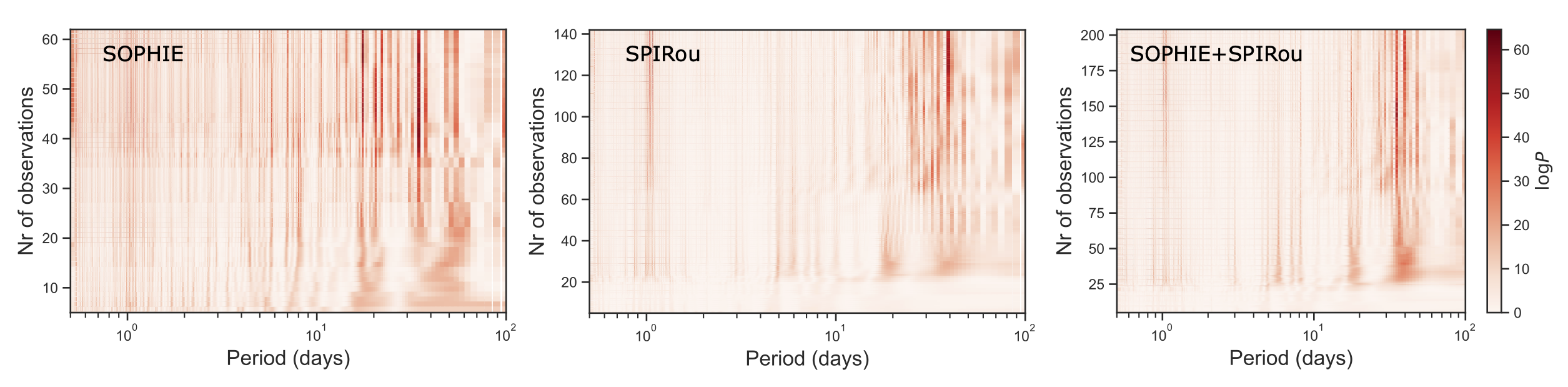}
  
    \caption{Stacked Bayesian generalized Lomb-Scargle periodogram of the SOPHIE, SPIRou and SOPHIE+SPIRou RVs. The periodogram of each observation is stacked along the y-axis. The color scale indicates the logarithm of the probability where dark-red indicates higher probability.}
      \label{fig:BLS}
\end{figure*} 

The Stacked-BGLS periodogram for the RVs from SOPHIE and SPIRou, together and independently, are shown in Figure~\ref{fig:BLS}. We observed that there are no strong signals in the periodograms at periods between 0 to 100 days. However, the activity close to 34 d\, is accompanied by other signals at close periods and show some level of variable probability, as expected for stellar activity signals.


\section{Discussion and Conclusions}\label{sec:discussion}

In this work, we present the long-term stellar activity and magnetic field characterization of the early, moderately-active M dwarf Gl~205 using quasi-simultaneous optical and near-infrared high-resolution spectra from the SOPHIE/OHP and SPIRou/CFHT spectrographs. We used the SPIRou circularly polarized spectra to measure the Zeeman signature produced by the presence of the magnetic field. The longitudinal magnetic field $B_{\ell}$ is modulated by the stellar rotation with a period of \prot, which is in agreement with previous works \citep{Kiraga2007,Bonfils2013,Hebrard2016}. The analysis of the periodicities found for the $B_{\ell}$ and activity indicators reinforces the efficiency and accuracy of the $B_{\ell}$ to constraint the $P_{\rm rot}$, over the standard activity indicators for early M dwarfs. 

We applied the Zeeman Doppler imaging technique to reconstruct the large-scale magnetic field map of the star over three seasons of observations. We observe a temporal evolution of the magnetic field topology while the strength of the large-scale field remains mostly constant. Moreover, we confirmed the differential rotation over the stellar surface of Gl~205 which could explain the disagreement between the stellar rotation period measured by previous studies and from different activity proxies. 

We derived the radial velocities using the CCF technique for the SOPHIE data and the line-by-line method for the SPIRou data. Both RVs data sets are comparable in amplitude and scatter with variations due to stellar activity. We applied a quasi-periodic GP regression in the SOPHIE and SPIRou radial velocities. Their periodicities are compatible with the result of the longitudinal magnetic field $B_\ell$. 

We filtered the activity-induced RV variations using a multi-dimensional Gaussian Process regression framework. For the optical RVs, we used the $\rm H\alpha$ index and the BIS as ancillary time series and, for the near-infrared, the $\rm FWHM_{\rm LBL}$. The obtained models fit the RVs down to the noise level with 2.7 \ms\, of scattering in the residuals for both instruments. Since we did not detect periodic signals left, we can rule out a priori the presence of Keplerian signatures in the given data set. Moreover, we used TESS photometry to discard the presence of transit events in the two available sectors. Nevertheless, we can not rule out completely the presence of planets around Gl~205 due to the flexibility of the GPs. At certain periods, in particular close to the stellar rotation period, the GP could still absorb the Keplerian signature in the RVs \cite{Rajpaul2015}.

\subsection{Magnetic field temporal evolution}

The long-term follow-up of more than two years of data presented in this work allows us to characterize the evolution of the large-scale magnetic field of Gl~205. The topology evolves significantly from a poloidal dipolar field with a strength of <B>=12 G in 2019 to a dominantly toroidal field in 2022 with a similar field strength of <B>=12.3 G. This change in the topology occurred quickly within two months during our last set of SPIRou observations. 

Previous studies of moderately-active early M dwarfs samples have shown the diversity of magnetic field topologies \citep{Donati2008,Hebrard2016,Martioli2022} proposing that there is no unique topology pattern for these stars. This diversity is especially seen in partly-convective, slowly rotating M dwarfs. Long-term follow-up campaigns of the early M dwarfs, such as for Gl~410 and Gl~846 \citep{Donati2008,Hebrard2016}, revealed a temporal evolution of the topology of the magnetic field. The variability of the magnetic field seen in early M dwarfs is at least partly attributable to their differential rotation over the stellar surface \citep{Donati2008,Morin2008,Hebrard2016}. 

Although most of the seasons observed in this study and \citet{Hebrard2016} exhibit poloidal fields for the case of Gl~205, the fact that at least during one of our seasons the field is dominantly toroidal is unexpected, since previous studies in partly-convective stars \citep{Petit2008,See2016} found that only fast rotators could generate toroidal dominated fields. 

Our results of the large-scale magnetic field during the four seasons of $B_{\ell}$ (see Table~\ref{tab:results_ZDI}) are mostly consistent with the previous magnetic field study by \citealt{Hebrard2016}, however, in this work we have been able to characterize the long-term evolution of the large-scale field. The long-term SPIRou monitoring was key to revealing the true nature of the magnetic field evolution and its intrinsic variability. 


\subsection{Diagnostics on optical and near-infrared activity indicators}

Previous works on photometric and spectroscopic/spectropolarimetric data of Gl~205 have measured similar rotation period values ($\rm P_{rot}\sim33$ d). Nevertheless, the rotational modulated variations are not always clear in all the activity indicators time series. \citet{Bonfils2013} showed that the strongest peaks at $P_{\rm rot}$ of the $\rm H\alpha$ and S index periodograms are seasonal, meaning that they become significant when the data set is restricted to one observation season of $\sim300$ d. \citet{Hebrard2016} measured the periodicity of $B_{\ell}$, RV, FWHM, and $\rm H\alpha$ finding values between 33.46 to 41.9 d, where the best result was given by the $B_{\ell}$ with a rotation period of $33.63\pm0.37$\, d. This discrepancy between activity indicators was explained due to the presence of differential rotation over the stellar surface but the HARPS-Pol and NARVAL data did not allow for confirming it. 

We studied stellar activity indicators of Gl~205 in the optical and near-infrared domains. In the optical domain, the $H\alpha$ index from SOPHIE exhibits a good correlation with the RVs and is modulated by the stellar rotation period as seen by \citet{Bonfils2013} in HARPS data. This is expected since the $\rm H\alpha$ index is a well-studied activity indicator for M dwarfs and its correlation with the stellar rotation is clear in large samples \citep[e.g.,][]{Delfosse1998,Newton2017,Jeffers2018}. In particular, the Carmencita sample of \citet{Jeffers2018} shows that $\rm H\alpha$ inactive (pEW($\rm H\alpha$) > -0.5\AA) early M dwarfs, which is the case of Gl~205, tend to exhibit modulations at larger periods from 10 to 100 d. 

The other optical indicators included in this work do not show clear hints of tracing activity. Even though most of them tend to have incipient peaks of periodicities close to the stellar rotation in their GLS periodograms, these peaks are not significant or they do not persist for longer than one season. For example, the BIS becomes significantly modulated by the stellar rotation during the season S$^{RV}_{2}$. However, the periodicity is slightly longer than the rotation period suggesting that the traced active features were closer to the poles. It is not surprising that the BIS is not anti-correlated with the RVs \citep{Desort2007} in the case of Gl~205 since the SOPHIE resolution is larger than the $v\rm{sin}i$ of $0.7$\,\kms.

In the near-infrared domain, we found that the $\rm FWHM_{\rm LBL}$ is the best tracer of activity in our data set as it is correlated with the RV with a Pearson's coefficient of 0.4, and its periodogram has the strongest peak at the stellar rotation period. Previous studies have found the same trend for other M dwarfs \citep[e.g.,][]{Klein2021,Zicher2022}. Moreover, we tested the performance of near-infrared spectral lines as possible activity tracers. The He \Romannum{1} triplet at 10833\r{A} has been extensively studied in the CARMENES sample of M dwarfs \citep{Fuhrmeister2019,Fuhrmeister2020} showing that its variability is correlated with $\rm H\alpha$ only for active cases. In our work, we obtained high error bars from the fitting of the He \Romannum{1} line due to the high amount of absorption lines around it making it difficult to obtain a good Voigt profile. The index we derived does not show to be modulated by the stellar rotation period.

The K line at 12435\r{A} is studied in detail by \citet{Fuhrmeister2022}. The authors find that the K \Romannum{1} is rarely correlated or anti-correlated with $\rm H\alpha$ since this line is less sensitive to chromospheric variability than $\rm H\alpha$. Our results agree with the findings of \citet{Fuhrmeister2022} seeing that the K \Romannum{1} line is not correlated with the RV variations and it is not modulated by $\rm P_{rot}$.

Only the Al \Romannum{1} line at 13154\r{A} shows several peaks of periodicity within the range of the expected differential rotation suggesting that this line could be tracing several active features across the stellar latitudes. Even though the Ti \Romannum{1} at 10499\r{A} and Fe \Romannum{1} at 11693\r{A} lines are not modulated by the stellar rotation, they appear to be affected by the magnetic field since they follow more or less the same changes in phase. Al \Romannum{1} and Ti \Romannum{1} have not been previously studied as activity tracers for Sun-like or M dwarf stars, further analysis in a larger sample would be of great benefit to prove their capacity of activity tracer as they are located in relatively clear parts of the spectrum with a small presence of telluric lines.

Interestingly, the analysis of several activity indicators over time reveals changes in the periodicities over different seasons of observations which may indicate an evolution in the distribution of active features over the stellar surface of Gl~205. All the activity indicators studied in this work reveal different information as may be tracing various activity processes in the stellar surface and disclose the complexity of the stellar activity of Gl~205.

As explained before, the lack of correlation or anti-correlation between the RVs and the activity indicators can be a result of time lags between the time series. For that reason, we can not completely discard as activity tracers the activity indicators that are not correlated with the RVs. Nevertheless, the existence of the correlation is a hint of the stellar nature of the signal.

On the other hand, the GLS periodogram is the standard tool to detect periodicities in time series, however, its efficiency depends on the number of data points, time span, and sampling, among other characteristics. If a significant peak of periodicity is not found by the periodogram, it does not directly imply that the indicator is not rotationally modulated. In this work, we compared the periodicities found in the GLS periodograms versus the period of the quasi-periodic GP applied in all the activity indicators. In some cases where the GLS periodogram does not find periodicities in the dataset, the GP succeeded in constraining a periodicity close to the expected rotation period. However, in this particular case, we know beforehand the expected stellar rotation and then, we can constraint properly the period of the GP. In blind searches of the stellar rotation period, the results may differ.

The periods found in the GP models are all within the ranges of the differential rotation of the star, between 32 d to 45.5 d, except for He \Romannum{1}. In particular, these periods tend to be longer than the period measured in the longitudinal magnetic field. This may be a geometrical effect due to the inclination of Gl~205 which is $~60\deg$. From the observer's point of view, there is one pole hidden producing that we observe preferably active regions at longer latitudes whose rotation periods are longer.

Finding near-infrared spectral lines that could work as stellar activity tracers remains an open challenge for stellar characterization and exoplanet searches in this domain. This is particularly important for less active M dwarfs since chromospheric lines may be better tracers of stellar activity than CCF-based activity indicators \citep{Lafarga2021}. A complementary route to face this challenge consists in identifying subsets of lines that are more or less sensitive to the activity jitter \citep{Bellotti2022}.

\subsection{Optical and near-infrared activity-induced RV jitter}

In this study, we confirm that the observed RV variations of Gl~205 in the optical and near-infrared are due to stellar activity as previously stated. The RV periodicities over the three RVs sub-datasets and during the whole time series are always linked to one or more activity indicators, in both domains.

\citealt{Hebrard2016} defined the total RV jitter as the sum of the jitter due to the rotational modulation and the jitter that comes from random components (e.g., short-lived spots). Even though we are close to reaching the noise level of the data set with the multi-dimensional GP modeling, the remaining jitter of $2.7\pm0.3$ m/s for SPIRou and $2.7\pm0.6$ m/s for SOPHIE is comparable with the random component contribution proposed for Gl~205 of the order of $J_{\rm r}=2.7$ m/s. This jitter contribution is greater than the average error bars of 1.9 m/s. 


In the GP models applied to the radial velocities, we observe that the model of the optical RVs is consistent over time and resembles a sinusoidal with its first harmonic. The near-infrared RVs, on the other hand, show a more complex behavior with a rapidly evolving signal. The amplitudes measured from the GP model show a slightly larger value for the near-infrared RVs.\citet{SuarezMascareno2018} measured the optical RV variation due to the rotational modulation and obtained a mean value of 4.4 \ms for early M dwarfs and 5.6 \ms if only targets from HADES are considered. Our results are in agreement since the SOPHIE RV semi-amplitudes through all the seasons are between 4.3 to 5.4 \ms\footnote{We approximate the semi-amplitude as the half of the amplitudes measured in Table \ref{table:season_GP_RV}.}. 

We have found that the optical and near-infrared RVs of Gl~205 are similar in amplitude and jitter levels for the whole time series, meaning that there is at least a little gain in observing at near-infrared wavelengths which is expected for early, moderately, or low active, M dwarfs. \citet{Reiners2010} showed that the gain in precision towards the near-infrared is seen especially for M4 and later spectral types. In the case of early M dwarfs, the best precision is obtained in the $V$ band even though the highest S/N is reached in the $J$ band. Moreover, this effect may not be exclusively dependent on the spectral type only but on levels of activity too. 


\citealt{Robertson2020} analyzed a small sample of fast-rotating M dwarfs to compare their optical and near-infrared RVs. The late M dwarf Gl~3959 exhibits comparable optical and near-infrared RVs from the HIRES and HPF spectrographs, while some other targets showed comparable RVs only during one season of observations. Active M dwarfs such as Gl~388 or AU Mic, show stronger RV jitter in the optical than in the near-infrared while less-active stars, such as Gl~205, do not. The strength of the magnetic field may play a key role in this phenomenon since it likely impacts the brightness contrast between magnetic and non-magnetic features. We have shown that Gl~205 has a much weaker field than those on active M dwarfs.

We plan to extend the work of quasi-simultaneous observations with SOPHIE and SPIRou for more M dwarfs to help understand the multi-wavelength stellar activity contribution in RVs studies.



\begin{acknowledgements}
    PCZ thanks Oscar Barrag\'an for the helpful discussions about \texttt{pyaneti} and the multi-dimensional GPs.
     This work is based on observations collected with the SOPHIE spectrograph on the 1.93 m telescope at the Observatoire de Haute-Provence (CNRS), France. We thank the staff of the Observatoire de Haute-Provence for their support at the 1.93 m telescope and on SOPHIE.
     This work is  based on observations obtained at the Canada-France-Hawaii Telescope (CFHT) which is 
          operated from the summit of Maunakea by the National Research Council of Canada, the Institut National des Sciences de l'Univers of the Centre National de la Recherche Scientifique of France, and the University of Hawaii. The observations at the Canada-France-Hawaii Telescope were performed with care and respect from the summit of Maunakea which is a significant cultural and historic site. 
          The authors wish to recognize and acknowledge the very significant cultural role and reverence that the summit of
MaunaKea has always had within the indigenous Hawaiian community. We are most fortunate to have the opportunity to conduct observations from this mountain.
          Based on observations obtained with SPIRou, an international project led by Institut de Recherche en Astrophysique et Plan\'etologie, Toulouse, France.
          This paper includes data collected by the TESS mission that are publicly available from the Mikulski Archive for Space Telescopes (MAST).
     We acknowledge funding from the French National Research Agency (ANR) under contract number ANR-18-CE31-0019 (SPlaSH). PCZ thanks the LSSTC Data Science Fellowship Program, which is funded by LSSTC, NSF Cybertraining Grant $ \rm \#1829740$, the Brinson Foundation, and the Moore Foundation; her participation in the program has benefited this work. 
     BK acknowledges funding from the European Research Council (ERC) under the European Union’s Horizon 2020 research and innovation programme (Grant agreement No. 865624, GPRV). 
     AAK acknowledges that this work was supported by FCT - Fundação para a Ciência e a Tecnologia through national funds and by FEDER through COMPETE2020 - Programa Operacional Competitividade e Internacionalização by these grants: UIDB/04434/2020 \& UIDP/04434/2020; PTDC/FIS-AST/32113/2017 \& POCI-01-0145-FEDER-032113; PTDC/FIS-AST/28953/2017 \& POCI-01-0145-FEDER-028953; IF/00849/2015/CP1273/CT0003. 
     JFD, CM and PIC acknowledge funding from the European Research Council under the H2020 \& innovation program (grant $\#740651$ NewWorlds). 
     XD, XB, TF, and AC acknoweldge a funding from  the French National Research Agency in the framework of the Investissements d Avenir program (ANR-15-IDEX-02), through the funding of the “Origin of Life" project of the Grenoble-Alpes University. SH acknowledges CNES funding through the grant 837319.
     
\end{acknowledgements}

%
%

\bibliographystyle{aa}
\bibliography{references}

\begin{appendix}
\onecolumn
\FloatBarrier
\section{Radial velocities and activity indicators}\label{App:RVs}

The log of the SOPHIE and SPIRou observations with their activity indicators are listed in Table \ref{table:SOPHIEdata} and \ref{table:SPIRoudata}, respectively.

\FloatBarrier
\begin{table*}[h!]
\caption{SOPHIE RVs and activity indicators. The values of the RVs are given as the variation from the mean $\rm RV = 8727.9$\,m\,s$^{-1}$.}         
\label{table:SOPHIEdata}      
\centering
\footnotesize
\begin{tabular}{ccccccccccc} 
\hline
BJD-2400000 & $\Delta$RV & $\sigma_{\rm RV}$ & FWHM &  contrast &  BIS &  $\sigma_{\rm BIS}$ &     $\rm S_{MW}$ &    $\sigma_{S}$ &   $\rm H_{\alpha}$ &   $\sigma_{\rm H_{\alpha}}$ \\ 
d & m\,s$^{-1}$ &  m\,s$^{-1}$ &  km\,s$^{-1}$ & \%CCF &  m\,s$^{-1}$ &  m\,s$^{-1}$ & - & - & - & -\\ 
\hline
58794.5891 &       -1.5 &    2.1 &  4.77 &     21.87 &     13.07 &       2.87 &  1.61 &   0.02 & 0.2519 &  0.0007 \\ 58796.5420 &        4.7 &    2.3 &  4.80 &     21.16 &      8.53 &       3.14 &  1.63 &   0.01 & 0.2499 &  0.0009 \\ 58805.5787 &       -6.9 &    2.5 &  4.80 &     21.53 &      9.48 &       3.40 &  1.62 &   0.03 & 0.2659 &  0.0010 \\ 58820.5323 &       -7.8 &    1.7 &  4.75 &     21.14 &     10.76 &       2.29 &  1.54 &   0.02 & 0.2457 &  0.0006 \\ 58821.5421 &       -9.2 &    2.4 &  4.82 &     20.87 &      9.65 &       3.30 &  1.66 &   0.03 & 0.2445 &  0.0009 \\ 58825.5023 &       -6.0 &    1.8 &  4.79 &     21.38 &     10.43 &       2.33 &  1.59 &   0.02 & 0.2522 &  0.0007 \\ 58832.5041 &        0.7 &    1.6 &  4.70 &     20.93 &     14.07 &       2.10 &  1.52 &   0.02 & 0.2514 &  0.0005 \\ 58851.4754 &       -4.0 &    2.8 &  4.83 &     20.24 &     24.34 &       3.81 &  1.49 &   0.03 & 0.2539 &  0.0011 \\ 58854.4361 &       -3.0 &    1.7 &  4.72 &     20.96 &     10.73 &       2.27 &  1.51 &   0.02 & 0.2461 &  0.0006 \\ 58859.4193 &       -9.0 &    2.1 &  4.81 &     21.25 &      6.20 &       2.84 &  1.43 &   0.02 & 0.2394 &  0.0008 \\ 58860.4112 &       -4.8 &    1.8 &  4.76 &     21.25 &      5.56 &       2.37 &  1.48 &   0.02 & 0.2401 &  0.0007 \\ 58878.3905 &        3.1 &    2.5 &  4.83 &     20.56 &     14.43 &       3.42 &  1.71 &   0.03 & 0.2745 &  0.0010 \\ 58892.3602 &       -3.0 &    2.3 &  4.81 &     20.95 &      8.26 &       3.12 &  1.66 &   0.03 & 0.2559 &  0.0009 \\ 58894.3182 &       -1.4 &    1.6 &  4.74 &     21.36 &      6.25 &       2.10 &  1.59 &   0.02 & 0.2434 &  0.0006 \\ 58895.2799 &        3.7 &    1.7 &  4.71 &     20.99 &     11.49 &       2.23 &  1.63 &   0.03 & 0.2456 &  0.0006 \\ 58903.2928 &        0.6 &    1.8 &  4.79 &     21.38 &      9.51 &       2.38 &  1.68 &   0.02 & 0.2521 &  0.0007 \\ 58904.3100 &       -0.7 &    1.7 &  4.76 &     21.27 &      9.18 &       2.23 &  2.02 &   0.02 & 0.2593 &  0.0006 \\ 58905.3351 &       -1.0 &    2.7 &  4.86 &     20.87 &      9.21 &       3.71 &  1.67 &   0.03 & 0.2571 &  0.0012 \\ 58911.2803 &        9.1 &    2.8 &  4.87 &     20.62 &     12.12 &       3.89 &  1.71 &   0.02 & 0.2629 &  0.0012 \\ 58920.2862 &       -6.6 &    1.7 &  4.77 &     21.35 &     19.21 &       2.24 &  1.60 &   0.02 & 0.2732 &  0.0006 \\ 58921.3072 &      -10.1 &    2.6 &  4.79 &     20.38 &      9.59 &       3.56 &  1.64 &   0.03 & 0.2707 &  0.0010 \\ 58922.2891 &       -4.1 &    2.3 &  4.84 &     21.12 &     17.30 &       3.11 &  1.71 &   0.02 & 0.2680 &  0.0009 \\ 59147.5761 &        3.2 &    1.6 &  4.78 &     21.84 &     16.70 &       2.15 &  1.67 &   0.02 & 0.2645 &  0.0005 \\ 59149.5806 &       -0.7 &    2.0 &  4.87 &     22.49 &     11.77 &       2.70 &  1.85 &   0.02 & 0.2664 &  0.0008 \\ 59152.5861 &       -4.6 &    1.7 &  4.83 &     22.35 &     12.68 &       2.27 &  1.70 &   0.01 & 0.2649 &  0.0006 \\ 59154.5745 &       -4.4 &    1.8 &  4.82 &     22.11 &     13.54 &       2.37 &  1.77 &   0.02 & 0.2658 &  0.0007 \\ 59159.6351 &       -3.3 &    1.6 &  4.76 &     22.02 &      9.15 &       2.18 &  1.63 &   0.02 & 0.2481 &  0.0005 \\ 59171.5482 &        6.6 &    1.6 &  4.80 &     22.13 &     11.01 &       2.15 &  1.74 &   0.02 & 0.2688 &  0.0006 \\ 59186.5302 &       -1.5 &    1.8 &  4.84 &     22.13 &     16.25 &       2.36 &  1.72 &   0.02 & 0.2665 &  0.0007 \\ 59194.4799 &       -4.1 &    1.4 &  4.81 &     21.93 &      4.33 &       1.80 &  1.82 &   0.01 & 0.2718 &  0.0005 \\ 59196.4731 &        1.4 &    2.5 &  4.89 &     20.80 &      9.19 &       3.43 &  1.72 &   0.02 & 0.2668 &  0.0010 \\ 59206.4607 &       -2.8 &    3.4 &  4.86 &     19.43 &      5.55 &       4.69 &  1.60 &   0.04 & 0.2788 &  0.0013 \\ 59240.3519 &       -3.3 &    2.3 &  4.94 &     21.16 &      5.49 &       3.14 &  1.80 &   0.01 & 0.2750 &  0.0010 \\ 59245.4175 &        1.8 &    2.8 &  4.87 &     20.12 &     13.67 &       3.90 &  1.99 &   0.03 & 0.2948 &  0.0011 \\ 59249.3216 &        8.7 &    1.6 &  4.84 &     21.50 &     10.24 &       2.15 &  1.96 &   0.01 & 0.2875 &  0.0007 \\ 59263.3251 &       -2.7 &    1.4 &  4.82 &     21.99 &     15.39 &       1.72 &  1.72 &   0.01 & 0.2498 &  0.0005 \\ 59271.3509 &        5.4 &    1.4 &  4.83 &     21.90 &     10.31 &       1.77 &  1.62 &   0.01 & 0.2656 &  0.0005 \\ 59272.3046 &        5.8 &    1.7 &  4.83 &     21.27 &     10.45 &       2.25 &  1.72 &   0.02 & 0.2718 &  0.0007 \\ 59279.3161 &        4.0 &    1.6 &  4.82 &     21.31 &      6.78 &       2.14 &  1.89 &   0.02 & 0.2859 &  0.0006 \\ 59475.6741 &        0.8 &    1.6 &  4.89 &     22.60 &     15.41 &       2.09 &  1.90 &   0.01 & 0.2731 &  0.0006 \\ 59486.6681 &        8.8 &    1.6 &  4.91 &     22.31 &     11.11 &       2.09 &  2.07 &   0.01 & 0.2984 &  0.0006 \\ 59488.6528 &        3.8 &    1.4 &  4.90 &     22.42 &     15.67 &       1.87 &  2.11 &   0.01 & 0.3043 &  0.0005 \\ 59500.6615 &        0.4 &    1.9 &  4.87 &     22.00 &     10.30 &       2.54 &  1.70 &   0.01 & 0.2534 &  0.0008 \\ 59502.6818 &        8.9 &    1.6 &  4.85 &     22.28 &     12.14 &       2.12 &  1.72 &   0.01 & 0.2512 &  0.0006 \\ 59505.5964 &        3.1 &    1.6 &  4.84 &     22.15 &     13.09 &       2.12 &  1.85 &   0.01 & 0.2820 &  0.0006 \\ 59506.6446 &        3.8 &    1.5 &  4.85 &     22.45 &     18.09 &       1.98 &  1.71 &   0.01 & 0.2632 &  0.0006 \\ 59511.6283 &       -0.6 &    1.8 &  4.86 &     22.02 &     13.26 &       2.35 &  1.76 &   0.01 & 0.2600 &  0.0007 \\ 59513.6577 &        2.5 &    1.9 &  4.87 &     21.83 &      7.76 &       2.61 &  1.70 &   0.01 & 0.2606 &  0.0008 \\ 59525.5697 &        1.3 &    1.8 &  4.89 &     21.76 &     10.40 &       2.45 &  1.85 &   0.01 & 0.2800 &  0.0008 \\ 59532.5356 &       -0.5 &    2.0 &  4.87 &     21.86 &     17.64 &       2.65 &  1.88 &   0.01 & 0.2808 &  0.0008 \\ 59536.5829 &       -2.0 &    1.5 &  4.81 &     22.22 &     13.27 &       1.98 &  1.71 &   0.01 & 0.2598 &  0.0008 \\ 59537.4909 &       -0.2 &    1.6 &  4.80 &     21.84 &     12.61 &       2.19 &  1.62 &   0.01 & 0.2499 &  0.0005 \\ 59538.5451 &        3.0 &    1.5 &  4.80 &     22.21 &     12.87 &       1.92 &  1.64 &   0.01 & 0.2668 &  0.0005 \\ 59547.5166 &       -6.2 &    2.7 &  4.88 &     20.98 &     12.45 &       3.68 &  1.77 &   0.02 & 0.2550 &  0.0011 \\ 59554.5152 &        9.0 &    2.0 &  4.87 &     21.38 &     12.03 &       2.65 &  1.85 &   0.02 & 0.2845 &  0.0008 \\ 59562.5315 &       -6.0 &    1.5 &  4.75 &     21.09 &     14.40 &       1.96 &  1.64 &   0.02 & 0.2650 &  0.0005 \\ 59587.4143 &        4.6 &    2.3 &  4.87 &     20.92 &      8.97 &       3.09 &  1.55 &   0.02 & 0.2683 &  0.0010 \\ 59594.3647 &        4.1 &    1.4 &  4.77 &     21.41 &      9.97 &       1.83 &  1.59 &   0.02 & 0.2686 &  0.0005 \\ 59595.3839 &       -1.8 &    1.6 &  4.81 &     21.35 &      6.96 &       2.18 &  1.68 &   0.02 & 0.2671 &  0.0006 \\ 59597.4072 &        3.0 &    1.6 &  4.80 &     21.54 &     15.26 &       2.04 &  1.66 &   0.01 & 0.2644 &  0.0006 \\ 59601.3991 &        3.9 &    1.9 &  4.83 &     21.42 &     12.38 &       2.59 &  1.75 &   0.02 & 0.2634 &  0.0008 \\ 59607.3530 &        3.7 &    1.6 &  4.79 &     21.67 &      9.40 &       2.04 &  1.69 &   0.02 & 0.2641 &  0.0006 \\ 
\hline
\end{tabular}
\end{table*}


\onecolumn
\longtab[2]{
\begin{longtable}{ccccccccccc}     
\caption{\label{table:SPIRoudata} SPIRou RVs and activity indicators. The values of the RVs are given as the variation from the mean $\rm RV = 9229.7$\,m\,s$^{-1}$.}\\
\hline
BJD-2400000 & $\Delta$RV & $\sigma_{\rm RV}$ & $\rm FWHM_{\rm LBL}$ & $\sigma_{\rm FWHM}$ & chrom. velocity slope & $\sigma_{\rm chrom.}$ &  $B_{\ell}$ & $\sigma_{B}$  & Phase & Season \\
d & m\,s$^{-1}$ &  m\,s$^{-1}$ &  km\,s$^{-1}$ &  km\,s$^{-1}$ & m\,s$^{-1}$ &m\,s$^{-1}$   & G & G & & \\ 
\hline
\endfirsthead
\caption{SPIRou RVs and activity indicators continued.}\\
\hline
BJD-2400000 & $\Delta$RV & $\sigma_{\rm RV}$ & $\rm FWHM_{\rm LBL}$ & $\sigma_{\rm FWHM}$ &  chrom. velocity slope & $\sigma_{\rm chrom.}$ &  $B_{\ell}$ & $\sigma_{B}$ & Phase & Season \\ 
d & m\,s$^{-1}$ &  m\,s$^{-1}$ &  km\,s$^{-1}$ &  km\,s$^{-1}$ &m\,s$^{-1}$  &m\,s$^{-1}$   & G & G & & \\ 
\hline
\endhead
\hline
\endfoot
58738.1365\tablefootmark{*} &   -1.3 &     1.6 &  6.006 &      0.009 &                 -16.0 &                    3.0 &   5.70 &       1.19 &   0.00 &      1 \\ 58745.1192 &    2.2 &     2.2 &  5.941 &      0.009 &                  -8.1 &                    4.6 &   0.06 &       1.16 &   0.20 &      1 \\ 58746.0224 &    7.8 &     2.2 &  5.946 &      0.010 &                  -9.7 &                    4.7 &  -1.18 &       1.33 &   0.23 &      1 \\ 58751.1257 &   -3.6 &     1.9 &  5.951 &      0.008 &                 -18.5 &                    4.0 &  -4.12 &       1.14 &   0.38 &      1 \\ 58752.1188 &   -5.6 &     2.1 &  5.929 &      0.009 &                 -17.7 &                    4.5 &   1.28 &       1.76 &   0.41 &      1 \\ 58753.1103 &    1.3 &     2.1 &  5.972 &      0.009 &                 -19.9 &                    4.1 &   0.07 &       1.22 &   0.44 &      1 \\ 58760.1240 &   -1.1 &     1.9 &  5.961 &      0.008 &                 -26.3 &                    3.9 &   5.55 &       1.04 &   0.64 &      1 \\ 58762.1541 &    5.6 &     2.1 &  5.928 &      0.009 &                 -19.2 &                    4.5 &   5.61 &       1.30 &   0.70 &      1 \\ 58763.1257 &    3.1 &     2.1 &  5.942 &      0.009 &                 -21.7 &                    4.4 &   4.51 &       1.18 &   0.73 &      1 \\ 58766.1035 &   -1.8 &     2.3 &  5.905 &      0.010 &                  -9.9 &                    4.8 &   2.23 &       1.21 &   0.81 &      1 \\ 58769.0896 &   -0.2 &     2.1 &  5.935 &      0.009 &                  -4.4 &                    4.5 &   5.79 &       1.36 &   0.90 &      1 \\ 58770.1405 &    0.3 &     2.0 &  5.945 &      0.009 &                 -10.7 &                    4.3 &   7.59 &       1.14 &   0.93 &      1 \\ 58771.0739 &   -1.1 &     2.3 &  5.932 &      0.010 &                  -1.7 &                    4.9 &   5.75 &       1.49 &   0.96 &      1 \\ 58772.1394 &    2.0 &     2.0 &  5.972 &      0.009 &                  -7.0 &                    4.1 &   4.25 &       1.26 &   0.99 &      1 \\ 58773.0344 &    0.8 &     2.1 &  5.959 &      0.009 &                  -6.0 &                    4.4 &   5.78 &       1.37 &   1.01 &      1 \\ 58788.1049 &   -2.3 &     2.0 &  5.922 &      0.009 &                  -9.5 &                    4.2 &  -2.68 &       1.19 &   1.45 &      1 \\ 58789.1097 &   -6.3 &     1.9 &  5.938 &      0.008 &                 -19.7 &                    3.9 &   1.73 &       1.05 &   1.48 &      1 \\ 58791.0882 &    0.8 &     2.1 &  5.898 &      0.009 &                 -24.1 &                    4.6 &   4.03 &       1.18 &   1.54 &      1 \\ 58792.1294 &   -0.4 &     2.1 &  5.930 &      0.009 &                 -12.7 &                    4.5 &   4.26 &       1.32 &   1.57 &      1 \\ 58793.1331 &    2.0 &     2.1 &  5.957 &      0.009 &                  -4.0 &                    4.3 &   5.60 &       1.12 &   1.60 &      1 \\ 58794.1219 &   -1.7 &     2.3 &  5.976 &      0.010 &                 -32.0 &                    5.1 &   5.15 &       1.41 &   1.63 &      1 \\ 58795.1056 &    6.6 &     1.9 &  5.980 &      0.008 &                 -10.7 &                    4.0 &   4.69 &       1.07 &   1.66 &      1 \\ 58796.1638\tablefootmark{*} &   10.7 &     1.8 &  5.994 &      0.011 &                  -9.7 &                    3.8 &   5.66 &       2.13 &   1.69 &      1 \\ 58796.9823 &    5.2 &     2.0 &  6.018 &      0.009 &                 -14.1 &                    4.5 &   5.73 &       1.24 &   1.71 &      1 \\ 58801.0872 &    9.7 &     2.0 &  5.972 &      0.009 &                 -10.9 &                    4.4 &   2.77 &       1.18 &   1.83 &      1 \\ 58802.0572 &   14.6 &     2.0 &  5.965 &      0.009 &                 -13.1 &                    4.5 &   2.16 &       1.22 &   1.86 &      1 \\ 58823.0087 &   -3.7 &     2.2 &  5.943 &      0.009 &                  -3.2 &                    4.7 &   1.32 &       1.42 &   2.47 &      1 \\ 58825.0963 &   -1.4 &     2.0 &  5.975 &      0.009 &                  -0.8 &                    4.2 &   4.52 &       1.15 &   2.53 &      1 \\ 58826.0288 &   -1.9 &     2.0 &  5.976 &      0.009 &                  -4.5 &                    4.1 &   5.20 &       1.15 &   2.56 &      1 \\ 58827.0595 &   -5.3 &     2.2 &  5.930 &      0.010 &                 -18.0 &                    4.6 &   5.81 &       1.11 &   2.58 &      1 \\ 58827.9807 &   -2.2 &     2.1 &  5.943 &      0.009 &                 -18.2 &                    4.3 &   2.76 &       1.15 &   2.61 &      1 \\ 58829.0723 &   -4.4 &     2.1 &  5.959 &      0.009 &                  -2.7 &                    4.4 &   3.00 &       1.18 &   2.64 &      1 \\ 58829.9988 &    4.1 &     2.0 &  5.947 &      0.008 &                  -3.6 &                    3.9 &   4.29 &       1.02 &   2.67 &      1 \\ 59088.1459 &   -0.0 &     1.6 &  5.986 &      0.007 &                   3.8 &                    2.9 &   1.11 &       0.91 &  10.17 &      2 \\ 59089.0973 &    0.3 &     1.6 &  5.959 &      0.007 &                   1.6 &                    3.0 &   2.59 &       0.87 &  10.20 &      2 \\ 59090.0986 &   -1.5 &     1.7 &  5.949 &      0.008 &                  -0.7 &                    3.3 &   2.87 &       0.86 &  10.23 &      2 \\ 59091.1334 &    0.6 &     1.9 &  5.925 &      0.008 &                  20.8 &                    3.7 &   1.87 &       0.89 &  10.26 &      2 \\ 59092.1231 &   -0.9 &     1.7 &  5.952 &      0.007 &                   4.4 &                    3.1 &   3.62 &       0.86 &  10.29 &      2 \\ 59093.1251 &   -1.0 &     2.0 &  5.897 &      0.009 &                   8.2 &                    4.0 &   2.71 &       0.82 &  10.32 &      2 \\ 59094.1339 &   -2.5 &     1.9 &  5.900 &      0.008 &                  14.4 &                    3.9 &   3.80 &       1.02 &  10.35 &      2 \\ 59095.1265 &    2.8 &     1.7 &  5.945 &      0.008 &                   3.2 &                    3.3 &   2.57 &       0.89 &  10.38 &      2 \\ 59096.1231 &    0.2 &     1.7 &  5.988 &      0.007 &                   4.3 &                    3.2 &   1.37 &       1.02 &  10.41 &      2 \\ 59097.1475 &   -2.6 &     2.1 &  5.903 &      0.009 &                  -0.0 &                    4.1 &   1.94 &       1.32 &  10.44 &      2 \\ 59098.1493 &   -6.5 &     1.6 &  5.968 &      0.007 &                  -7.4 &                    3.0 &   0.74 &       0.84 &  10.47 &      2 \\ 59099.1228 &   -6.3 &     1.9 &  5.914 &      0.008 &                   9.6 &                    3.8 &   1.47 &       1.02 &  10.49 &      2 \\ 59101.1095 &   -3.0 &     1.9 &  5.923 &      0.008 &                   0.0 &                    3.9 &  -0.89 &       0.98 &  10.55 &      2 \\ 59102.0900 &   -0.0 &     1.7 &  5.954 &      0.007 &                  -7.3 &                    3.3 &   1.75 &       0.99 &  10.58 &      2 \\ 59103.0880 &   -2.5 &     1.9 &  5.920 &      0.008 &                   2.0 &                    3.7 &   0.68 &       0.92 &  10.61 &      2 \\ 59111.1241 &   -1.0 &     1.9 &  5.960 &      0.008 &                  -4.6 &                    4.0 &   0.85 &       1.14 &  10.84 &      2 \\ 59112.1125 &   -0.1 &     1.8 &  5.974 &      0.008 &                   6.5 &                    3.3 &   2.15 &       1.04 &  10.87 &      2 \\ 59114.0475 &    2.5 &     2.1 &  5.933 &      0.009 &                   2.9 &                    4.2 &   0.27 &       0.92 &  10.93 &      2 \\ 59115.1229 &    0.9 &     1.8 &  5.983 &      0.008 &                 -11.9 &                    3.2 &  -0.84 &       0.99 &  10.96 &      2 \\ 59116.0896 &    4.1 &     1.8 &  5.957 &      0.008 &                   0.1 &                    3.5 &  -4.41 &       1.11 &  10.99 &      2 \\ 59118.0749\tablefootmark{*} &    3.7 &     1.8 &  5.914 &      0.008 &                   4.3 &                    3.6 &  -4.26 &       0.92 &  11.04 &      2 \\ 59120.1385 &    5.5 &     1.9 &  5.964 &      0.012 &                  -3.6 &                    4.2 &  -2.66 &       2.06 &  11.10 &      2 \\ 59121.0153 &   -4.6 &     1.8 &  5.931 &      0.008 &                   6.0 &                    3.5 &  -2.10 &       0.90 &  11.13 &      2 \\ 59122.1230 &   -0.4 &     1.8 &  5.929 &      0.008 &                   6.3 &                    3.4 &   1.95 &       1.17 &  11.16 &      2 \\ 59123.0174 &   -5.4 &     2.0 &  5.903 &      0.009 &                   1.1 &                    4.0 &   3.13 &       1.01 &  11.19 &      2 \\ 59127.0178 &   -4.9 &     1.8 &  5.921 &      0.008 &                   3.6 &                    3.4 &   2.37 &       0.90 &  11.30 &      2 \\ 59128.0016 &   -4.3 &     1.8 &  5.918 &      0.008 &                   1.7 &                    3.3 &   2.55 &       0.84 &  11.33 &      2 \\ 59129.0533 &   -3.7 &     1.8 &  5.945 &      0.008 &                   7.0 &                    3.4 &   2.72 &       1.05 &  11.36 &      2 \\ 59129.9930 &   -1.9 &     1.9 &  5.897 &      0.008 &                   7.1 &                    3.7 &   2.36 &       0.86 &  11.39 &      2 \\ 59131.1038 &   -1.1 &     1.8 &  5.917 &      0.008 &                   5.7 &                    3.4 &   1.31 &       0.92 &  11.42 &      2 \\ 59154.1087 &   -9.4 &     1.7 &  5.933 &      0.007 &                   0.4 &                    3.0 &   1.26 &       0.75 &  12.09 &      \\ 59154.9730 &   -8.1 &     1.7 &  5.925 &      0.007 &                  -9.5 &                    3.0 &   3.66 &       0.80 &  12.12 &      \\ 59156.1429 &   -6.5 &     1.8 &  6.000 &      0.013 &                 -10.2 &                    3.6 &   4.49 &       1.15 &  12.15 &      \\ 59157.0273 &   -8.2 &     1.8 &  5.924 &      0.008 &                  -1.0 &                    3.6 &   3.42 &       1.24 &  12.18 &      \\ 59158.0274 &   -9.3 &     1.8 &  5.921 &      0.008 &                  -2.9 &                    3.5 &   6.36 &       1.18 &  12.21 &      \\ 59159.0282 &   -3.0 &     1.7 &  5.937 &      0.007 &                   4.5 &                    3.2 &   7.64 &       0.84 &  12.24 &      \\ 59207.9651 &    6.7 &     2.0 &  5.900 &      0.009 &                   1.5 &                    4.0 &   1.63 &       1.15 &  13.66 &      \\ 59208.9774 &    4.2 &     1.8 &  5.942 &      0.008 &                   0.6 &                    3.4 &   0.71 &       0.95 &  13.69 &      \\ 59209.9278 &    1.3 &     1.6 &  5.961 &      0.007 &                   7.3 &                    3.0 &   2.87 &       0.80 &  13.71 &      \\ 59211.8520 &    1.4 &     1.8 &  5.962 &      0.007 &                  14.1 &                    3.1 &   4.12 &       0.92 &  13.77 &      \\ 59212.9753 &    4.1 &     1.7 &  5.961 &      0.007 &                   2.7 &                    3.2 &   1.49 &       1.01 &  13.80 &      \\ 59214.0266 &    2.1 &     1.7 &  5.979 &      0.007 &                   5.8 &                    3.2 &   2.87 &       0.93 &  13.83 &      \\ 59214.9368 &   -1.7 &     1.7 &  5.953 &      0.007 &                  -0.1 &                    3.1 &   3.91 &       0.91 &  13.86 &      \\ 59215.8867 &    1.1 &     2.0 &  5.985 &      0.009 &                   0.4 &                    4.2 &  -0.06 &       1.39 &  13.89 &      \\ 59216.8987 &    5.4 &     1.6 &  5.938 &      0.007 &                  -0.4 &                    3.0 &   1.14 &       0.77 &  13.92 &      \\ 59217.9291 &    4.9 &     1.6 &  5.963 &      0.007 &                  10.5 &                    2.9 &  -0.82 &       0.83 &  13.95 &      \\ 59218.9165 &    1.9 &     1.7 &  5.945 &      0.007 &                   5.4 &                    3.2 &  -0.64 &       0.84 &  13.98 &      \\ 59219.9436 &    5.0 &     1.7 &  5.943 &      0.007 &                  11.0 &                    3.1 &  -0.10 &       0.84 &  14.01 &      \\ 59220.9034 &    0.8 &     1.6 &  5.947 &      0.007 &                   6.6 &                    3.0 &   2.01 &       0.85 &  14.03 &      \\ 59221.9382 &    1.8 &     1.6 &  5.945 &      0.007 &                  22.3 &                    2.9 &   0.75 &       0.83 &  14.06 &      \\ 59222.9549 &   -1.7 &     1.6 &  5.953 &      0.007 &                  20.6 &                    2.9 &   3.87 &       0.84 &  14.09 &      \\ 59267.7684 &   -3.6 &     1.7 &  5.942 &      0.007 &                   7.3 &                    3.1 &   0.02 &       0.97 &  15.40 &      \\ 59271.7394 &   -5.7 &     1.6 &  5.957 &      0.007 &                   9.2 &                    2.8 &  -0.86 &       0.84 &  15.51 &      \\ 59273.8932 &   -2.6 &     1.8 &  5.956 &      0.008 &                  -0.9 &                    3.4 &  -0.40 &       1.06 &  15.57 &      \\ 59275.8955 &   -5.6 &     1.9 &  5.994 &      0.008 &                  -5.5 &                    3.8 &   0.19 &       1.25 &  15.63 &      \\ 59276.8024 &   -5.4 &     1.5 &  5.964 &      0.007 &                  16.5 &                    2.9 &  -2.27 &       0.77 &  15.66 &      \\ 59277.7671 &   -5.3 &     1.6 &  5.951 &      0.007 &                  -1.4 &                    3.0 &  -1.68 &       0.74 &  15.69 &      \\ 59297.7410 &   -6.3 &     1.7 &  5.896 &      0.008 &                   4.7 &                    3.4 &   2.83 &       0.81 &  16.27 &      \\ 59440.1423 &    0.5 &     1.8 &  5.930 &      0.008 &                   0.6 &                    3.4 &   0.33 &       1.07 &  20.41 &      3 \\ 59442.1181 &   -0.4 &     1.8 &  5.934 &      0.008 &                   3.8 &                    3.5 &   1.99 &       0.94 &  20.46 &      3 \\ 59443.1094 &   -1.9 &     1.6 &  5.975 &      0.007 &                   4.2 &                    3.0 &   2.30 &       0.87 &  20.49 &      3 \\ 59444.1392 &   -0.4 &     1.7 &  5.949 &      0.007 &                  -4.1 &                    3.2 &   3.95 &       0.93 &  20.52 &      3 \\ 59445.1268 &    2.1 &     1.6 &  5.959 &      0.007 &                  -6.8 &                    2.9 &   6.81 &       0.87 &  20.55 &      3 \\ 59447.1009 &    1.5 &     1.7 &  5.972 &      0.007 &                  -6.5 &                    3.1 &   7.99 &       0.87 &  20.61 &      3 \\ 59448.1311 &    6.5 &     1.8 &  5.942 &      0.008 &                  -9.3 &                    3.4 &   5.45 &       1.00 &  20.64 &      3 \\ 59449.1436 &    0.6 &     1.8 &  5.954 &      0.008 &                 -11.1 &                    3.4 &   8.44 &       0.90 &  20.67 &      3 \\ 59450.1248 &    3.1 &     1.7 &  5.973 &      0.007 &                  -2.1 &                    3.1 &   4.68 &       0.98 &  20.70 &      3 \\ 59451.1236 &    2.6 &     1.9 &  5.951 &      0.008 &                  -8.3 &                    3.4 &   4.54 &       0.95 &  20.73 &      3 \\ 59452.1097 &    1.7 &     1.7 &  5.971 &      0.007 &                   0.7 &                    3.0 &   3.13 &       0.86 &  20.76 &      3 \\ 59453.1165 &    9.1 &     1.7 &  5.981 &      0.007 &                   1.5 &                    3.1 &   6.72 &       0.92 &  20.78 &      3 \\ 59473.0971 &    3.0 &     1.6 &  5.964 &      0.007 &                   0.7 &                    3.0 &  -2.72 &       0.89 &  21.37 &      3 \\ 59475.0965 &    6.6 &     1.8 &  5.932 &      0.008 &                  -0.5 &                    3.4 &  -2.60 &       1.09 &  21.42 &      3 \\ 59477.1203 &    6.5 &     1.9 &  5.922 &      0.008 &                   2.4 &                    3.7 &   1.47 &       0.86 &  21.48 &      3 \\ 59478.0777 &   -0.2 &     1.9 &  5.919 &      0.008 &                   7.0 &                    3.8 &   3.05 &       0.93 &  21.51 &      3 \\ 59480.0106 &    7.1 &     1.7 &  5.955 &      0.007 &                   0.0 &                    3.1 &   4.14 &       0.85 &  21.57 &      3 \\ 59481.0283 &    8.0 &     1.7 &  5.954 &      0.008 &                  -4.3 &                    3.2 &   4.29 &       0.97 &  21.60 &      3 \\ 59482.0279 &    9.6 &     1.8 &  5.945 &      0.008 &                  -3.3 &                    3.6 &   6.76 &       1.34 &  21.62 &      3 \\ 59502.1252 &   -1.7 &     1.9 &  5.892 &      0.008 &                  -9.2 &                    3.5 &  -3.36 &       0.99 &  22.21 &      \\ 59504.1295 &   -3.5 &     1.9 &  5.971 &      0.008 &                   2.1 &                    3.9 &  -1.66 &       1.19 &  22.27 &      \\ 59506.1212 &  -16.5 &     2.5 &  5.904 &      0.011 &                   8.9 &                    5.4 &  -2.18 &       1.46 &  22.33 &      \\ 59507.1298 &   -2.2 &     1.8 &  5.931 &      0.008 &                  -4.6 &                    3.5 &  -2.24 &       0.91 &  22.35 &      \\ 59509.1337 &  -13.4 &     2.0 &  5.962 &      0.009 &                   3.2 &                    4.3 &  -0.33 &       1.19 &  22.41 &      \\ 59510.1171 &   -5.5 &     2.1 &  5.950 &      0.009 &                  15.0 &                    4.4 &  -1.46 &       1.22 &  22.44 &      \\ 59511.1047 &   -5.4 &     1.7 &  5.918 &      0.007 &                   0.7 &                    3.2 &   0.15 &       0.93 &  22.47 &      \\ 59513.0782 &   -6.6 &     1.5 &  5.961 &      0.007 &                  -5.2 &                    2.7 &   0.50 &       0.78 &  22.53 &      \\ 59514.0972 &   -1.4 &     1.6 &  5.946 &      0.007 &                   1.1 &                    2.9 &   0.82 &       0.88 &  22.56 &      \\ 59515.1549 &   -2.8 &     1.8 &  5.924 &      0.008 &                  -3.0 &                    3.2 &   0.33 &       0.82 &  22.59 &      \\ 59516.0071 &    2.5 &     1.7 &  5.970 &      0.007 &                   0.6 &                    2.8 &   0.83 &       0.78 &  22.61 &     \\ 59535.0426 &   -7.0 &     2.0 &  5.926 &      0.009 &                   0.4 &                    4.1 &  -6.65 &       1.31 &  23.17 &      4 \\ 59537.9797 &   -1.3 &     1.7 &  5.906 &      0.007 &                 -15.9 &                    3.2 &  -3.38 &       1.05 &  23.25 &      4 \\ 59539.0540 &   -4.0 &     1.7 &  5.925 &      0.007 &                  -3.6 &                    3.3 &  -1.83 &       0.93 &  23.28 &      4 \\ 59540.0405 &   -0.1 &     1.7 &  5.925 &      0.007 &                  -9.8 &                    3.1 &  -1.41 &       0.91 &  23.31 &      4 \\ 59541.0569 &    0.1 &     1.6 &  5.923 &      0.007 &                  -3.0 &                    3.1 &  -0.24 &       0.76 &  23.34 &      4 \\ 59558.0071 &    7.6 &     1.7 &  5.985 &      0.007 &                   3.7 &                    2.9 &   0.96 &       0.80 &  23.83 &      4 \\ 59560.0476 &    5.2 &     1.6 &  5.996 &      0.007 &                  14.7 &                    3.0 &   1.01 &       0.93 &  23.89 &      4 \\ 59561.0577 &    9.7 &     1.6 &  5.977 &      0.007 &                  14.2 &                    3.1 &   0.82 &       0.94 &  23.92 &      4 \\ 59562.0182 &    3.4 &     1.7 &  5.974 &      0.008 &                  14.3 &                    3.3 &   0.48 &       1.06 &  23.95 &      4 \\ 59563.8546 &    2.7 &     1.6 &  5.952 &      0.007 &                   2.3 &                    2.8 &   0.55 &       0.76 &  24.00 &      4 \\ 59564.9371 &    3.8 &     1.7 &  5.950 &      0.007 &                   7.4 &                    3.1 &   0.19 &       0.89 &  24.03 &      4 \\ 59566.9795 &   -0.3 &     1.8 &  5.896 &      0.008 &                   7.8 &                    3.4 &  -1.53 &       0.92 &  24.09 &      4 \\ 59585.9682 &    1.3 &     1.9 &  5.916 &      0.008 &                   6.9 &                    3.2 &  -3.13 &       0.77 &  24.65 &      4 \\ 59587.9584 &    2.1 &     1.8 &  5.919 &      0.008 &                   4.8 &                    3.5 &  -0.20 &       0.89 &  24.70 &      4 \\ 59588.9581 &    2.1 &     1.8 &  5.925 &      0.008 &                   1.2 &                    3.4 &  -1.16 &       1.07 &  24.73 &      4 \\ 59589.8836 &   -0.7 &     1.9 &  5.912 &      0.008 &                   5.8 &                    3.8 &  -3.13 &       0.99 &  24.76 &      4 \\ 59590.7756 &    2.5 &     1.6 &  5.954 &      0.007 &                  13.7 &                    2.8 &  -0.73 &       0.81 &  24.79 &      4 \\ 59591.8296 &    2.5 &     1.6 &  5.960 &      0.007 &                  12.8 &                    2.9 &   1.69 &       0.86 &  24.82 &      4 \\ 59592.7883 &    3.3 &     1.7 &  5.930 &      0.007 &                  14.1 &                    3.2 &   1.57 &       0.86 &  24.84 &      4 \\ 59593.9250 &    7.8 &     1.5 &  5.961 &      0.007 &                  10.1 &                    2.7 &   2.23 &       0.74 &  24.88 &      4 \\ 59594.9754 &    5.8 &     1.6 &  5.952 &      0.007 &                  21.2 &                    2.8 &   1.81 &       0.80 &  24.91 &      4 \\ 59596.9323 &   -1.5 &     1.6 &  5.939 &      0.007 &                  11.8 &                    2.9 &   2.60 &       0.81 &  24.96 &      4 \\ 59597.9986 &   -0.9 &     1.6 &  5.923 &      0.007 &                  16.1 &                    3.1 &   2.66 &       0.93 &  25.00 &      4 \\ 59598.9765 &    0.2 &     1.6 &  5.929 &      0.007 &                   6.8 &                    3.0 &   1.15 &       0.84 &  25.02 &      4 \\ 59599.9542 &   -3.6 &     1.7 &  5.913 &      0.007 &                  -8.2 &                    3.2 &   0.40 &       0.82 &  25.05 &      4 \\ 59601.9558 &    0.4 &     1.7 &  5.973 &      0.007 &                   4.3 &                    3.3 &  -1.90 &       1.13 &  25.11 &      4 \\ 59603.9449 &   -3.7 &     1.7 &  5.909 &      0.007 &                   1.1 &                    3.2 &  -3.85 &       0.84 &  25.17 &      4 \\ 59604.9396 &   -1.2 &     1.6 &  5.953 &      0.007 &                  -3.3 &                    2.9 &  -1.71 &       0.87 &  25.20 &      4 \\ 59606.8864 &   -2.4 &     1.7 &  5.938 &      0.007 &                   0.5 &                    3.0 &  -1.20 &       0.92 &  25.25 &      4 \\
59607.9038 & -4.6 & 1.7 & 5.978 & 0.007 & -1.7 & 2.9 & 1.27 & 1.60 & 25.28 & 4 \\
59609.8721 & -4.1 & 1.5 & 5.969 & 0.008 & -5.7 & 2.8 & 6.21 & 1.46 & 25.34 & 4\\
\hline                  
\end{longtable}

\tablefoot{
\tablefoottext{*}{Night with more than one spectropolarimetric sequence. The RV and activity indicators are computed with the nightly average spectra.}
}
}
\FloatBarrier
\section{Longitudinal magnetic field $B_{\ell}$ GP posteriors}

\begin{figure*}[ht]
    \centering
    \includegraphics[width=0.99\hsize]{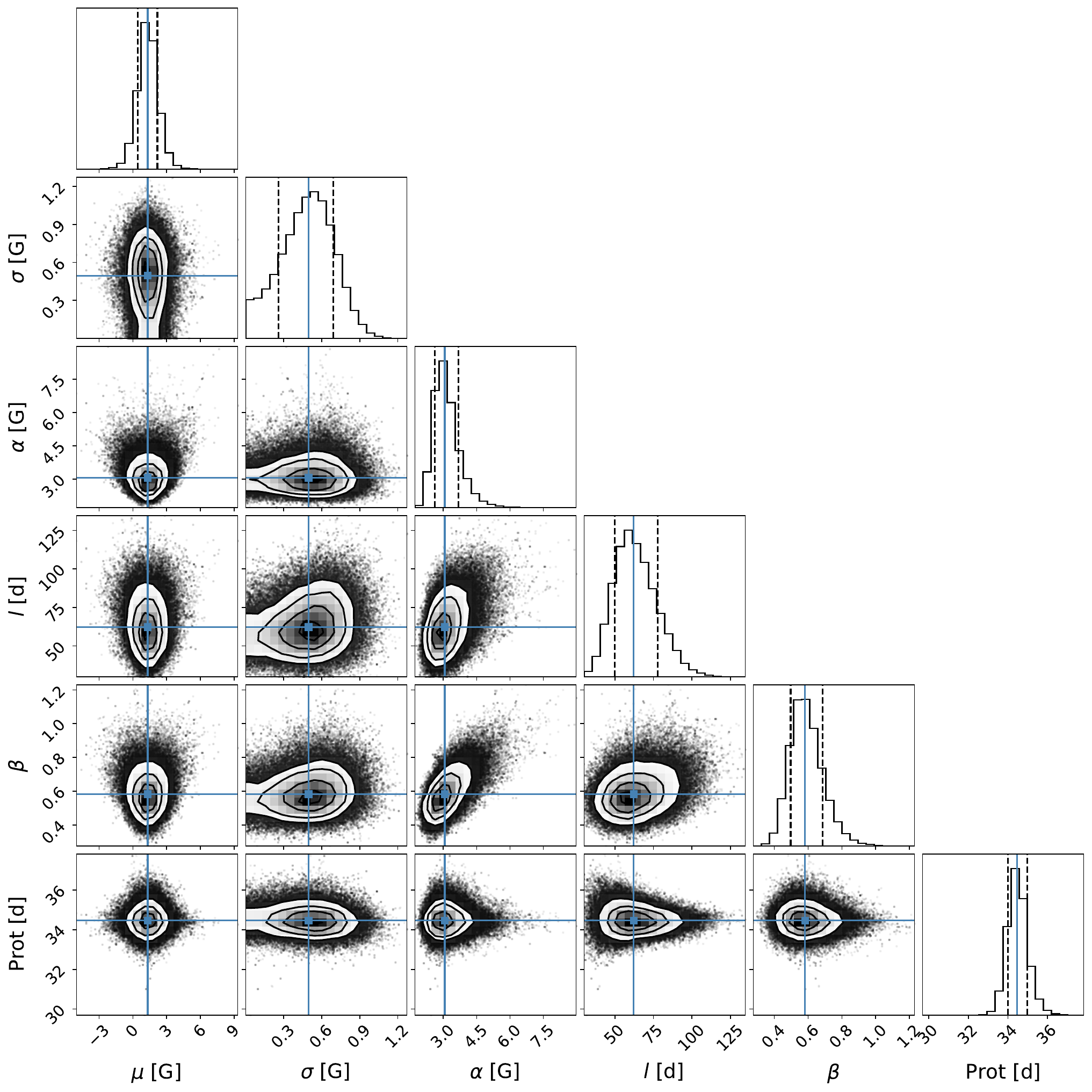}
    \caption{Corner plot of the GP hyper-parameters on the $B_\ell$ time series sampled from our MCMC framework generated with the \texttt{corner} Python package \citep{corner}. The blue line over the posterior distributions indicates the mean and the dashed lines, the 0.16 and 0.84 quantiles uncertainties.} 
    \label{Blong_GP_corner}
\end{figure*}

The posterior distributions of the GP model applied in the longitudinal magnetic field are shown in Figure \ref{Blong_GP_corner}.

\FloatBarrier
\section{ZDI inversion of the Stokes $V$ LSD profiles}\label{App:StokesV_fits}

\begin{figure*}
    \centering
    \includegraphics[width=0.4\linewidth]{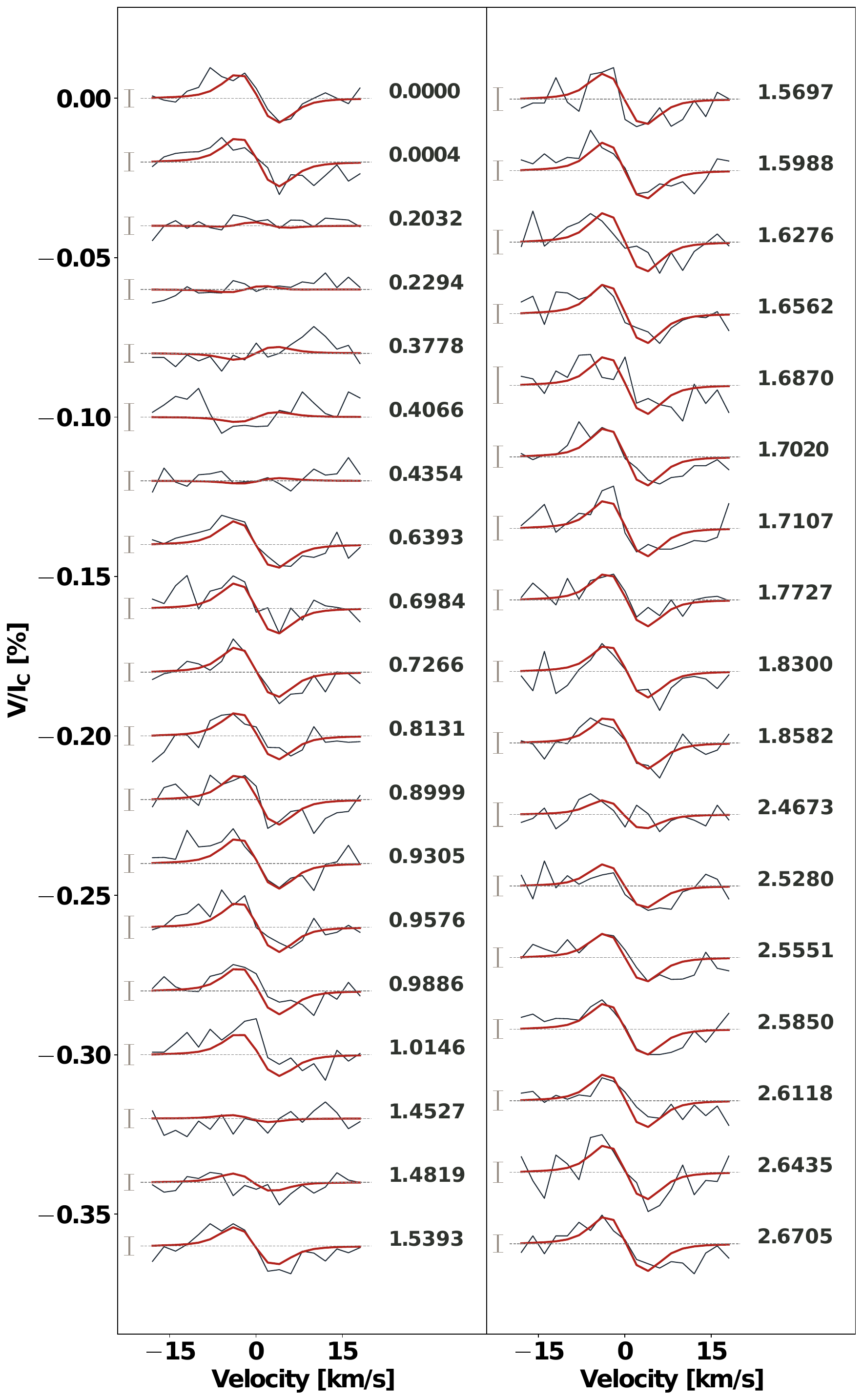}
    \hspace{0.05\linewidth}
    \includegraphics[width=0.4\linewidth]{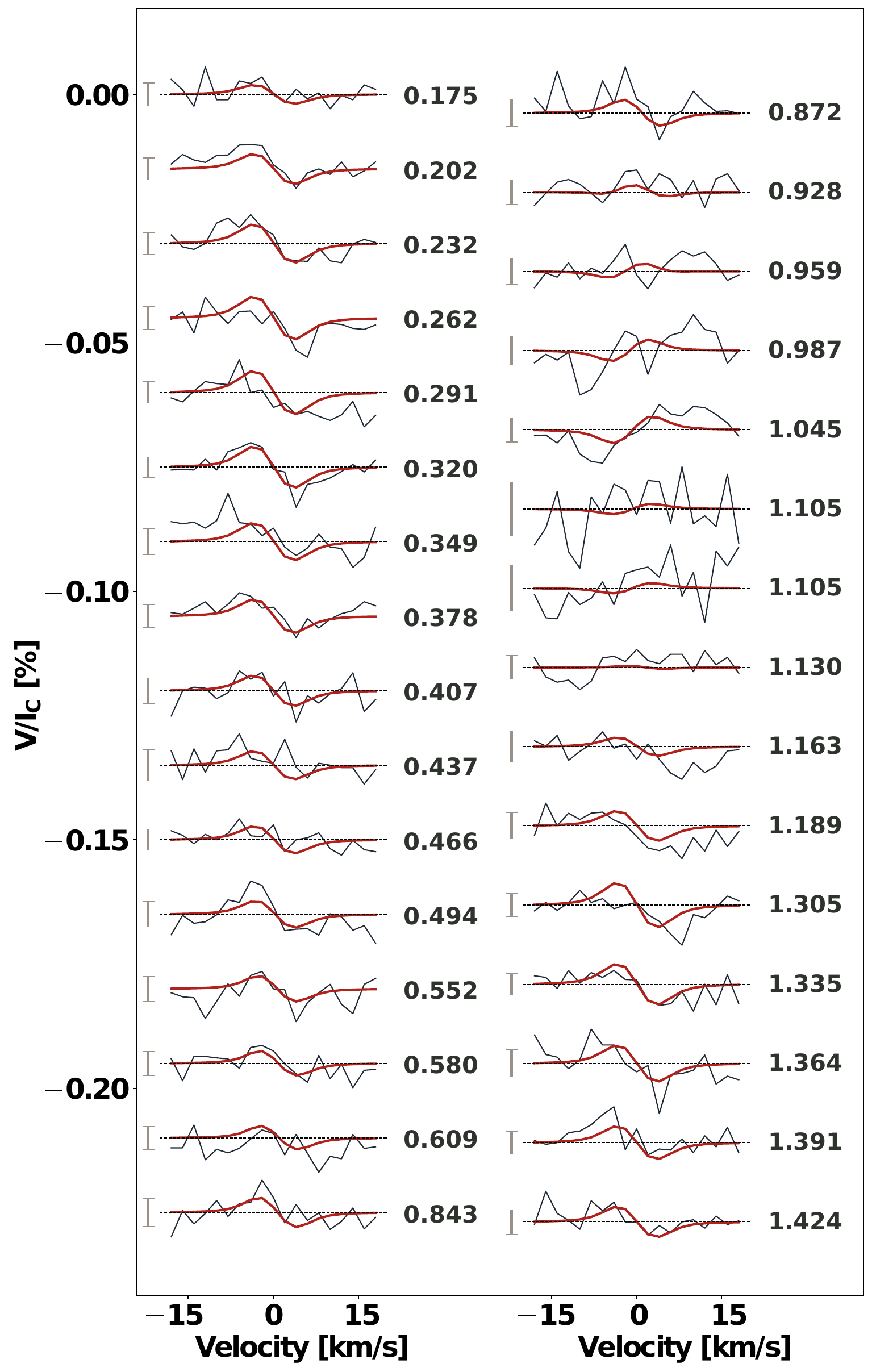}    
    \caption{Stokes $V$ LSD profiles (thin black lines) and best-fitting ZDI profiles (thick red lines). Profiles observed between September and December 2019 (Season S$_{1}$ in Table~\ref{tab:results_ZDI}) are shown on the left-hand panel whereas the profiles observed between August and October 2020 (Season S$_{2}$ in Table~\ref{tab:results_ZDI}) are shown in the right-hand panel.  The stellar rotation cycle, computed using the first observation (BJD = 2\,458\,738.128) as a reference time and assuming a rotation period of 34.4\,d (see Table~\ref{table:GP}), are given on the right-hand side of each observation. The $\pm$1$\sigma$ error bars are indicated on the left-hand side of each profile.}
    \label{fig:StokesV_fit12}
\end{figure*}

\begin{figure*}
    \centering
    \includegraphics[width=0.4\linewidth]{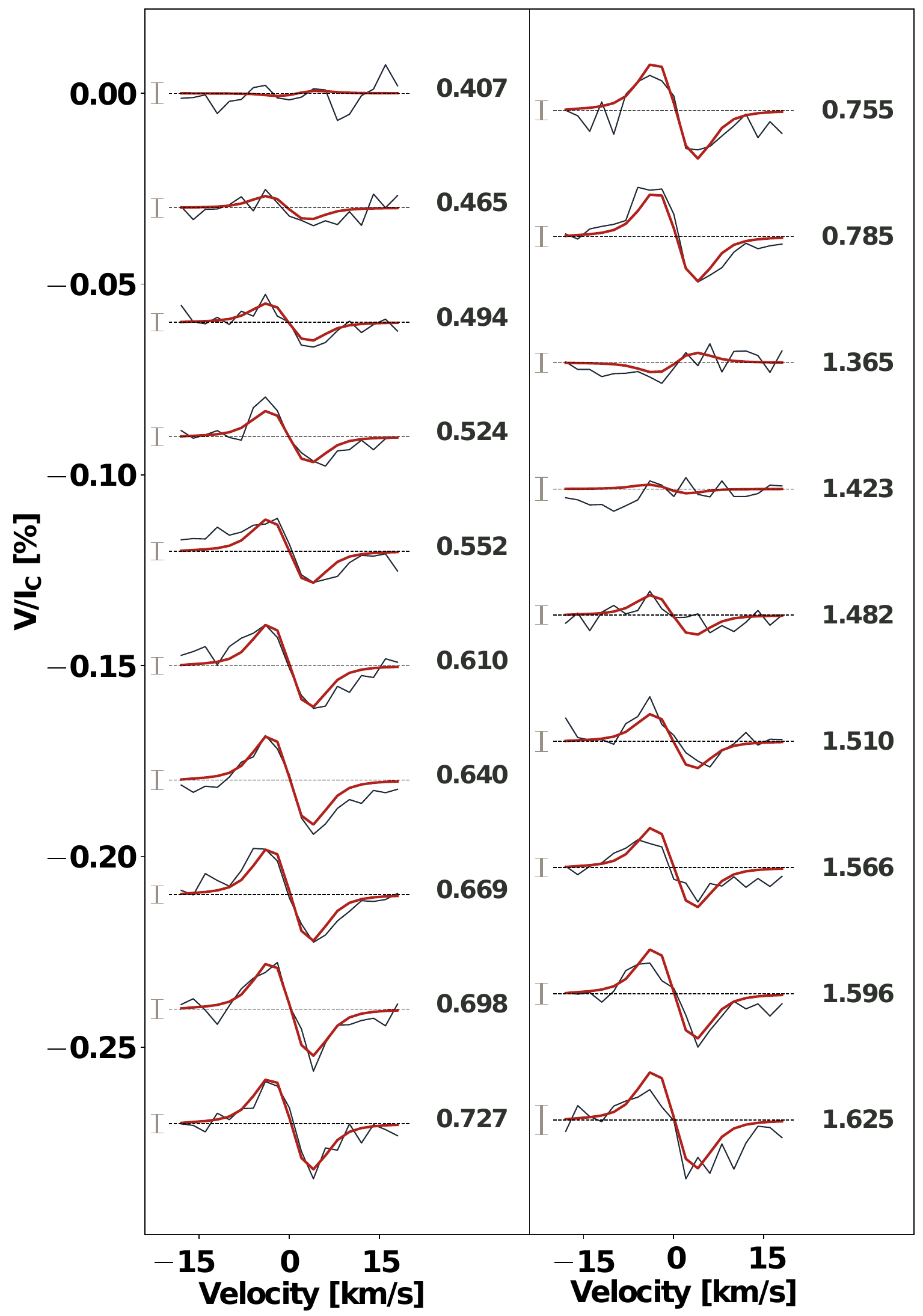}
    \hspace{0.05\linewidth}
    \includegraphics[width=0.4\linewidth]{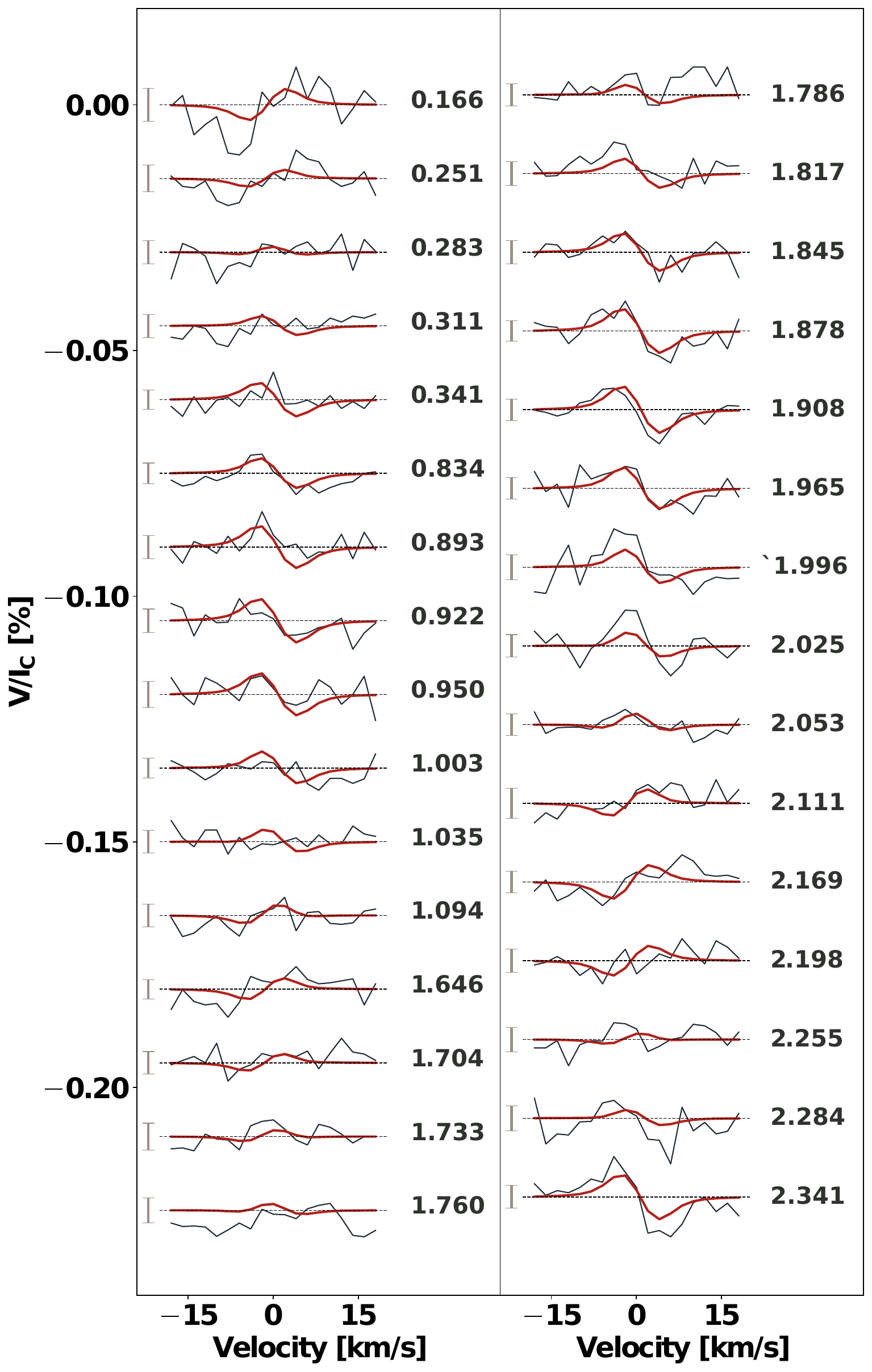}    
    \caption{Same as Figure~\ref{fig:StokesV_fit12}, but for the observations collected in Season~3 (August to September 2021, left-hand panel), and Season 4 
    (November 2021 and January 2022, right-hand panel). Note that we use the same rotation period and reference time as in Figure~\ref{fig:StokesV_fit12} to compute the stellar rotation phases.}
    \label{fig:StokesV_fit34}
\end{figure*}

The best ZDI fit to the time series of Stokes $V$ LSD profiles observed in September--December 2019, August--October 2020, August--September 2021, and November 2021 -- January 2022, are shown in Figure~\ref{fig:StokesV_fit12} and Figure~\ref{fig:StokesV_fit34}.

\FloatBarrier
\section{Activity indicators periodograms}

\begin{figure*}[h!]
\centering
\includegraphics[width=0.99\hsize]{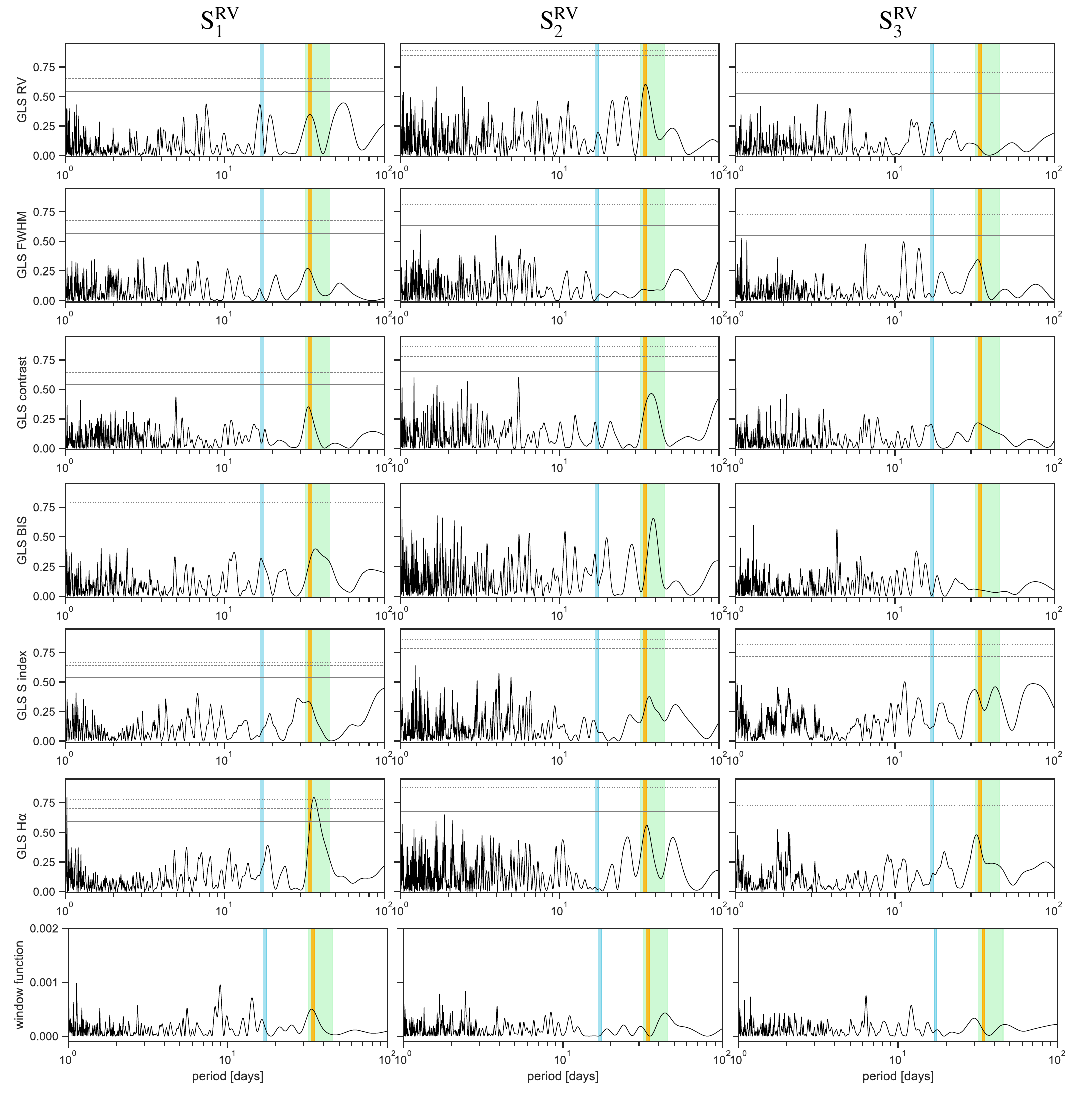}
\caption{Generalized Lomb-Scargle periodograms of the SOPHIE RVs, activity indicators, and window function of the three seasons of RVs observations: S$^{RV}_{1}$, S$^{RV}_{2}$, and S$^{RV}_{3}$. The vertical color lines mark the $P_{\rm rot}$ at 34.4\,d in yellow, the first harmonic at $P_{\rm rot}/2$ in light blue and the time range of differential rotation in light green between $P_{\rm eq} = 32.0$\,d and $P_{\rm pol}=45.5$\, d. The black horizontal lines indicate the false-alarm probability (FAP) levels of 10\% (solid), 1\% (dashed), and 0.1\% (dotted).}
\label{SOPHIE_seasons}
\end{figure*}

\begin{figure*}
\includegraphics[width=\hsize]{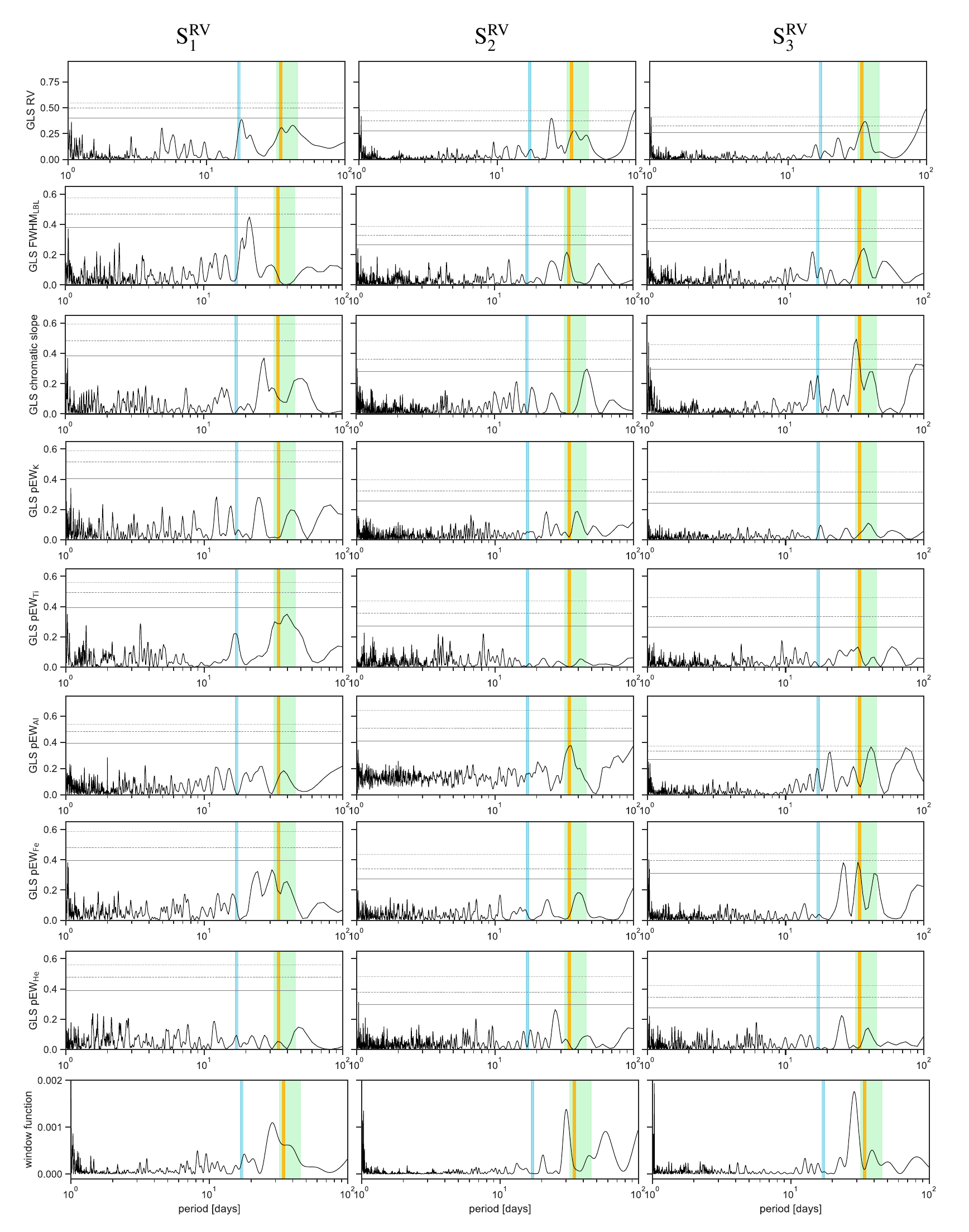}
\caption{Same as Figure~\ref{SOPHIE_seasons} for SPIRou data.}
\label{SPIRou_seasons}
\end{figure*}

The GLS periodograms of all the SOPHIE and SPIRou activity indicators, for each of the three subset of observations S$^{RV}_{1}$, S$^{RV}_{2}$, and S$^{RV}_{3}$, are shown in Figures \ref{SOPHIE_seasons} and \ref{SPIRou_seasons}.
\FloatBarrier

\FloatBarrier
\section{GP model of the activity indicators time series}

\begin{figure*}[h!]
\centering
    \includegraphics[width=0.6\textwidth]{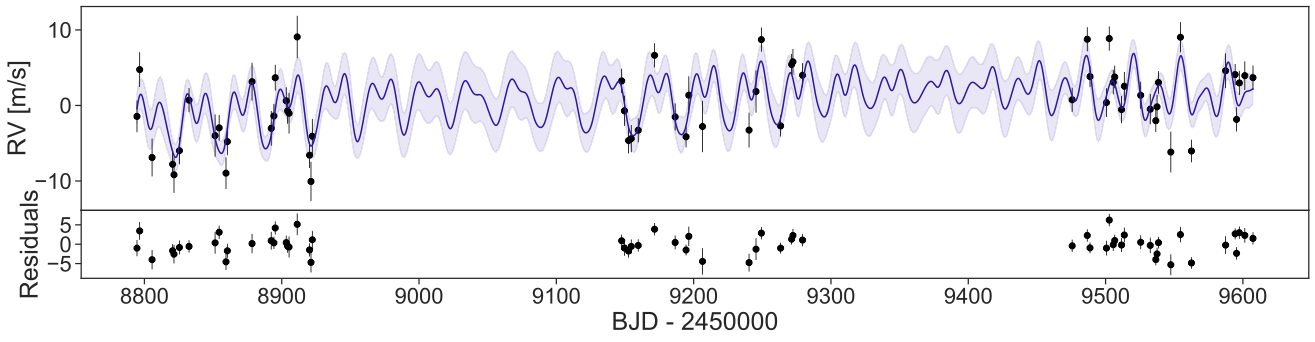} \\
    \includegraphics[width=0.6\textwidth]{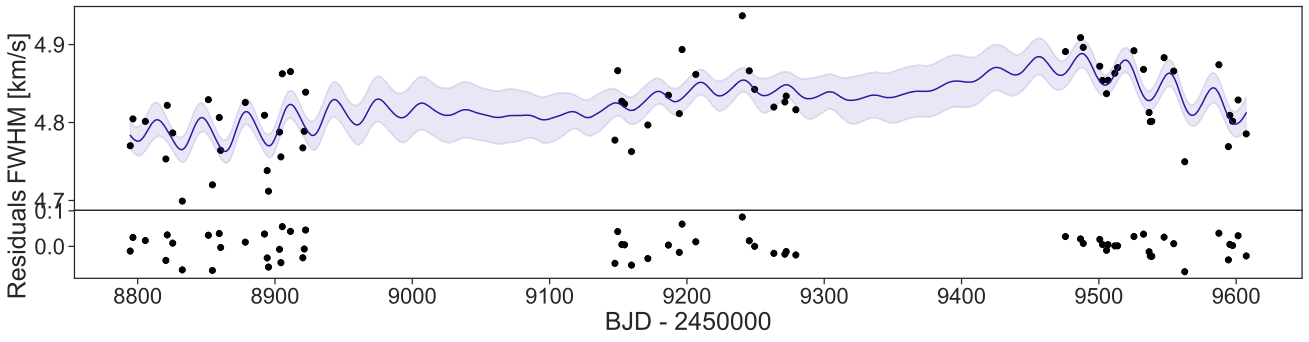} \\
    \includegraphics[width=0.6\textwidth]{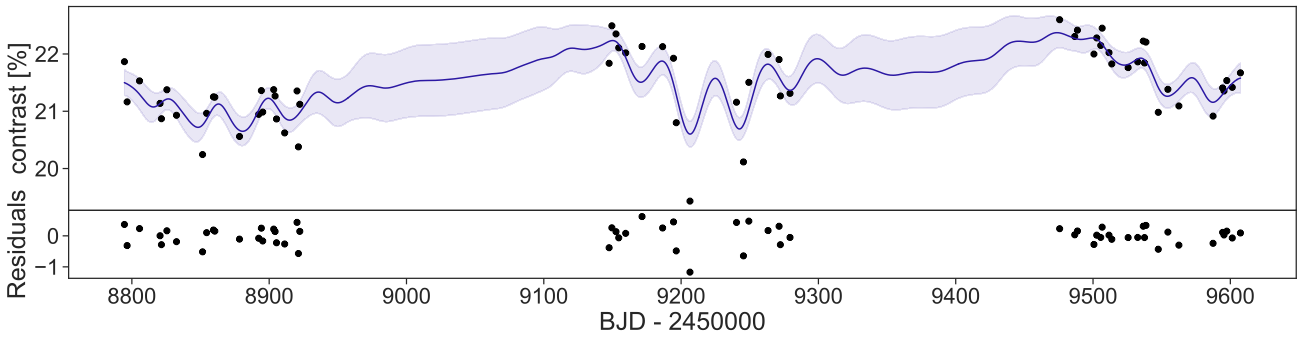} \\
    \includegraphics[width=0.6\textwidth]{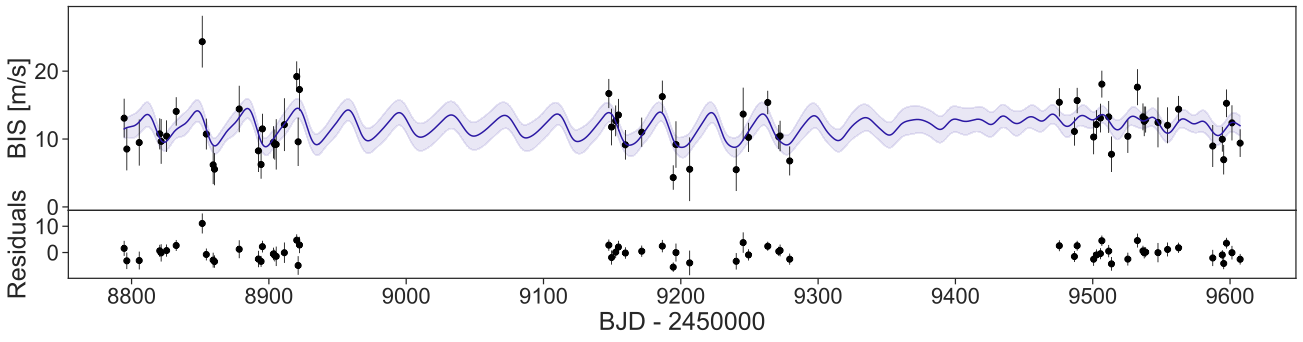} \\
    \includegraphics[width=0.6\textwidth]{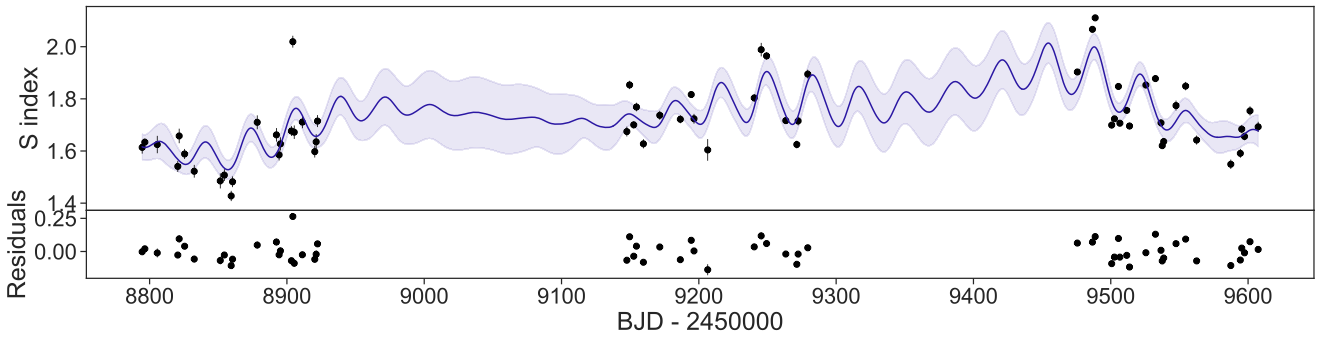} \\
    \includegraphics[width=0.6\textwidth]{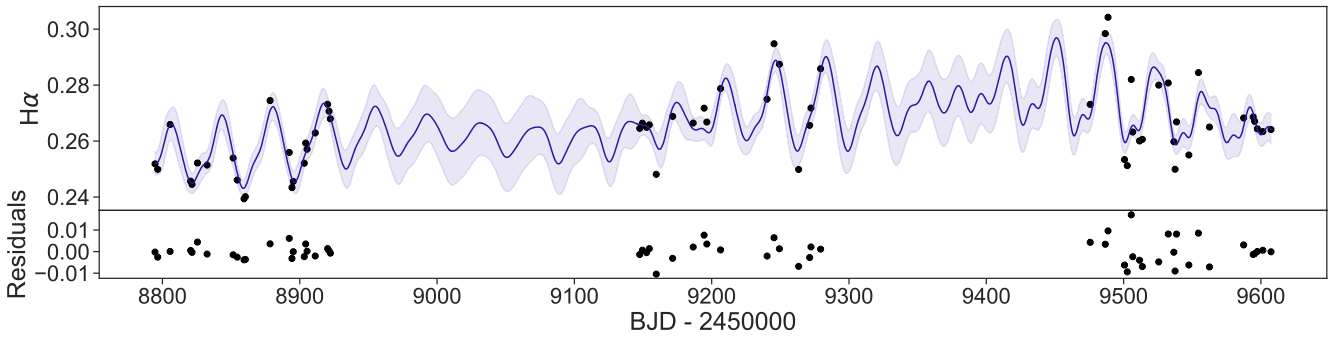} 
    \caption{Time series of the SOPHIE RVs, FWHM, CCF contrast, BIS, S index, and H$\alpha$. The GP model of each time series is shown in color blue with the 3$\sigma$ uncertainties in light-blue.}
\label{fig:GP_SOPHIE_activity}
\end{figure*}

\begin{figure*}[h!]
\centering
    \includegraphics[width=0.55\textwidth]{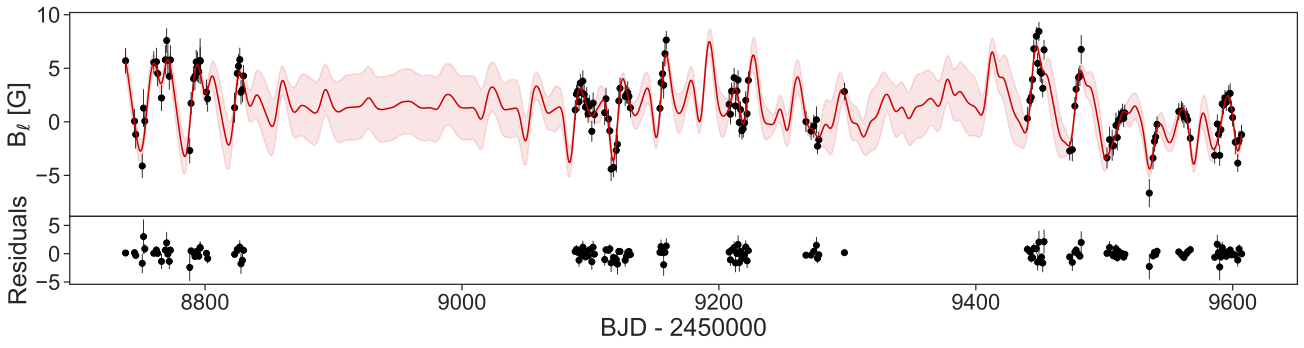}\\
    \includegraphics[width=0.55\textwidth]{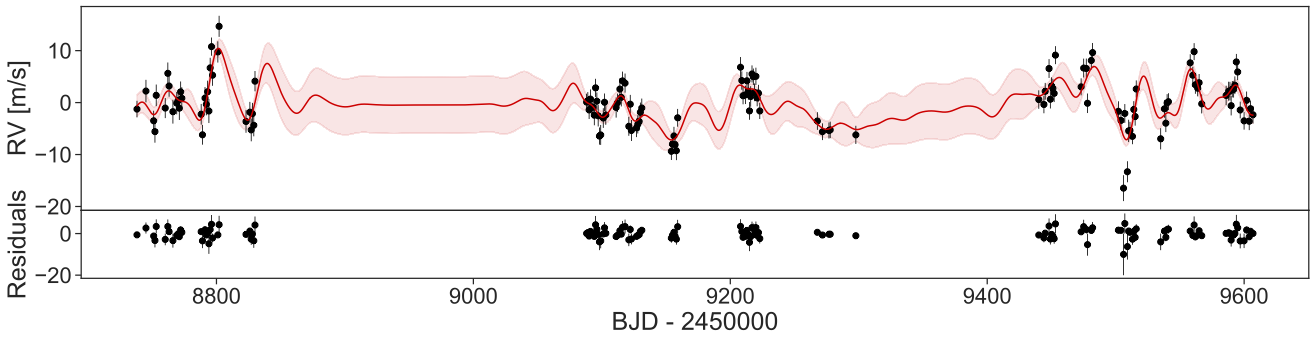} \\
    \includegraphics[width=0.55\textwidth]{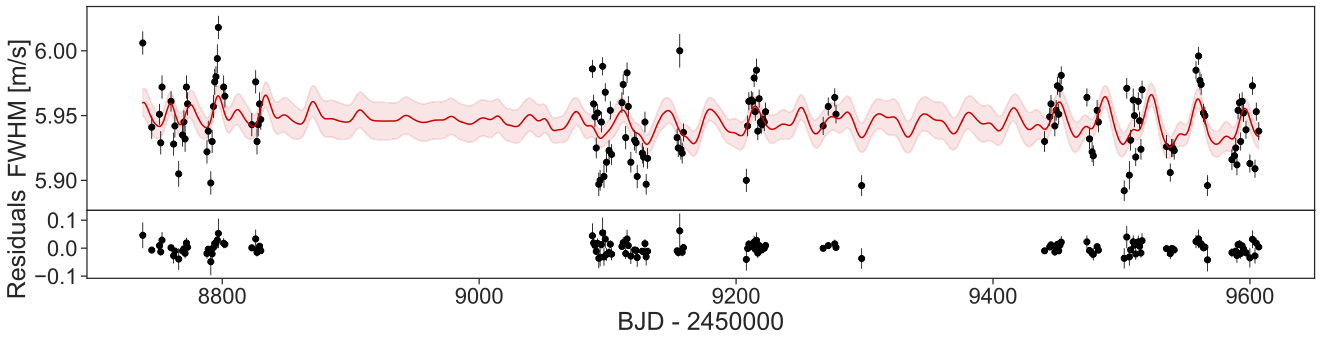} \\
    \includegraphics[width=0.55\textwidth]{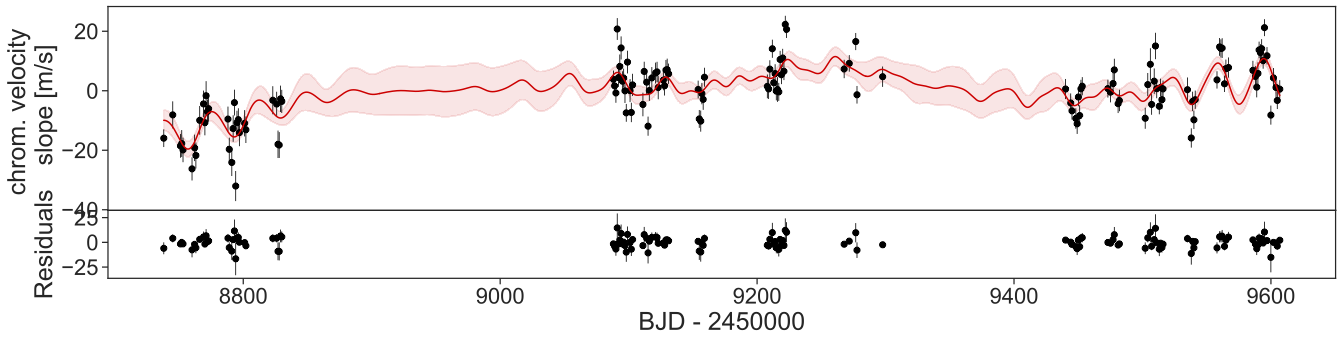} \\
    \includegraphics[width=0.55\textwidth]{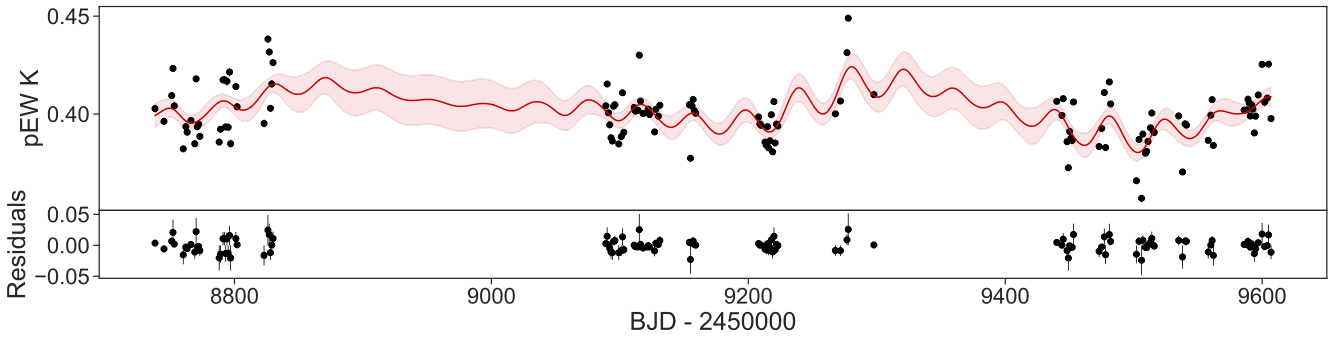} \\
    \includegraphics[width=0.55\textwidth]{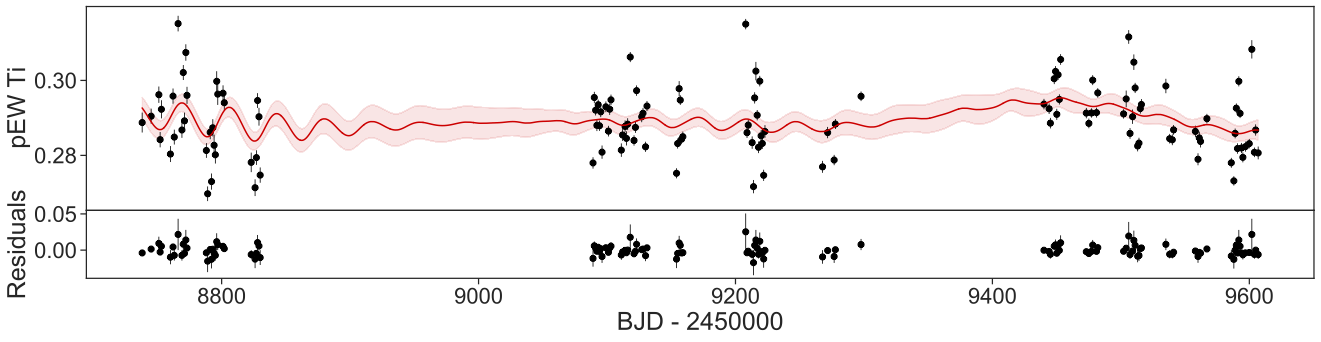} \\
    \includegraphics[width=0.55\textwidth]{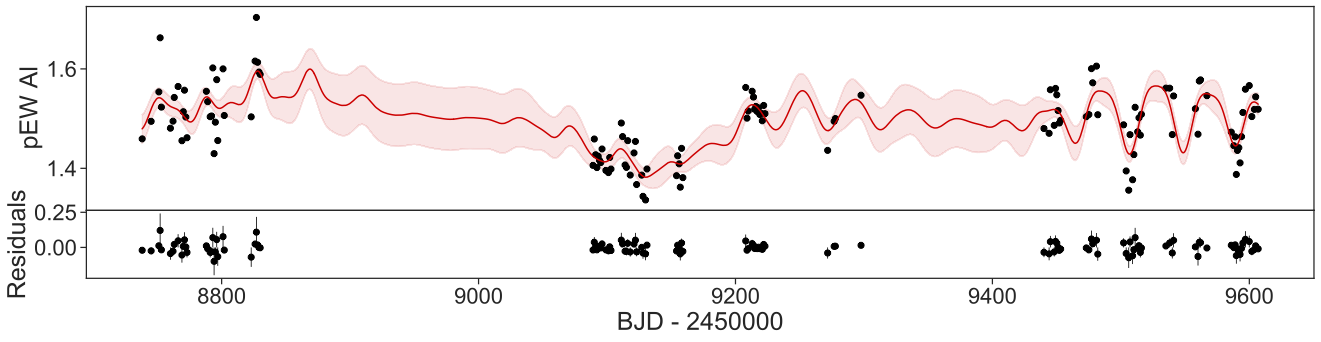} \\
    \includegraphics[width=0.55\textwidth]{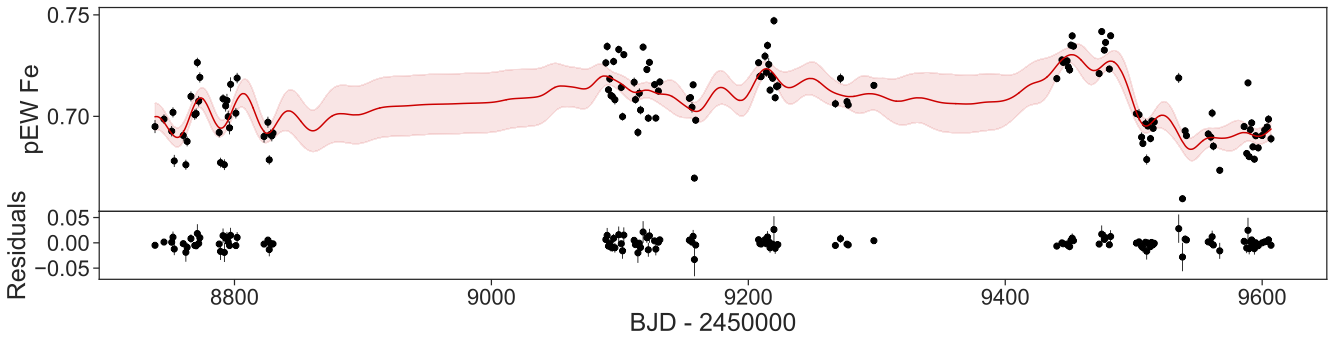} \\
    \includegraphics[width=0.55\textwidth]{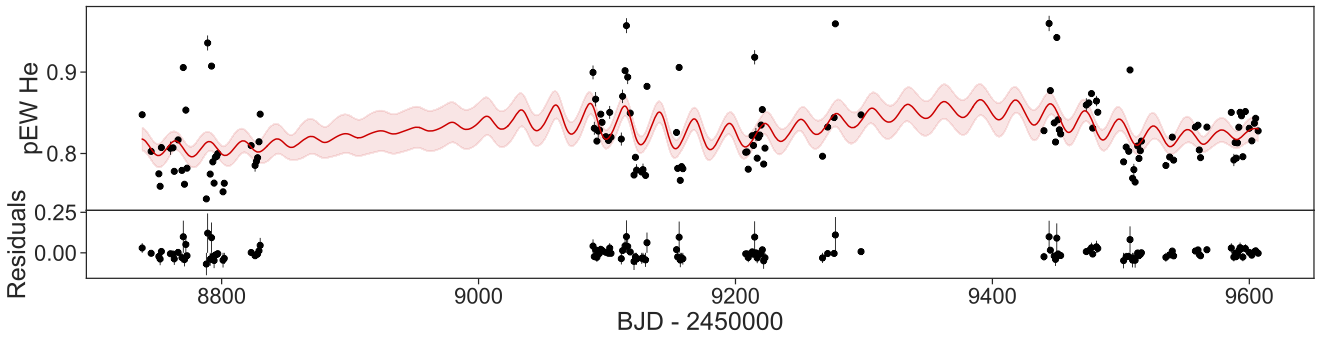} 
    \caption{Time series of the SPIRou longitudinal magnetic field $B_\ell$, RVs, FWHM$_{\rm LBL}$, chromatic velocity slope, and the pEW of K \Romannum{1}, Ti \Romannum{1}, Al \Romannum{1}, Fe \Romannum{1}, and He \Romannum{1}. The GP model of each time series is shown in color red with the 3$\sigma$ uncertainties in ligth-red.}
\label{fig:GP_SPIRou_activity}
\end{figure*}

\begin{figure*}[h!]
\centering
    \includegraphics[width=0.89\textwidth]{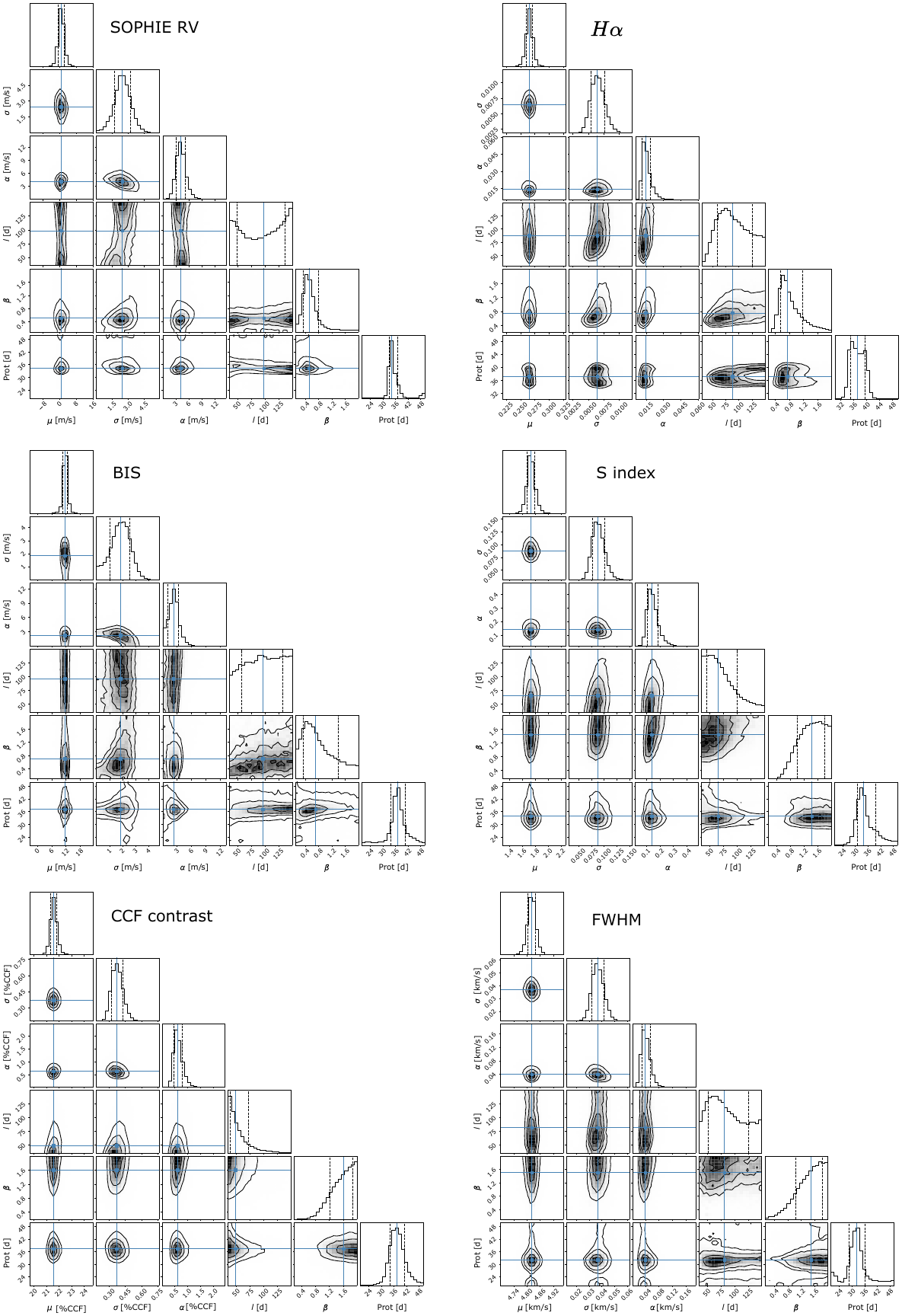}  
    \caption{Corner plot of the posterior distributions of the GP regression applied in the SOPHIE RVs, H$\alpha$, BIS, S index, CCF contrast, and FWHM, independently.}
\label{fig:GP_SOPHIE_activity_corner}
\end{figure*}

\begin{figure*}
\centering
    \includegraphics[width=0.9\textwidth]{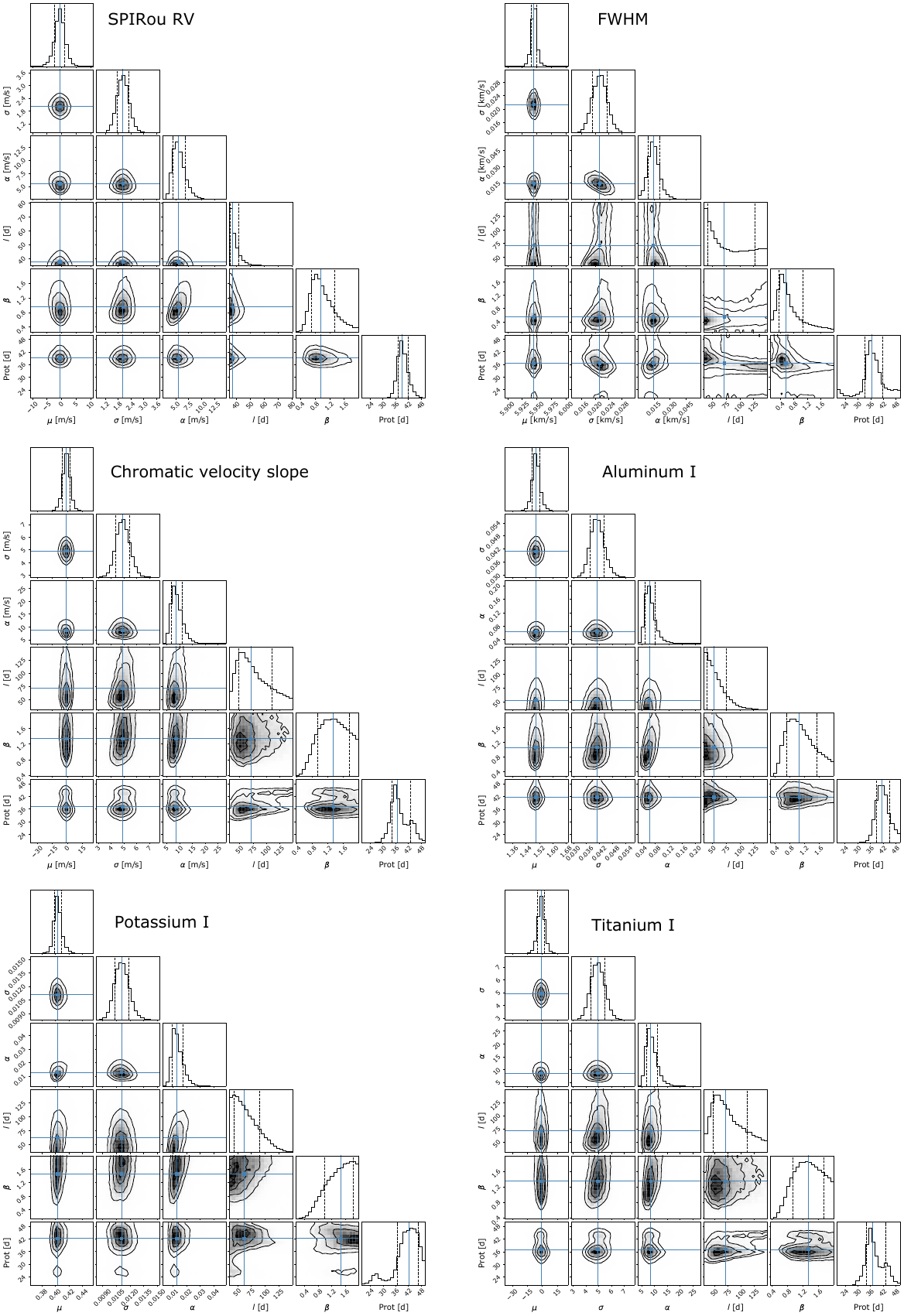}  
    \caption{Corner plot of the posterior distributions of the GP regression applied in the SPIRou RVs, FWHM, chromatic velocity slope, Al \Romannum{1}, Ti \Romannum{1},  independently.}
\label{fig:GP_SPIRou_activity_corner1}
\end{figure*}

\begin{figure*}
\centering
    \includegraphics[width=0.9\textwidth]{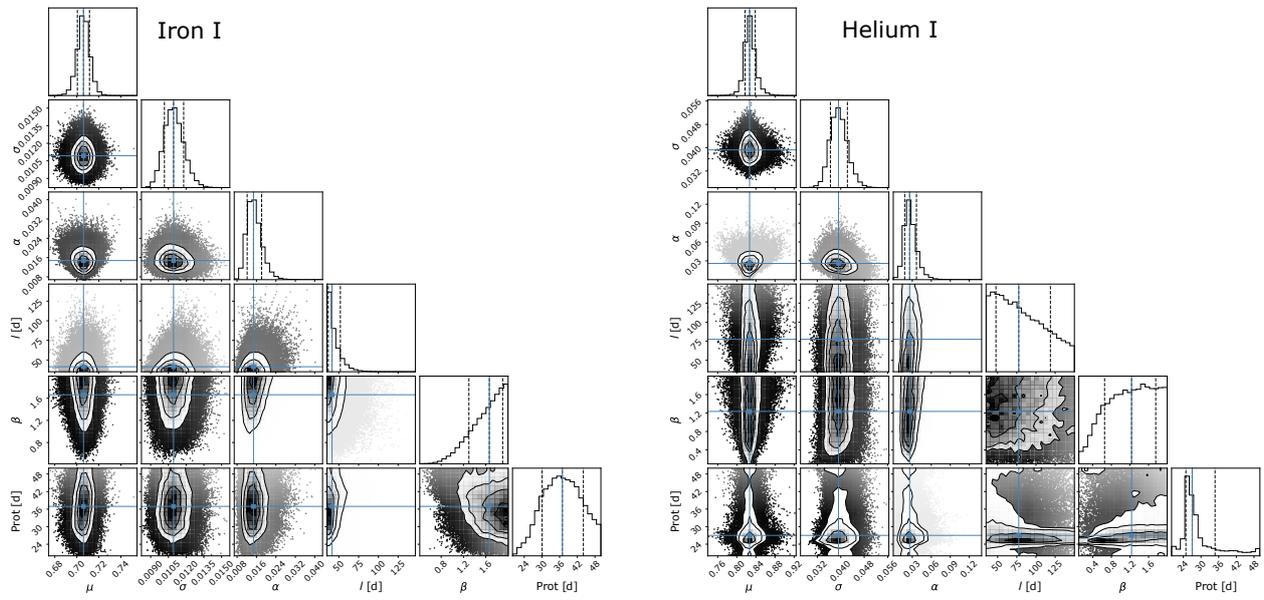}  
    \caption{Same as Figure \ref{fig:GP_SPIRou_activity_corner1}, but for Fe \Romannum{1} and He \Romannum{1}.}
\label{fig:GP_SPIRou_activity_corner2}
\end{figure*}

The best-fit hyper-parameters of the GP model applied in the SOPHIE and SPIRou activity indicators are listed in Table \ref{table:GP_activity}. In Figures \ref{fig:GP_SOPHIE_activity} and \ref{fig:GP_SPIRou_activity} are shown the best GP model for each activity indicators, and in Figures \ref{fig:GP_SOPHIE_activity_corner}, \ref{fig:GP_SPIRou_activity_corner1}, and \ref{fig:GP_SPIRou_activity_corner2} the corner plots of the posterior distributions.

\begin{landscape}
\begin{table}
\footnotesize
\setlength{\tabcolsep}{2.0pt}
\caption{Best-fit hyper-parameters of the GP model using a quasi-periodic kernel in the SOPHIE and SPIRou activity indicators.}         
\label{table:GP_activity}      
\centering
\renewcommand{\arraystretch}{1.5}
\begin{tabular}{ c | ccccc | ccccccc}     
\hline
          &    \multicolumn{5}{c}{SOPHIE Posteriors} & \multicolumn{7}{|c}{SPIRou Posteriors} \\
Parameter & FWHM & contrast & BIS & S index & H$\alpha$ & FWHM$_{\rm LBL}$ & chromatic & pEW K & pEW Ti & pEW Al & pEW Fe & pEW He \\  
           &     &           &   &         &             &                 &  velocity slope &   &  &  &  &  \\
  &  \kms & \% CCF & \ms & & &  \kms & \ms & \AA & \AA & \AA & \AA & \AA \\
\hline                    
   mean value $\mu $             & $4.82\pm0.02$  & $21.6\pm0.3$       & $11.7\pm0.9$ & $1.73\pm0.6$  & $0.265\pm0.005$  & $5.945\pm0.005$ &  $-1.1\pm3.6$   &  $0.403\pm0.005$  & $0.289\pm0.002$ & $1.48\pm0.02$ & $0.71\pm0.01$ & $0.83\pm0.01$\\  
   jitter $\sigma $              & $0.04\pm0.05$  & $0.37\pm0.05$      & $1.8\pm0.8$  & $0.09\pm0.01$ & $0.007\pm0.001$  & $0.021\pm0.002$  &  $4.9\pm0.5$  &  $0.0111\pm0.0008$ & $0.008\pm0.001$ & $0.043\pm0.004$ &  $0.011\pm0.001$  & $0.039\pm0.003$\\ 
   amplitude $\alpha $           &  $0.04\pm0.01$ & $0.63\pm0.16$      & $2.3\pm1.1$  &  $0.14\pm0.4$ &$0.014\pm0.003$  & $0.015\pm0.005$   &  $8.7\pm2.0$  &  $0.013\pm0.004$  & $0.005\pm0.002$ & $0.08\pm0.02$ & $0.015\pm0.003$ & $0.03\pm0.01$\\
   decay time $l$ [d]               &  $82\pm36$   & $48^{+24}_{-10}$ & $96\pm38$    &  $67^{+34}_{-21} $ &$88^{+38}_{-27}$ & $71^{+55}_{-29}$   &  $71^{+40}_{-22}$ &  $62\pm24$  & $81\pm35$ & $51^{+29}_{-12}$ &  $41\pm8$ & $77\pm35$\\
   smoothing factor $\beta$      & $1.5\pm0.4$    & $1.6\pm0.3$        & $0.7\pm0.5$  & $1.4\pm0.4$ &$0.7\pm0.3$      & $0.5\pm0.4$  &  $1.3\pm0.4$ &   $1.5\pm0.4$  & $1.3\pm0.5$ & $0.6\pm0.2$ & $1.7\pm0.3$  &  $1.2\pm0.6$\\
   rotation period $P_{\rm rot}$ [d] & $32.1\pm3.9$ & $37.4\pm3.5$     & $36.7\pm3.5$ &  $33.8\pm4.4$ &$37\pm2.3$       & $36.7\pm4.1$  &  $37.1\pm4.5$  &  $42.2\pm4.9$  &  $37.1\pm6.3$ & $37.9\pm2.1$ &  $36.9\pm7.0$ &  $27.0\pm4.9$\\
   RMS of residuals              & 0.03           & 0.31               & 2.8        & 0.07  & 0.005            & 0.02  & 5.6 & 0.01 & 0.01 & 0.04 & 0.01 & 0.04\\
   $\chi^2$ & 12.1  & 10.7  & 1.3 & 1.8 & 23.4  & 7.1   & 2.7 & 63.6 & 23.1 & 282 & 28.6 & 158.5\\
\hline                  
\end{tabular}
\end{table}
\end{landscape}

\FloatBarrier
\FloatBarrier

\FloatBarrier

\FloatBarrier

\FloatBarrier
\section{Multi-dimensional GP corner plots}

\begin{figure*}[h!]
\centering
    \includegraphics[width=0.99\textwidth]{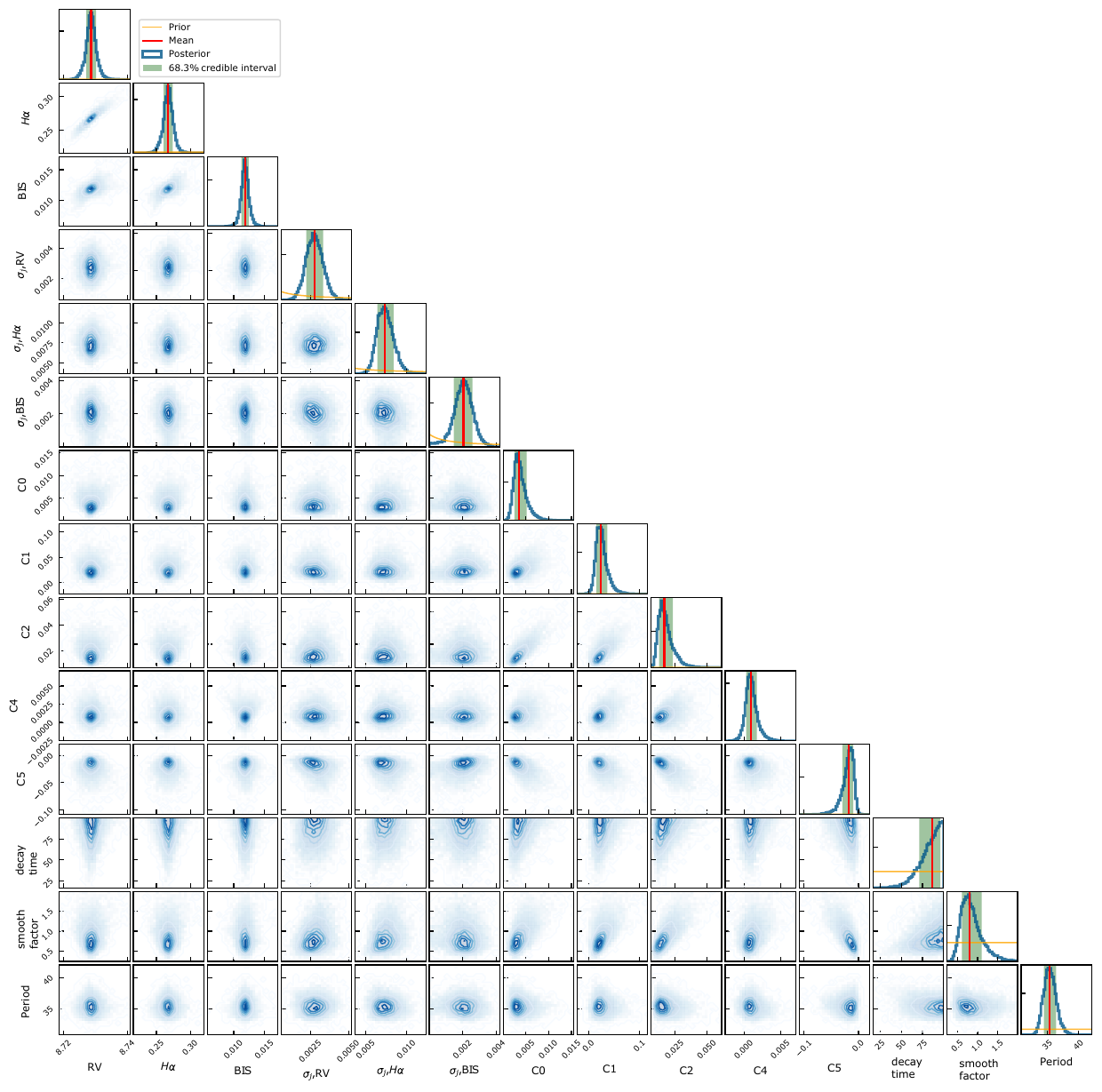}  
    \caption{Corner plot with the posterior distributions of the multi-dimensional GP applied in the SOPHIE RVs, H$\alpha$, and BIS time series.}
\label{fig:multiGP_SOPHIE_corner}
\end{figure*}

\begin{figure*}
\centering
    \includegraphics[width=0.99\textwidth]{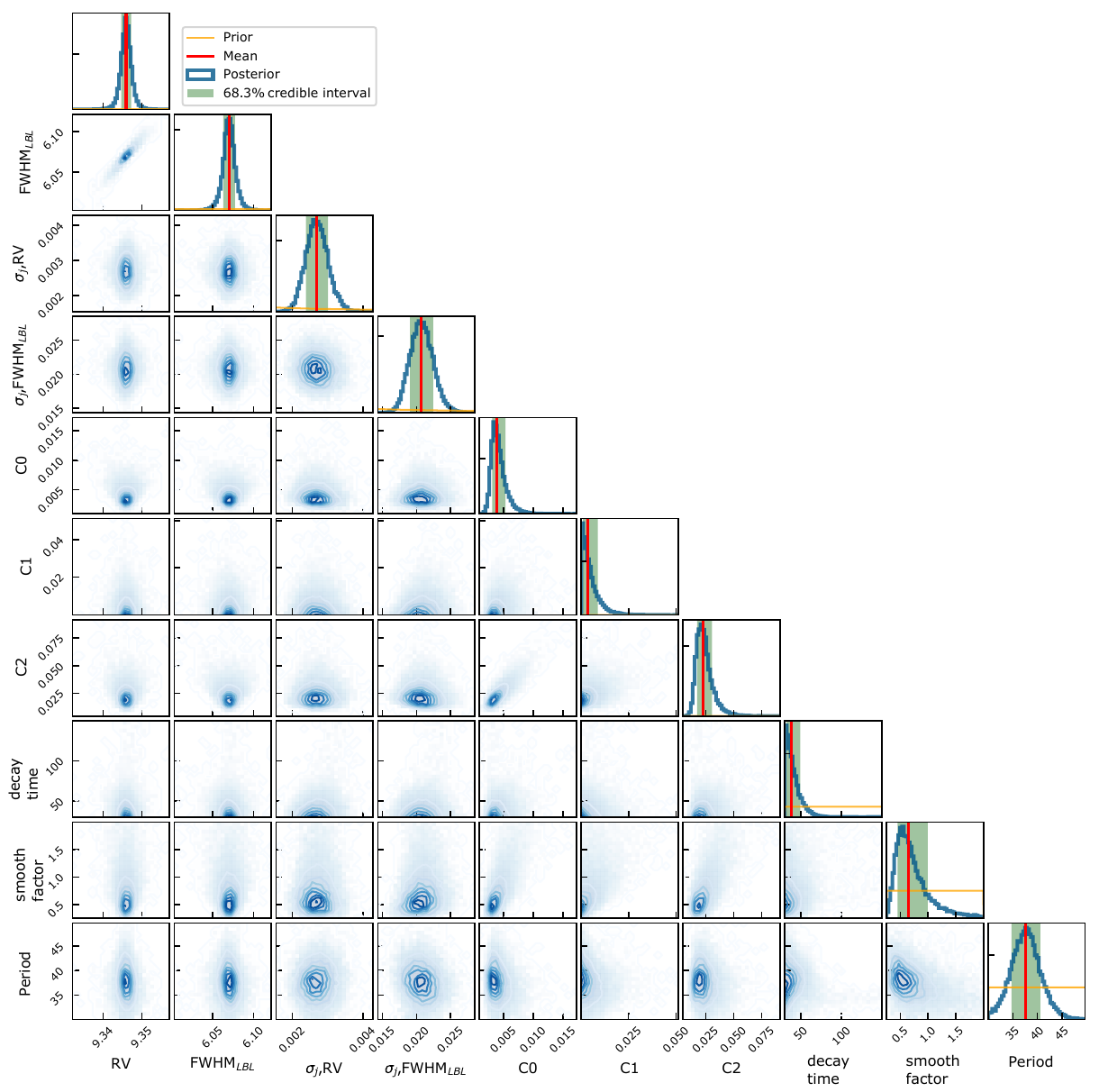}  
    \caption{Corner plot with the posterior distributions of the multi-dimensional GP applied in the SPIRou RVs and FWHM$_{\rm LBL}$ time series.}
\label{fig:multiGP_SPIRou_corner}
\end{figure*}

The posterior distributions of the multi-dimensional GP model applied in the SOPHIE and SPIRou data are shown in Figures \ref{fig:multiGP_SOPHIE_corner} and \ref{fig:multiGP_SPIRou_corner}.
\FloatBarrier
\FloatBarrier


\end{appendix}

\end{document}